\documentclass[twocolumn,aps,superscriptaddress,floatfix]{revtex4}

\usepackage{amsmath}
\usepackage{amssymb}
\usepackage{graphicx}
\usepackage{color}

\newcommand{\be}{\begin{equation}}
\newcommand{\ee}{\end{equation}}
\newcommand{\bea}{\begin{eqnarray}}
\newcommand{\eea}{\end{eqnarray}}
\newcommand{\bse}{\begin{subequations}}
\newcommand{\ese}{\end{subequations}}
\newcommand{\bma}{${\rm BaMn_2As_2}$}
\newcommand{\cma}{${\rm CaMn_2As_2}$}
\newcommand{\sma}{${\rm SrMn_2As_2}$}
\newcommand{\tcs}{${\rm ThCr_2Si_2}$}

\newcommand{\cca}{CaCo$_{2-y}$As$_2$}
\newcommand{\sca}{${\rm SrCo_2As_2}$}
\newcommand{\bca}{${\rm BaCo_2As_2}$}
\newcommand{\eca}{EuCo$_{2-y}$As$_2$}
\newcommand{\ecp}{${\rm EuCo_2P_2}$}

\newcommand{\ecaa}{${\rm EuCo_2As_2}$}

\begin{document}

\title{Enhanced moments of Eu in single crystals of the \\metallic helical antiferromagnet {\bf EuCo$_{2-y}$As$_2$}}

\author{N. S. Sangeetha}
\affiliation{Ames Laboratory, Iowa State University, Ames, Iowa 50011, USA}
\author{V. K. Anand}
\altaffiliation{Present address: Helmholtz-Zentrum Berlin f\"{u}r Materialien und Energie GmbH, Hahn-Meitner Platz 1, D-14109 Berlin, Germany}
\affiliation{Ames Laboratory, Iowa State University, Ames, Iowa 50011, USA}
\author{Eduardo Cuervo-Reyes}
\affiliation{Swiss Federal Laboratories for Materials Science and Technology (Empa), \"{U}berlandstrasse 129, CH-8600 D{\"u}bendorf, Switzerland}
\affiliation{Swiss Federal Institute of Technology (ETH), Vladimir-Prelog-Weg 1, CH-8093 Z{\"u}rich, Switzerland}
\altaffiliation{Present address: ABB Corporate Research Ltd., CH-5405, Baden-Daettwil, Switzerland}
\author{V. Smetana}
\affiliation{Department of Materials and Environmental Chemistry, Stockholm University, Svante Arrhenius v\"ag 16 C, 106 91 Stockholm, Sweden}
\author{A.-V. Mudring}
\affiliation{Department of Materials and Environmental Chemistry, Stockholm University, Svante Arrhenius v\"ag 16 C, 106 91 Stockholm, Sweden}
\author{D. C. Johnston}
\affiliation{Ames Laboratory, Iowa State University, Ames, Iowa 50011, USA}
\affiliation{Department of Physics and Astronomy, Iowa State University, Ames, Iowa 50011}

\date{\today}

\begin{abstract}

The compound \eca\ with the tetragonal \tcs\ structure is known to contain Eu$^{+2}$ ions with spin $S=7/2$ that order below a temperature $T_{\rm N} \approx 47$~K into an antiferromagnetic (AFM) proper helical structure with the ordered moments aligned in the tetragonal $ab$~plane, perpendicular to the helix axis along the $c$~axis, with no contribution from the Co atoms.  Here we carry out a detailed investigation of the properties of single crystals.  We consistently find about 5\% vacancies on the Co site from energy-dispersive x-ray analysis and x-ray diffraction refinements.   Enhanced ordered and effective moments of the Eu spins are found in most of our crystals.  Electronic structure calculations indicate that the enhanced moments arise from polarization of the $d$~bands, as occurs in ferromagnetic Gd metal. Electrical resistivity measurements indicate metallic behavior.  The low-field in-plane magnetic susceptibilities $\chi_{ab}(T<T_{\rm N})$ for several crystals are reported that are fitted well by unified molecular field theory (MFT), and the Eu--Eu exchange interactions $J_{ij}$ are extracted from the fits.  High-field magnetization~$M$ data for magnetic fields $H\parallel ab$ reveal what appears to be a first-order spin-flop transition followed at higher field by a second-order metamagnetic transition of unknown origin, and then by another second-order transition to the paramagnetic (PM) state.  For $H\parallel c$, the magnetization shows only a second-order transition from the canted AFM to the PM state, as expected.  The critical fields for the AFM to PM transition are in approximate agreement with the predictions of MFT\@.  Heat capacity $C_{\rm p}$ measurements in zero and high~$H$ are reported.  Phase diagrams for $H\parallel c$ and $H\parallel ab$ versus~$T$ are constructed from the high-field $M(H,T)$ and $C_{\rm p}(H,T)$ measurements.  The magnetic part $C_{\rm mag}(T,H=0)$ of $C_{\rm p}(T,H=0)$ is extracted and is fitted rather well below~$T_{\rm N}$ by MFT, although dynamic short-range AFM order is apparent in $C_{\rm mag}(T)$ up to about 70~K, where the molar entropy attains its high-$T$ limit of $R\ln8$.  

\end{abstract}

\maketitle

\section{\label{Sec:Intro} Introduction}

Many studies of iron-based layered pnictides and chalcogenides have appeared due to their unique lattice, electronic, magnetic and superconducting properties \cite{{Johnston2010},{Stewart2011},{Scalapino2012}, {Dagotto2013},{Fernandes2014},{Hosono2015},{Dai2015}, {Inosov2016},{Si2016}}. An important family of these materials consists of doped and undoped compounds $A$Fe$_2$As$_2$ ($A$ = Ca, Sr, Ba, Eu) with the body-centered tetragonal \tcs\ structure  with space group $I4/mmm$ (122-type compounds). Searches for novel physical properties in various 122-type compounds with other transition metals replacing Fe have been carried out, such as for Mn \cite{{An2009}, {Singh2009}, {Singh2009b}, {Johnston2011}, {Antal2012}, {Calder2014}, {Zhang2016}, Sangeetha2016a, Das2017, Sangeetha2017} and Cr \cite{{DJSingh2009}, {Filsinger2017}, {Nayak2017}, {Richard2017}, {Paramanik2017}, {Nandi2016}, {Pfisterer1980}, {Pfisterer1983}}.

Here we are concerned with $A$Co$_2$As$_2$ and $A$Co$_2$P$_2$ compounds (Co122 systems) with the \tcs\ structure that have also attracted much interest due to their rich magnetic behaviors, where the electronic states of the CoAs and CoP layers are sensitive to the crystal structure. By forming As--As and P--P bonds along the $c$~axis, their crystal structures can  collapse along this axis, resulting in the so-called collapsed-tetragonal (cT) structure which is to be distinguished from the uncollapsed-tetragonal (ucT) structure.  In contrast to the Fe122 compounds that exhibit a magnetic to nonmagnetic transition under pressure coincident with a ucT to cT transition, the Co-based compounds behave in the opposite manner, with the ambient-pressure ucT compounds being paramagnetic and the cT compounds exhibiting magnetic ordering \cite{Anand2012}.  For example, \cca\ has a cT structure at ambient pressure and manifests itinerant A-type antiferromagnetic (AFM) ordering with the ordered moments aligned along the $c$~axis \cite{Anand2014, Quirinale2013}, whereas the 122-type \sca\ and \bca\ compounds have ucT structures with no long-range magnetic ordering \cite{{Pandey2013},{Anand2014b}}. Inelastic neutron scattering and NMR studies on \sca\ have revealed  strong stripe-type AFM correlations at high energies whereas NMR measurements reveal strong FM correlations at low energies \cite {{Jayasekara2015},{Wiecki2015}}. On the other hand, the system SrCo$_2$(Ge$_{1-x}$P$_x)_2$ develops weak itinerant ferromagnetism during the course of the dimer breaking, and a quantum critical point (QCP) is observed at the onset of the FM phase, although both $\rm SrCo_2P_2$ (ucT) and $\rm SrCo_2Ge_2$ (cT) are paramagnetic (PM) \cite{Jia2011}. From first-principles calculations, it was shown that the degree of As-As covalent bonding in ${\rm CaFe_2As_2}$ and the magnitude of the spin on the Fe atoms are inversely related \cite{Yildirim2009, Yildirim2009b}.  Similarly, the magnetic properties of the cobalt pnictides were correlated with changes in the formal Co charge as determined by the estimated degree of P-P covalent bonding along the $c$~axis \cite{Reehuis1998}.

\ecp\ is an interesting ucT compound in the Co122 family. It shows AFM ordering of the Eu$^{2+}$ spins $S=7/2$ below $T_{\rm N}=66$~K \cite{Morsen1988}. Neutron diffraction studies demonstrated that the AFM structure is a planar helix with the Eu ordered moments aligned in the $ab$ plane of the tetragonal structure, and with the helix axis being the $c$~axis \cite{Reehuis1992}. This compound shows a pressure-induced first-order ucT to cT transition at $\approx 3$~GPa \cite{Huhnt1997} associated with the valence change of Eu from $\rm Eu^{2+}$ to nonmagnetic $\rm Eu^{3+}$ together with the emergence of itinerant 3$d$ magnetism in the Co sublattice, which orders AFM at $T\rm_{N}^{Co}$ = 260 K \cite{Chefki1998}. We showed that \ecp\ is a textbook example of a noncollinear helical antiferromagnet for which the thermodynamic properties in the antiferromagnetic state are well described by our unified molecular field theory (MFT) \cite{Sangeetha2016}.

\ecaa\ also has the ucT 122-type structure and hence is isostructural and isoelectronic to \ecp\ \cite{Tan2016, Marchand1978}.  It exhibits AFM ordering of the Eu$^{+2}$ spins-7/2 at $T_{\rm N} = 47$~K \cite{Raffius1993, Ballinger2012}. Neutron diffraction measurements showed that the AFM structure is the  same coplanar helical structure as in \ecp, with no participation by Co moments \cite{Tan2016}.  Here the reported helix propagation vector is ${\bf k}= (0, 0, 0.79)(2\pi/c)$ \cite{Tan2016}, very similar to that of \ecp\ which is ${\bf k}= (0, 0, 0.85)(2\pi/c)$ \cite{Reehuis1992}. The $c/a$ ratios of \ecp\ (3.01) and \ecaa\ (2.93) are also similar and both indicate a ucT structure. High-pressure measurements on \ecaa\ showed a continuous tetragonal to collapsed tetragonal crossover at a pressure $p\approx 5$~GPa \cite{Bishop2010} and a change in the associated valence state of Eu, achieving the average oxidation state of Eu$^{+2.25}$ at 12.6~GPa. As a result, ferromagnetic (FM) ordering arises from both Eu and Co moments with a Curie temperature $T_{\rm C} = 125$~K, which is confirmed by x-ray magnetic circular dichroism measurements and electronic structure calculations.

One reason for carrying out the present detailed study of \eca\ is that the reported effective magnetic moment in the paramagnetic (PM) state $\mu\rm_{eff} \approx 8.22~\mu\rm_B$/Eu is significantly larger than the value of $\mu\rm_{eff}$ = 7.94 $\mu\rm_B$ expected for Eu$^{2+}$ \cite{Marchand1978} (see also Table~\ref{Tab:ChemAnal} below). Normally, the effective and ordered moments of Eu$^{+2}$ and Gd$^{+3}$ are rather robust due to the spin-only electronic configurations of these $S=7/2$  ions (orbital angular momentum $L=0$).  The questions we wanted to address were how repeatable the large $\mu_{\rm eff}$ is in different samples, how it comes about, and to see if it correlates with other properties of the material.  In addition, we wanted to test our unified molecular field theory to fit the magnetic and thermal properties below~$T_{\rm N}$ for another helical AFM to complement our earlier studies of \ecp~\cite{Sangeetha2016}.  We grew single crystals of \eca\ with two different fluxes and report their properties.  We find that there is a rather large range of $\mu_{\rm eff}$ values as well as of low-temperature ordered (saturation) moments $\mu_{\rm sat}$ of the Eu spins in different crystals.  As in \cca\ \cite{Anand2014, Quirinale2013}, we also find a significant ($\sim 5\%$) vacancy concentration on the Co sites in most of our \eca\ crystals. 

The experimental details are given in Sec.~\ref{Sec:ExpDetails}.  In Sec.~\ref{Sec:XtalStruct} the crystal structure and composition analyses are presented for six crystals for which the physical properties are later studied in detail.  Our magnetic susceptibility~$\chi$ versus temperature~$T$ data and magnetization versus field $M(H)$ isotherms for the crystals are presented in Sec.~\ref{Sec:ChiMH}, where we find enhancements in both $\mu_{\rm eff}$ and $\mu_{\rm sat}$ compared to expectation for Eu$^{+2}$ spins with $S=7/2$ and spectroscopic splitting factor~$g=2$.  We also obtain an estimate of the amount of anisotropy in the system and fit the in-plane $\chi_{ab}(T)$ at temperatures~$T$ less than the AFM ordering temperature $T_{\rm N}$ by MFT\@.

Our zero-field and high-field heat capacity $C_{\rm p}(H,T)$ measurements are presented in Sec.~\ref{Cp}, where the magnetic contribution $C_{\rm mag}(T,H=0)$ is extracted and found to agree rather well with the prediction of MFT for $S=7/2$ at $T\leq T_{\rm N}$.  However, dynamic short-range AFM  ordering is found from $T_{\rm N}\approx 42$~K up to about 70~K, which is not accounted for by MFT\@.  The molar magnetic entropy $S_{\rm mag}$ is found to agree with expection for Eu spins $S=7/2$ at high~$T\gtrsim70$~K, $R\ln(2S+1)$, where $R$ is the molar gas constant.  From the high-field $C_{\rm p}(H,T)$ we extract $T_{\rm N}(H)$ for $H\parallel c$ and obtain a good fit by MFT\@.  Using the high-field data from the $M(H)$ and $C_{\rm p}(T)$ measurements, the phase diagrams in the $H\parallel c$ and $H\parallel ab$ versus $T$ planes are constructed for two different crystals in Sec.~\ref{Sec:PhaseDiags}.  Electrical resistivity data for currents in the $ab$~plane are presented in Sec.~\ref{Sec:Res} together with an analysis of these data in terms of the generic electron-electron scattering model at low~$T$ and the Bloch-Gr\"uneisen, parallel-resistor, and $s$-$d$ scattering models at higher $T$\@.

Our total-energy and electronic-structure calculations are presented in Sec.~\ref{Sec:ElecStruct}.  We find that the Eu spins ferromagnetically polarize the spins of the electrons deriving from the Co $3d$ $t_{2g}$ states near the Fermi level by an amount consistent with the observed enhancement of the Eu moments.  The calculations also indicate that the Co atoms make no contribution to the helical structure, again consistent with experiment.  In Sec.~\ref{Sec:HeisExchInts} we extract the Heisenberg exchange interactions $J_{ij}$ from the prevously-presented MFT fit to the  $\chi_{ab}(T\leq T_{\rm N})$ data.    A summary of our results is given in Sec.~\ref{Summary}.

\section{\label{Sec:ExpDetails} Experimental Details}

Single crystals of EuCo$_2$As$_2$ were grown in Sn flux and CoAs flux. The purity and sources of the elements used were Eu (Ames Lab), and Co (99.998\%), As (99.999 99\%) and Sn (99.9999\%) from Alfa Aesar. For some crystal growths, the Co powder was additionally heated under a flow of H$_2$ gas under a pressure of 12~bar at a temperature of 324~$^{\circ}$C for 12~h to remove possible surface oxidation. At this H$_2$~pressure and temperature, negligible H is absorbed by the Co \cite{Fukai2006}.  Single crystals were grown in both Sn flux and CoAs flux using both H$_2$-treated and as-received Co powder.

For Sn-flux growth, the starting materials were mixed in the molar ratio Eu:Co:As:Sn~=~1.05:2:2:15. Excess Eu was required in order to obtain crystals without impurity phases occluded on or embedded within the crystals. The mixture was placed in an alumina crucible and then sealed  in a silica tube under high-purity argon gas.  After prereacting the elements at 600~$^{\circ}$C for 6~h, the mixtures were placed in a box furnace and heated to 1050~$^{\circ}$C at a rate of 50~$^{\circ}$C/h, held there for 20~h, and then cooled to 600~$^{\circ}$C at a rate of 4~$^{\circ}$C/h. At this temperature the molten Sn flux was decanted using a centrifuge. Shiny platelike crystals of area 4--80~mm$^2$ by $\approx 0.4$~mm thick were obtained. 

For CoAs-flux growth, a mixture of Eu metal and prereacted CoAs powder taken in the molar ratio Eu:CoAs = 1:4 which was placed in an alumina crucible and then sealed in a quartz tube under high purity argon gas. The tube assembly was placed in a box furnace and heated to 1300~$^{\circ}$C  at a rate of 50~$^{\circ}$C/h, held there for 15~h, and then cooled to 1180~$^{\circ}$C at a rate of 6~$^{\circ}$C/h. At this temperature the excess CoAs flux was decanted using a centrifuge. For this crystal-growth method shiny platelike crystals of size 4--40 mm$^2$ by 0.3--0.4~mm thick were obtained.  

The phase purity and chemical composition of the \ecaa\ crystals were checked using energy dispersive x-ray (EDX) semiquantitative chemical analysis attachment to a JEOL scanning electron microscope (SEM). SEM scans were taken on  cleaved surfaces of the crystals which verified the single-phase nature of the crystals. The compositions of each side of a platelike crystal was measured at six or seven positions on each face, and the results were averaged. The EDX composition analysis revealed the presence of vacancies on the Co-site and an absense of Sn incorporated into the bulk of the crystals. The EDX data also showed no evidence for oxygen in any of the crystals.  We selected six crystals having different Co-site occupancies for further investigations. 

Single-crystal X-ray diffraction (XRD) measurements were performed at room temperature on a Bruker D8 Venture diffractometer operating at 50~kV and 1~mA equipped with a Photon 100 CMOS detector, a flat graphite monochromator and a Mo~K$\alpha$ I$\mu$S microfocus source ($\lambda = 0.71073$~\AA). The raw frame data were collected using the Bruker APEX3 program \cite{APEX2015}, while the frames were integrated with the Bruker SAINT software package \cite{SAINT2015} using a narrow-frame algorithm for  integration of the data and were corrected for absorption effects using the multiscan method (SADABS) \cite{Krause2015}. The occupancies of the  Co atomic sites were refined assuming random occupancy of the Co sites and assuming complete occupancy of the Eu and As sites.  The atomic thermal factors were refined anisotropically.  Initial models of the crystal structures were first obtained with the program SHELXT-2014 \cite{Sheldrick2015A} and refined using the program SHELXL-2014 \cite{Sheldrick2015C} within the APEX3 software package.

Magnetization data were obtained using a Quantum Design, Inc., magnetic properties measurement system (MPMS) and a vibrating sample magnetometer in a Quantum Design, Inc., physical properties measurement system (PPMS) for high-field measurements up to 14~T, where 1~T~$\equiv10^{4}$~Oe. The PPMS was  used for $C_{{\rm p}}(T)$ and $\rho(T)$ measurements. The $C_{{\rm p}}(T)$ was measured by the relaxation method and the $\rho(T)$  using the standard four-probe ac technique.

\section{\label{Sec:XtalStruct} Crystal Structures and Compositions}

The chemical compositions and crystallographic data are presented in Table~\ref{Tab:ChemAnal} for six crystals of \eca\ grown under different conditions with different Co vacancy concentrations as determined above, which are labeled \#1 to \#6, respectively.  The chemical compositions obtained from the EDX and single crystal XRD analyses for these six crystals of \ecaa\ are also listed in Table~\ref{Tab:ChemAnal} in comparison with the previous studies on this compound \cite{{Marchand1978},{Raffius1993},{Bishop2010},{Tan2016}}. The physical property measurements reported in this paper were carried out on these six crystals.

\begin{table*}
\caption{\label{Tab:ChemAnal}  The compositions of our six \eca\ single crystals, together with the error bars on the Co concentrations obtained from the combined EDX and XRD data, in comparison with previous studies on this compound. Also listed are crystallographic data for the single crystals at room temperature, including the fractional $c$-axis position $z_{\rm As}$ of the As site, the tetragonal lattice parameters $a$ and~$c$, the unit cell volume $V_{\rm cell}$ containing two formula units of \eca, and the $c/a$ ratio.  The AFM ordering temperature $T_{\rm{N}}$ are also shown.  The listed values of the effective moment $\mu_{\rm eff}$ obtained from the Curie constant in the Curie-Weiss law are averages of the $c$-axis and $ab$-plane values (see Table~\ref{Tab:CuriFit} below).  Most values are larger than the value obtained for $S=7/2$ and $g=2$, which is $\mu_{\rm eff} = 7.94~\mu_{\rm B}$/Eu.  The present work is denoted by PW.  Data from the literature are also shown.}

\begin{ruledtabular}
\begin{tabular}{lcllllccc}

Sample,								& $z_{\rm As}$ & $~~~a$ 		& $~~~~c$		& $~~V_{\rm cell}$ & $~~~c/a$ & $T_{\rm{N}}$ & $\mu_{\rm {eff}}$		&  Ref. 			\\
Composition							& 			& ~~(\AA)		& ~~~(\AA)		& ~~(\AA$^3$) 	&  		& (K) 		& $(\mu_{\rm B}$/Eu)	&				\\
\hline

\#1 EuCo$_{1.90(1)}$As$_2$\footnotemark[1] 	& 0.3601(4) 	& 3.922(9) 	& 11.370(3)	& 174.9(8) 	& 2.899(7)	& 45.1(8)		& 8.47				& 	PW				\\  
\#2 EuCo$_{1.99(2)}$As$_2$\footnotemark[2] 	& 0.3611(5) 	& 3.910(5) 	& 11.306(9)	& 172.8(6) 	& 2.891(6) 	& 44.9(5) 	& 8.62				& 	PW				\\
\#3 EuCo$_{1.92(4)}$As$_2$\footnotemark[3]	& 0.3603(6) 	& 3.926(7) 	& 11.137(18) 	& 171.6(8) 	& 2.836(9) 	& 40.8(7)		& 8.54				& 	PW				\\
\#4 EuCo$_{1.90(2)}$As$_2$\footnotemark[4] 	& 0.3607(1) 	& 3.9478(7)	& 11.232(2) 	& 175.05(7) 	& 2.845(1) 	& 40.6(7)		& 8.51				&	PW				\\ 
\#5 EuCo$_{1.92(1)}$As$_2$\footnotemark[4] 	& 0.3623(2) 	& 3.9505(2)	& 11.2257(7) 	& 175.19(2) 	& 2.8416(2) 	& 40.3(5)		& 8.61				&	PW				\\ 
\#6 EuCo$_{1.94(2)}$As$_2$\footnotemark[1] 	& 0.3683(3) 	& 3.9323(4)	& 11.402(1) 	& 176.32(3) 	& 2.8996(5) 	& 45.8(3)		& 					&	PW				\\ 
EuCo$_2$As$_2$ 						&           	& 3.964(2) 	& 11.111(6) 	& 174.6(2) 	& 2.803(3) 	&			& 					& \cite{Marchand1978} 	\\
EuCo$_2$As$_2$\footnotemark[5] 			&              & 3.934(1) 	& 11.511(6) 	& 178.1(2) 	& 2.926(2) 	& 47(2) 		& 7.4(1)				& \cite{Raffius1993} 	\\
EuCo$_2$As$_2$\footnotemark[4] 			&  0.36        & 3.9671(1)	& 11.0632(5) 	& 174.11(1) 	& 2.7887(2) 	& 			& 					& \cite{Bishop2010} 	\\
EuCo$_2$As$_2$\footnotemark[6] 			&  0.36109(5)  & 3.929(1)	& 11.512(4) 	& 177.7(1) 	& 2.930(2) 	& 47			& 8.00\footnotemark[7]	& \cite{Tan2016} 		\\
EuCo$_2$As$_2$\footnotemark[4] 			&              &  			&  			&  			&  			& 38.5		& 8.27				& \cite{Ballinger2012} 	\\

\end{tabular}

\end{ruledtabular}

\footnotetext[1]{Grown in Sn flux}
\footnotetext[2]{Grown in Sn flux with H$_2$-treated Co powder}
\footnotetext[3]{Grown in CoAs flux with H$_2$-treated Co powder}
\footnotetext[4]{Grown in CoAs flux}
\footnotetext[5]{Polycrystalline sample}
\footnotetext[6]{Grown in Bi flux}
\footnotetext[7]{Obained by us by fitting the published $\chi(T)$ data}

\end{table*}

\section{\label{Sec:ChiMH} Magnetic Susceptibility and High-Field Magnetization}

\subsection{\label{Sec:Chi} Magnetic Susceptibility}

\begin{figure}
\includegraphics[width=3in]{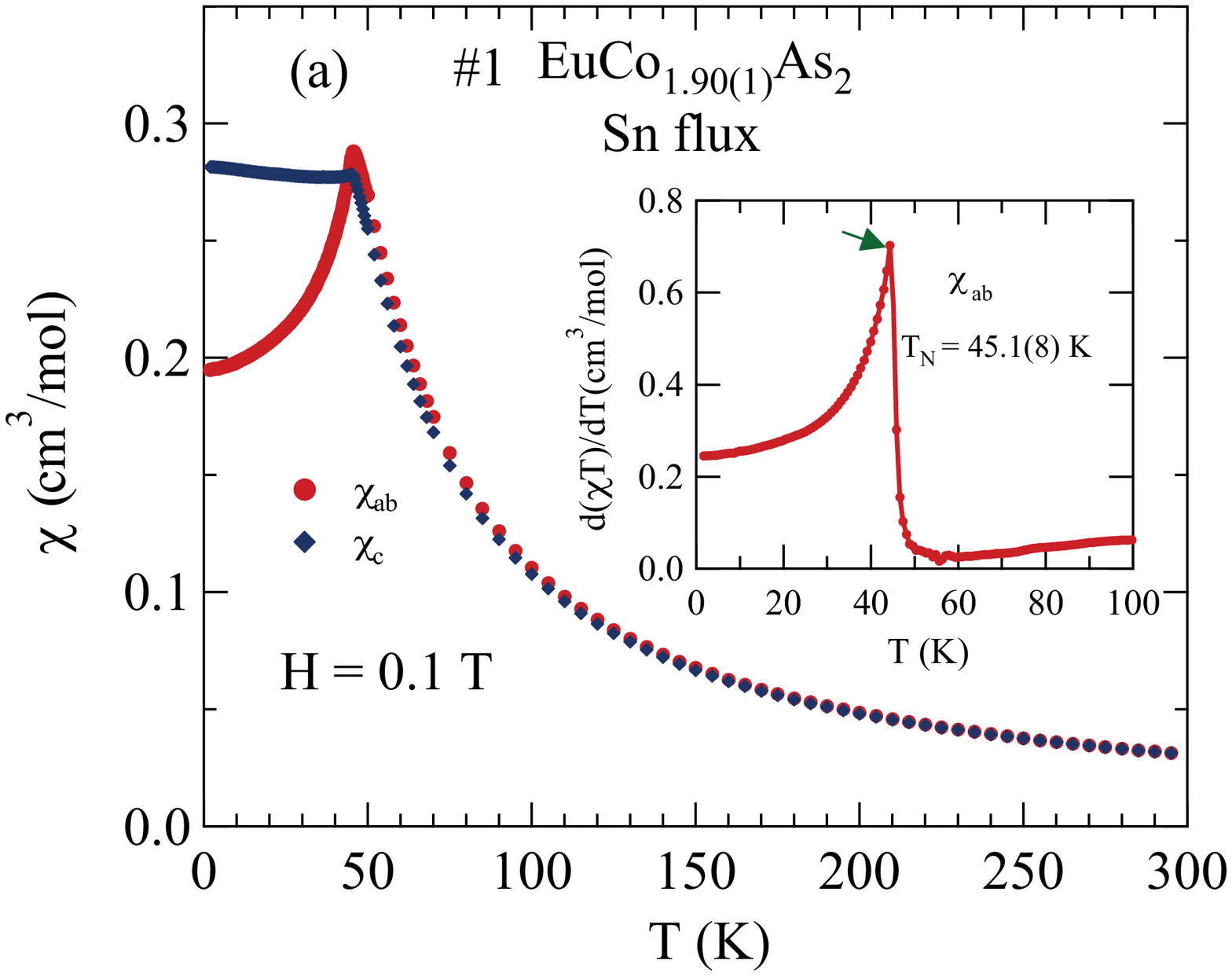}\vspace{0.2in}
\includegraphics[width=3in]{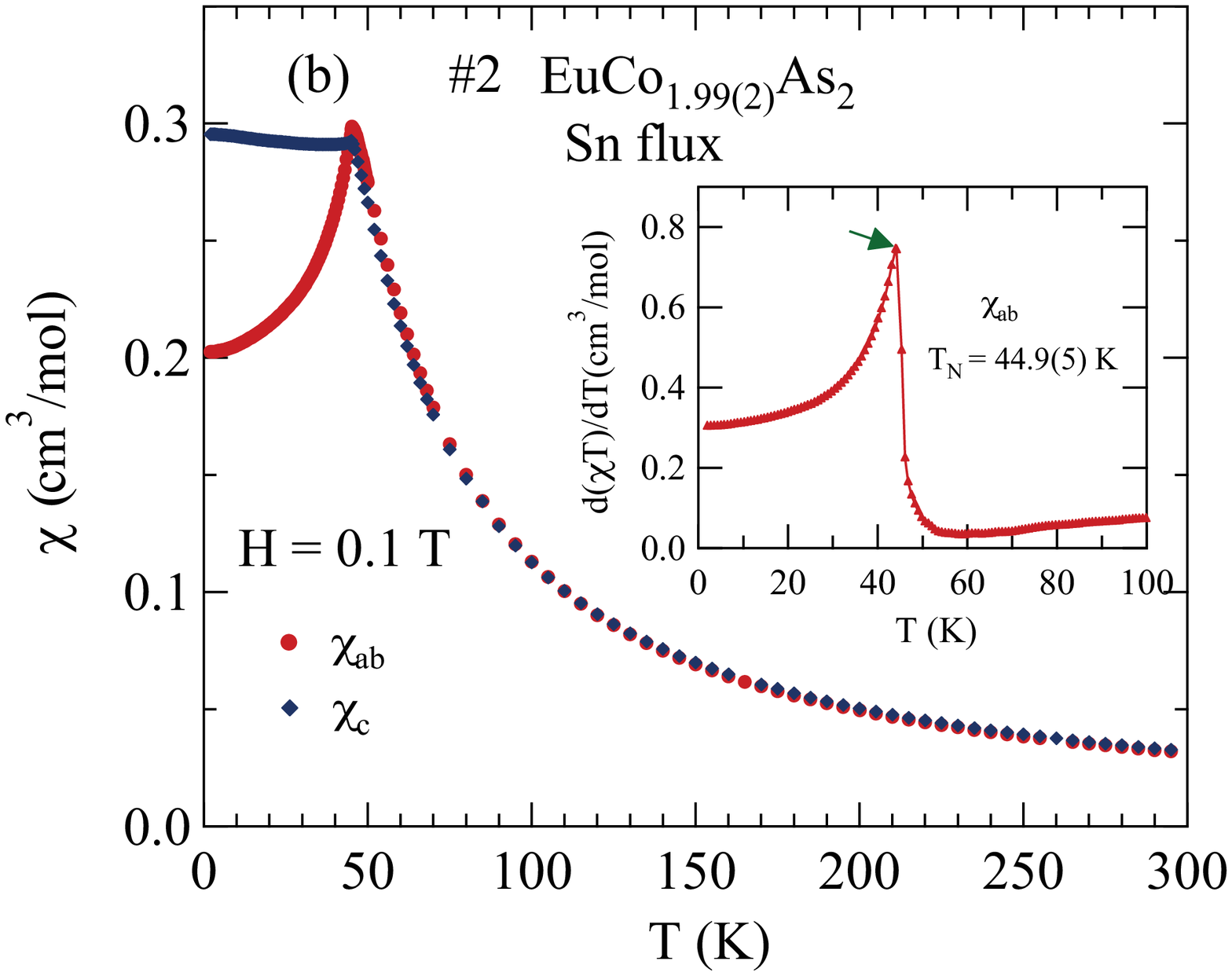}
\caption{Zero-field-cooled (ZFC) magnetic susceptibility $\chi \equiv M/H$ of Sn-flux-grown crystals (a)~\#1~EuCo$_{1.90(1)}$As$_2$ and (b)~\#2~EuCo$_{1.99(2)}$As$_2$ as a function of temperature~$T$ measured in magnetic fields~$H = 0.1$~T applied in the $ab$~plane ($\chi_{ab}$) and along the $c$~axis ($\chi_c$).  Insets:~The respective derivative $d(\chi_{ab}T)/dT$ versus~$T$.}
\label{Fig_MT_1kOe_Snflux}
\end{figure}

\begin{figure}
\includegraphics[width=3in]{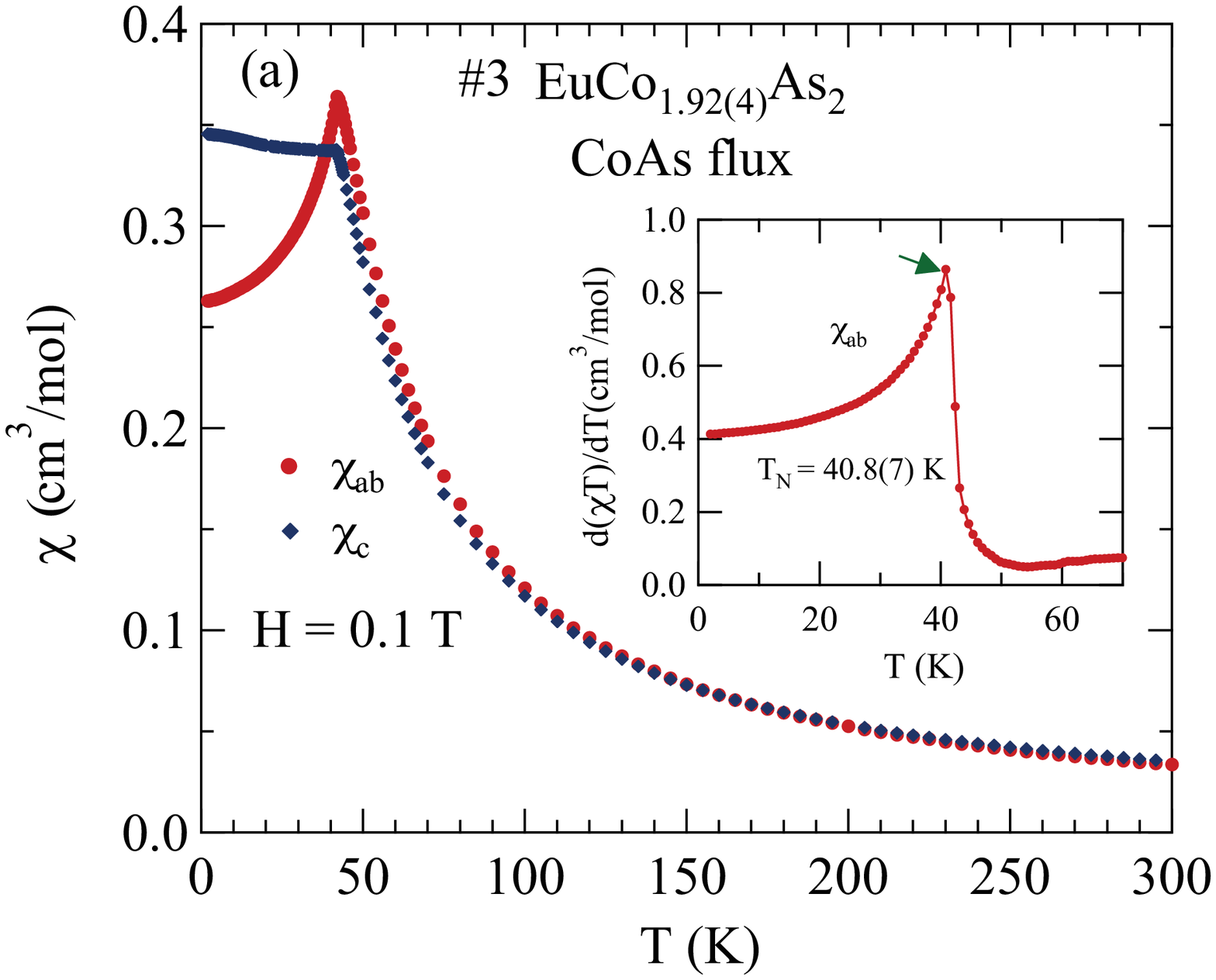}
\includegraphics[width=3in]{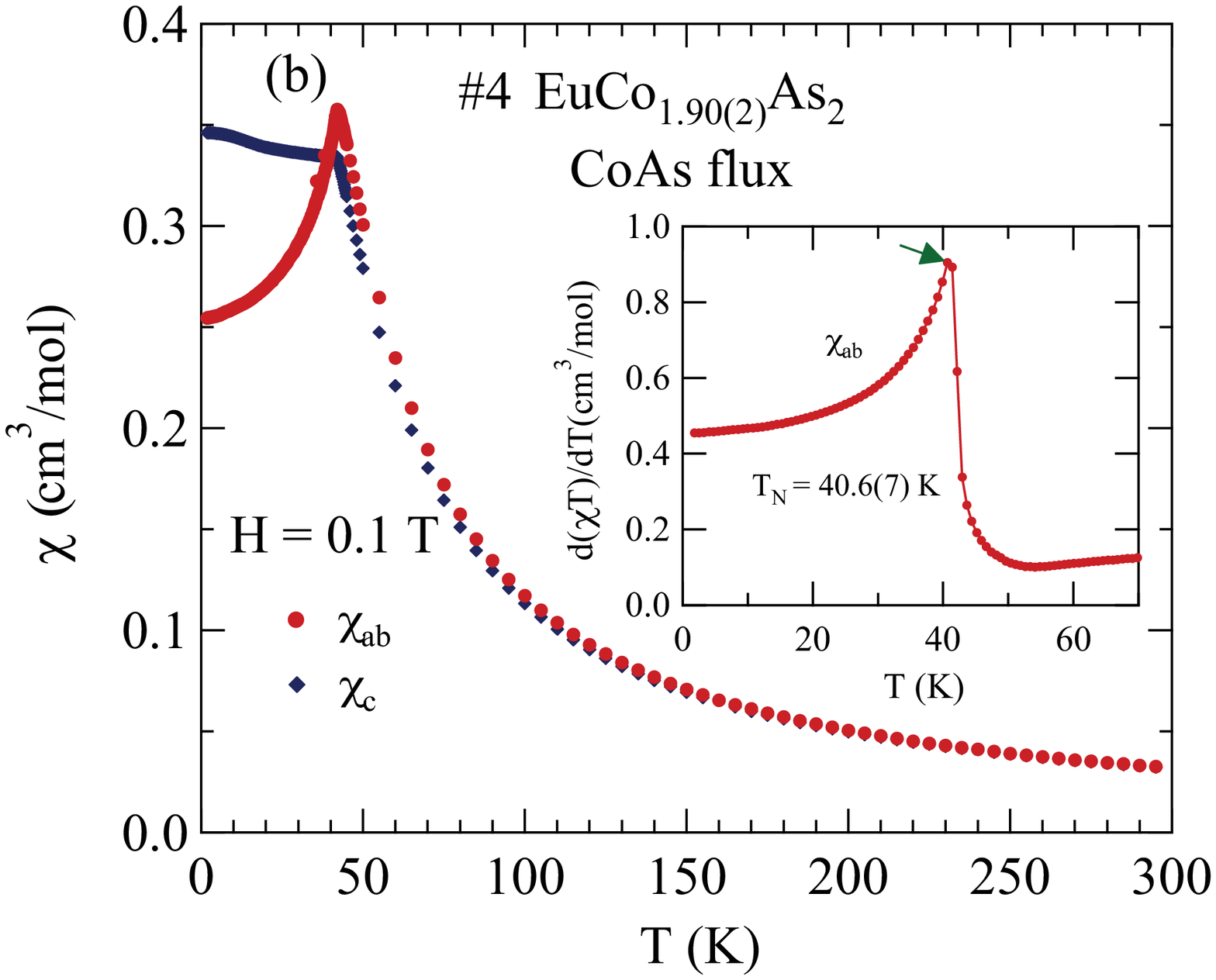}
\caption{Zero-field-cooled (ZFC) magnetic susceptibility  $\chi \equiv M/H$ of CoAs-flux-grown crystals (a)~\#3~EuCo$_{1.92(4)}$As$_2$ and (b)~\#4~EuCo$_{1.90(2)}$As$_2$ versus temperature~$T$ measured in a magnetic field~$H = 0.1$~T applied in the $ab$~plane ($\chi_{ab}$) and along the $c$~axis ($\chi_c$).  Insets: The derivative $d(\chi_{ab}T)/dT$ versus~$T$.}
\label{Fig_MT_1kOe_CoAsflux}
\end{figure}

Figures~\ref{Fig_MT_1kOe_Snflux} and~\ref{Fig_MT_1kOe_CoAsflux} display the zero-field-cooled (ZFC) magnetic susceptibility $\chi \equiv M/H$ of Sn-flux-grown crystals and CoAs-flux-grown crystals, respectively, as a function of $T$ with $H=~$0.1~T applied along the $c$~axis ($\chi_c$, $H \parallel c$) and in the $ab$~plane ($\chi_{ab}$, $H \parallel ab$). The $T_{\rm N}$ of a collinear AFM is given by the temperature of the maximum slope of $\chi T$ versus~$T$ for the easy axis direction \cite{Fisher1962}; here, the corresponding field direction is within the easy $ab$~plane of the helical magnetic structure.  The inset of each figure shows $d(\chi_{ab}T)/dT$ versus~$T$ in the $T$ range 2~to 100~K, with the peak temperature being~$T_{\rm N}$.  The $T_{\rm N}$ obtained in this way for each crystal is shown in the insets of Figs.~\ref{Fig_MT_1kOe_Snflux} and~\ref{Fig_MT_1kOe_CoAsflux} as well as in Table~\ref{Tab:ChemAnal} and in Table~\ref{Tab:CuriFit} below.  From Table~\ref{Tab:ChemAnal} one sees that the $T_{\rm N}$ values correlate with the crystallographic $c/a$ ratio and with the flux used to grow the crystals, but not with the Co-site occupancy. The $T_{\rm N}$ values from previous reports on \eca\ are also listed in Table~\ref{Tab:ChemAnal} \cite{Tan2016, Raffius1993, Ballinger2012}.


For all four crystals, from the main panels in Figs.~\ref{Fig_MT_1kOe_Snflux} and~\ref{Fig_MT_1kOe_CoAsflux} one sees that $\chi_{ab}>\chi_c$ in the paramagnetic regime ($T > T_{\rm{N}}$), indicating the presense of a magnetic anisotropy favoring the $ab$~plane. This is consistent with the data for $T\ll T_{\rm N}$ which indicates that the crystallographic $ab$-plane is an AFM easy~plane. For $T<T_{\rm N}$, one sees that $\chi_c$ is nearly independent of~$T$, consistent with the molecular-field theory prediction for a field perpendicular to the ordering axis or plane of a Heisenberg AFM \cite{Johnston2012, Johnston2015}.  Magnetocrystalline anisotropy determines the ordering axis or plane such as for a Heisenberg AFM with dipolar \cite{Johnston2016}, uniaxial single-ion $DS_z^2$ \cite{Johnston2017}, and classical field \cite{Johnston2017b} anisotropies.  The observation that $\chi_{ab}$ for $T\to0$ is a large fraction of $\chi_c(T\to 0)$ indicates that \eca\ is either a collinear AFM with multiple domains in the $ab$~plane or a coplanar noncollinear $ab$~plane AFM structure. The previous neutron diffraction study on EuCo$_2$As$_2$ indeed showed an incommensurate AFM helical structure in which Eu spins are aligned ferromagnetically within the $ab$~plane, where the helix axis is the $c$-axis with an AFM propagation vector of ${\bf k} = (0, 0, 0.79)\pi /c$ where $c$ is the tetragonal $c$-axis lattice parameter \cite{Tan2016}. An incommensurate helical spin structure with almost the same propagation vector was found in the isostructural compound $\rm EuCo_2P_2$ \cite{Reehuis1992, Sangeetha2016}.

\begin{figure}
\includegraphics[width=3in]{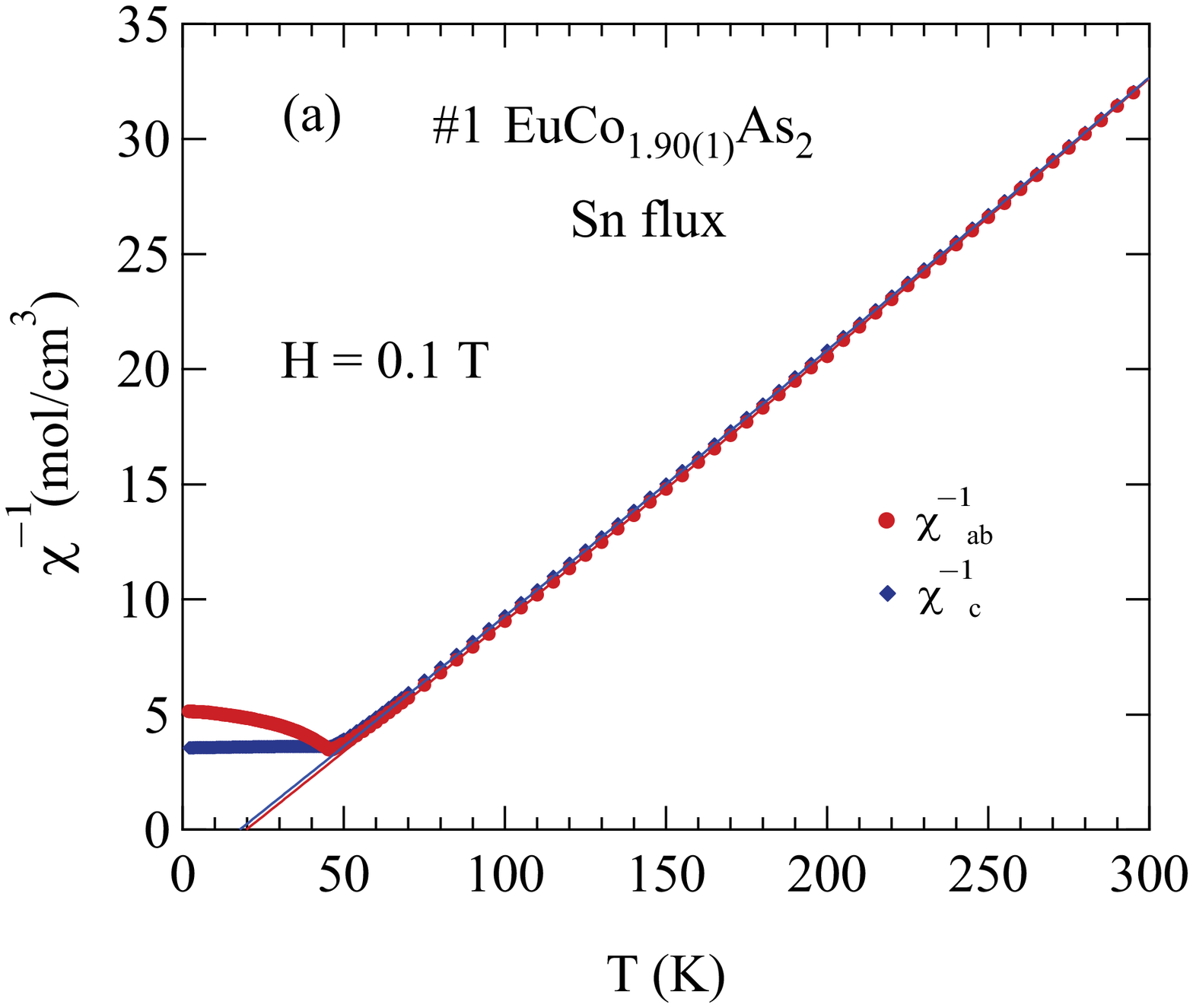}
\includegraphics[width=3in]{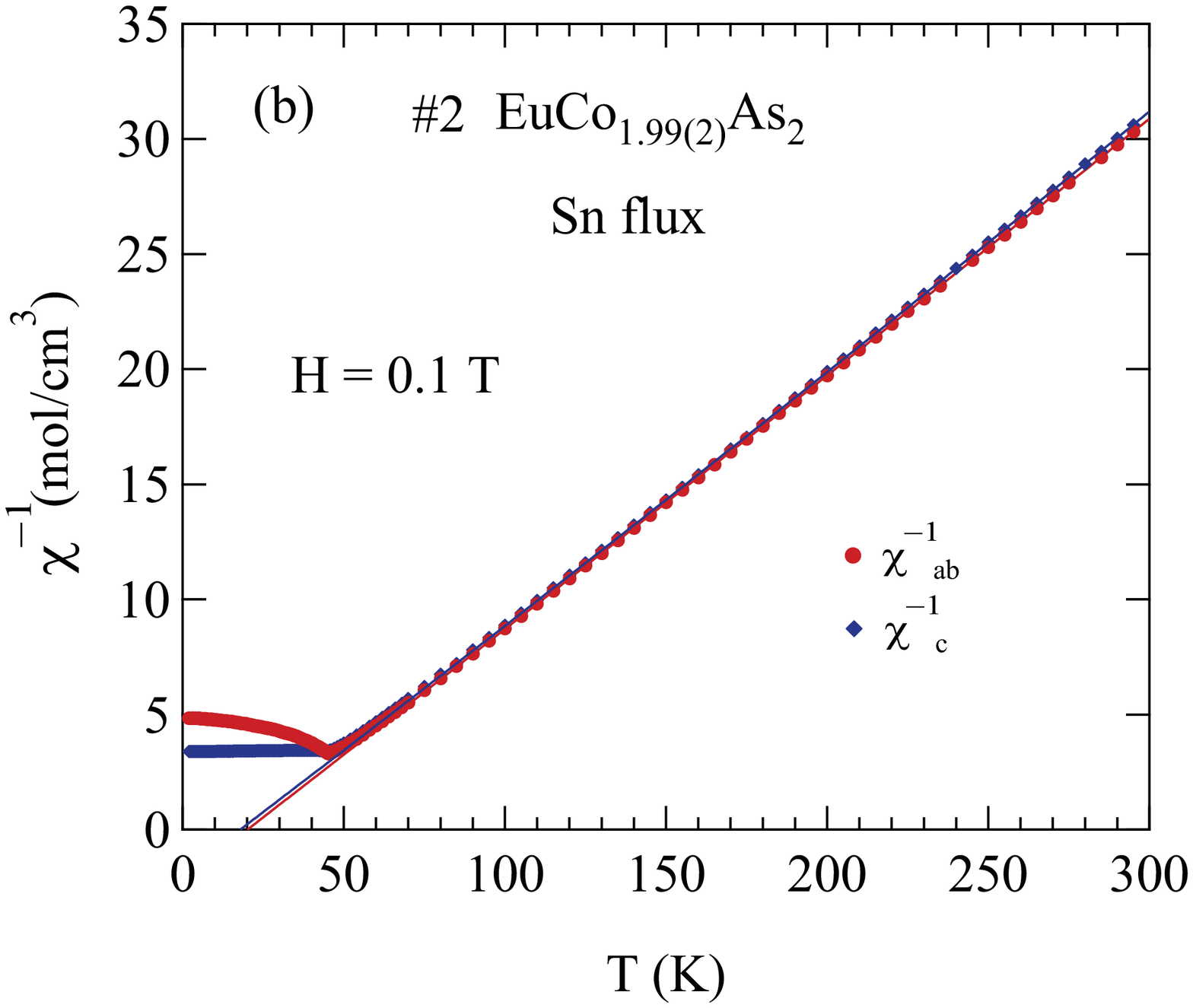}
\caption{(a)~Inverse susceptibility  $\chi ^{-1}$ versus temperature~$T$ of Sn-flux-grown crystals (a)~\#1~EuCo$_{1.90(1)}$As$_2$  and (b)~\#2~EuCo$_{1.99(2)}$As$_2$ for $H = 0.1$~T applied in the $ab$~plane ($H\parallel ab,\ \chi ^{-1}_{ab}$) and along the $c$~axis ($H\parallel c,\ \chi ^{-1}_c$).  The solid curves are fits by the modified Curie-Weiss law~(\ref{Eq:ModCWLaw}) with parameters given in Table~\ref{Tab:CuriFit}.}
\label{Fig:InvChi_Snflux}
\end{figure}

\begin{figure}
\includegraphics[width=3in]{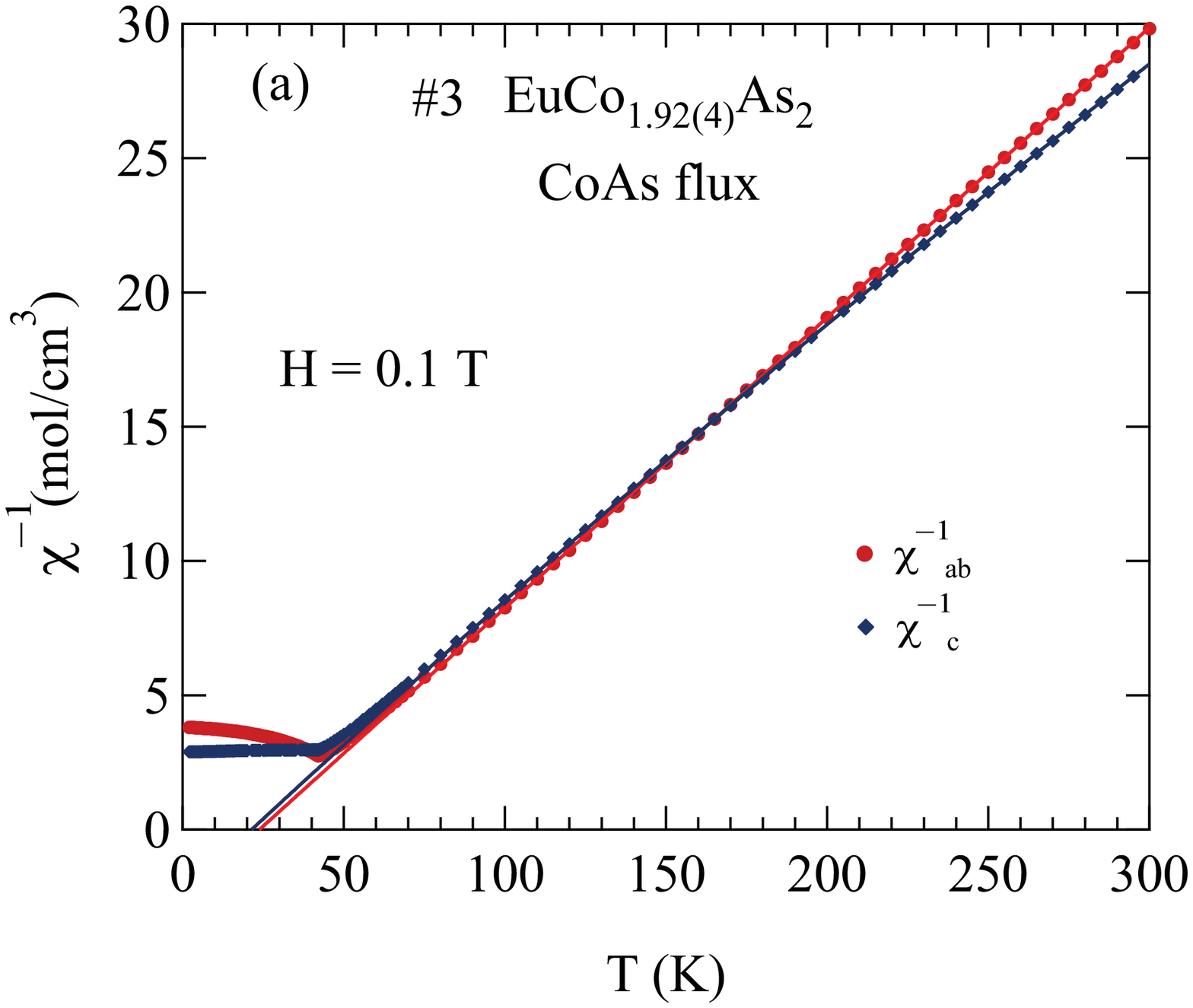}
\includegraphics[width=3in]{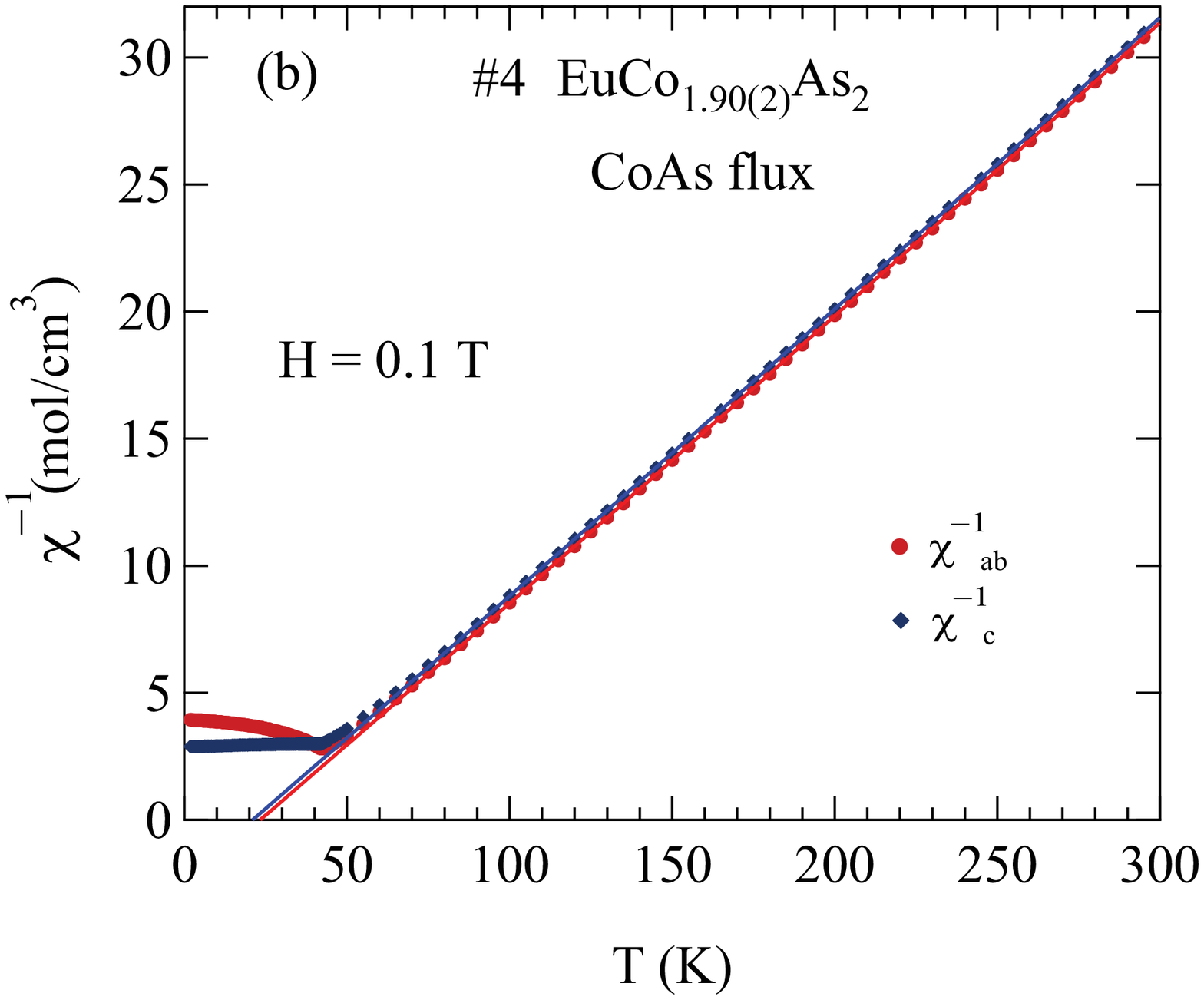}
\caption{(a)~Inverse susceptibility $\chi^{-1}$ of CoAs-flux-grown crystals (a)~$\rm EuCo_{1.92(4)}As_2$ and (b)~$\rm EuCo_{1.90(2)}As_2$ as a function of temperature~$T$ for $H = 0.1$~T applied along the $c$~axis ($H\parallel c$) and along the $ab$~plane ($H\parallel ab$).  The solid curves are fits by the modified Curie-Weiss law~(\ref{Eq:ModCWLaw}) with parameters given in Table~\ref{Tab:CuriFit}.}
\label{Fig:InvChi_CoAsflux}
\end{figure}

The inverse susceptibility $\chi^{-1}(T)$ measured in $H = 0.1$~T applied along the $c$~axis ($\chi ^{-1}_c$) and in the $ab$~plane ($\chi ^{-1}_{ab}$) for Sn-flux- and CoAs-flux-grown crystals are shown in Figs.~\ref{Fig:InvChi_Snflux} and~\ref{Fig:InvChi_CoAsflux}, respectively.  As one can see from the figures, the  $\chi ^{-1}$($T$) plots are slightly curved. One can fit this curvature by including a $T$-independent term $\chi_0$ in addition to the Curie-Weiss law, giving a so-called modified Curie-Weiss law 
\bse
\be
\chi_\alpha =\chi_0 + \frac{C_\alpha}{T - \theta_{\rm p\alpha}} \qquad (\alpha = ab,\ c),
\label{Eq:ModCWLaw}
\ee
where $\chi_0$ is an isotropic temperature-independent term given by
\be
\chi_0 = \chi^{\rm dia} + \chi^{\rm para} = \chi^{\rm core} + \chi^{\rm Landau} + \chi^{\rm Pauli},
\label{Eq:Chi0Parts}
\ee
which is comprised of the diamagnetic (negative) atomic core ($\chi^{\rm core}$) and conduction-electron orbital Landau ($\chi^{\rm Landau}$) contributions and the paramagnetic (positive) contribution from the Pauli spin susceptibility ($\chi^{\rm Pauli}$) of the conduction electrons and/or holes.  The Curie constant per mole of spins is given by \cite{Kittel2005}
\be
C_\alpha =  \frac{N_{\rm A}g_\alpha^2S(S+1)\mu_B^2}{3k_B} \equiv \frac{N_{\rm A}\mu_{\rm eff}^2\mu_{\rm B}^2}{3k_{\rm B}},
\label{eq:Curie-constant}
\ee
where $N_{\rm A}$ is Avogadro's number and $\mu_{\rm eff}$ is the ``effective moment'' of a spin in units of Bohr magnetons. From Eq.~(\ref{eq:Curie-constant}) one obtains
\be
\mu_{\rm eff} = g\sqrt{S(S+1)} = \sqrt{\frac{3k_{\rm B}C}{N_{\rm A}\mu_{\rm B}^2}}.
\label{Eq:mueffDef}
\ee
Inserting the Gaussian cgs values of the fundamental constants into Eq.~(\ref{Eq:mueffDef}) gives
\be
\mu_{\rm eff} \approx \sqrt{7.99\,684\,C} \approx \sqrt{8\,C}.
\label{Eq:mueffFromC}
\ee
\ese

\begin{table}
\caption{\label{Tab:ChiFitChi00} Parameters obtained by fitting the $\chi(T)$ data in Figs.~\ref{Fig_MT_1kOe_Snflux} and~\ref{Fig_MT_1kOe_CoAsflux} for our crystals by Eq.~(\ref{Eq:ModCWLaw}) assuming $\chi_0=0$. Shown for each crystal are the Curie constant~$C$, Weiss temperature~$\theta_{\rm p}$, and effective moment $\mu_{\rm eff}$ obtained from $C$ using Eq.~(\ref{Eq:mueffDef}).  For reference, for a spin $S=7/2$ with $g=2$, Eqs.~(\ref{Eq:ModCWLaw}) and~(\ref{Eq:mueffFromC}) yield $C = 7.878~{\rm {cm^3\,K/mol~Eu}}$ and $\mu_{\rm eff} = 7.937~\mu_{\rm B}$/Eu.}

\begin{ruledtabular}
\begin{tabular}{ccccccc}
Crystal								& Field			&  $C$						& $\theta_{\rm p}$	& $\mu_{\rm eff}$	\\
               						& Direction& $\left({\rm\frac{cm^3\,K}{mol}}\right)$	& (K)			&($\mu_{\rm B}$/Eu)	\\
\hline
\#1 EuCo$_{1.90(1)}$As$_2$\footnotemark[1]	& $H\parallel ab$ 	& 8.477(5) 					& 24.4(1)			& 8.233			\\
									& $H\parallel c$ 	& 8.543(4)					& 21.66(9)		& 8.265			\\
\#2 EuCo$_{1.99(2)}$As$_2$\footnotemark[2]	& $H\parallel ab$ 	& 9.020(2)					& 21.81(4)		& 8.493			\\
									& $H\parallel c$ 	& 8.948(5)					& 21.70(9)		& 8.459		 	\\
\#3 EuCo$_{1.92(4)}$As$_2$\footnotemark[3]	& $H\parallel ab$ 	& 9.251(2)					& 23.61(4)		& 8.601			\\
									& $H\parallel c$ 	& 10.01(1)					& 12.6(2)			& 8.947			\\
\#4 EuCo$_{1.90(2)}$As$_2$\footnotemark[4]	& $H\parallel ab$ 	& 8.753(3)					& 26.05(7)		& 8.366			\\
									& $H\parallel c$ 	& 8.784(2)					& 23.1(5)			& 8.381			\\
\#5 EuCo$_{1.92(1)}$As$_2$\footnotemark[4]	& $H\parallel ab$ 	& 8.68(5)						& 28.9(1)			& 8.33			\\
									& $H\parallel c$ 	& 8.97(1)						& 27.2(1)			& 8.47			\\
\end{tabular}
\end{ruledtabular}

\footnotetext[1]{Grown in Sn flux}
\footnotetext[2]{Grown in Sn flux with H$_2$-treated Co powder}
\footnotetext[3]{Grown in CoAs flux with H$_2$-treated Co powder}
\footnotetext[4]{Grown in CoAs flux}

\end{table}

As a baseline, we fitted the $\chi_\alpha(T)$ data by Eq.~(\ref{Eq:ModCWLaw}) from 100 to 300~K with $\chi_0=0$ for each of five of our crystals for each of the two field directions, and the fitted~$C_\alpha$ and~$\theta_{\rm p\alpha}$ values are shown in Table~\ref{Tab:ChiFitChi00} together with $\mu_{\rm eff}$ calculated from $C$ using Eq.~(\ref{Eq:mueffFromC}).  One sees that the values of~$\mu_{\rm eff}$ are~4\% to~7\% larger than the value for $S=7/2$ with $g=2$ given in the table caption, not including the data for outlier crystal~\#3.  These differences are outside the experimental error of $\sim 1$\%.  Our enhanced values of $\mu_{\rm eff}$ are in qualitative agreement with the previous value in Table~\ref{Tab:ChemAnal} reported in Ref.~\cite{Ballinger2012}.  The positive values of $\theta_{\rm p\alpha}$ indicate a net FM exchange interaction between the Eu$^{+2}$ spins-7/2.

 The value of $\theta_{\rm p\alpha}$ obtained from a fit of experimental $\chi_\alpha(T)$ data in the paramagnetic regime at $T > T_{\rm N}$ by Eq.~(\ref{Eq:ModCWLaw}) can be affected by crystal-shape (demagnetization) effects if $\chi_\alpha$ is large such as for compounds containing high concentrations of large-spin species such as Eu$^{+2}$ with spin $S=7/2$ in \eca.  From the treatment in Ref.~\cite{Johnston2016}, for $\chi_0=0$ these affect the Weiss temperature according to
\bse
\be
\theta_{\rm p\alpha} = \theta_{\rm p\alpha0} - \frac{4\pi C_\alpha N_{\rm d\alpha}}{V_{\rm M}},
\label{Eq:thetapalpha}
\ee
where $\theta_{\rm p\alpha}$ is the fitted value as above, $C_\alpha$ is the Curie constant per mole of magnetic atoms, $\theta_{\rm p\alpha0}$ is the Weiss temperature that would have been obtained in the absence of demagnetization effects, $N_{\rm d\alpha}$ is the magnetometric demagnetization factor in SI units $(0\leq N_{\rm d\alpha}\leq 1)$ of a crystal with the applied field in the $\alpha$ direction, and $V_{\rm M}$ is the volume per mole of magnetic atoms in the crystal.  For spins-7/2 with $g=2$ one has isotropic $C_\alpha = 7.88~{\rm cm^3\,K/mol}$ and using the crystal data in Table~\ref{Tab:ChemAnal} one obtains $V_{\rm M}\approx 53~{\rm cm^3/mol}$ for \eca.  Then for \eca, Eq.~(\ref{Eq:thetapalpha}) gives
\be
\theta_{\rm p\alpha} = \theta_{\rm p\alpha0} -  ({\rm 1.9~K})N_{\rm d\alpha}.
\label{Eq:thetapalphvalue}
\ee
\ese
Since $0\leq N_{\rm d\alpha}\leq 1$, a fitted positive value of $\theta_{\rm p\alpha}$ in Table~\ref{Tab:ChiFitChi00} can thus be decreased by up to 1.9~K due to demagnetization effects, which is a maximum of  $\sim 10$\% of the $\theta_{\rm p\alpha}$ values.

The data for $C$, $\mu_{\rm eff}$, and $\theta_{\rm p}$ for crystal~\#3 in Table~\ref{Tab:ChiFitChi00} are outliers.  We infer that these erroneous values arise from the contribution of a small amount of a ferromagnetic impurity to the magnetization.  In particular, including a $\chi_0$ in the fits below yields a positive value that includes the FM impurity contribution and leads to $C$, $\mu_{\rm eff}$, and $\theta_{\rm p}$ values in better alignment with those for the other four crystals. From the value of $\chi_0$ obtained for crystal~\#3 below we estimate the contribution of the FM impurity to the magnetization of the crystal in the measuring field of 0.1~T to be \mbox{$\sim 5\times10^{-4}~\mu_{\rm B}$/f.u.}

\begin{figure}
\includegraphics[width=3in]{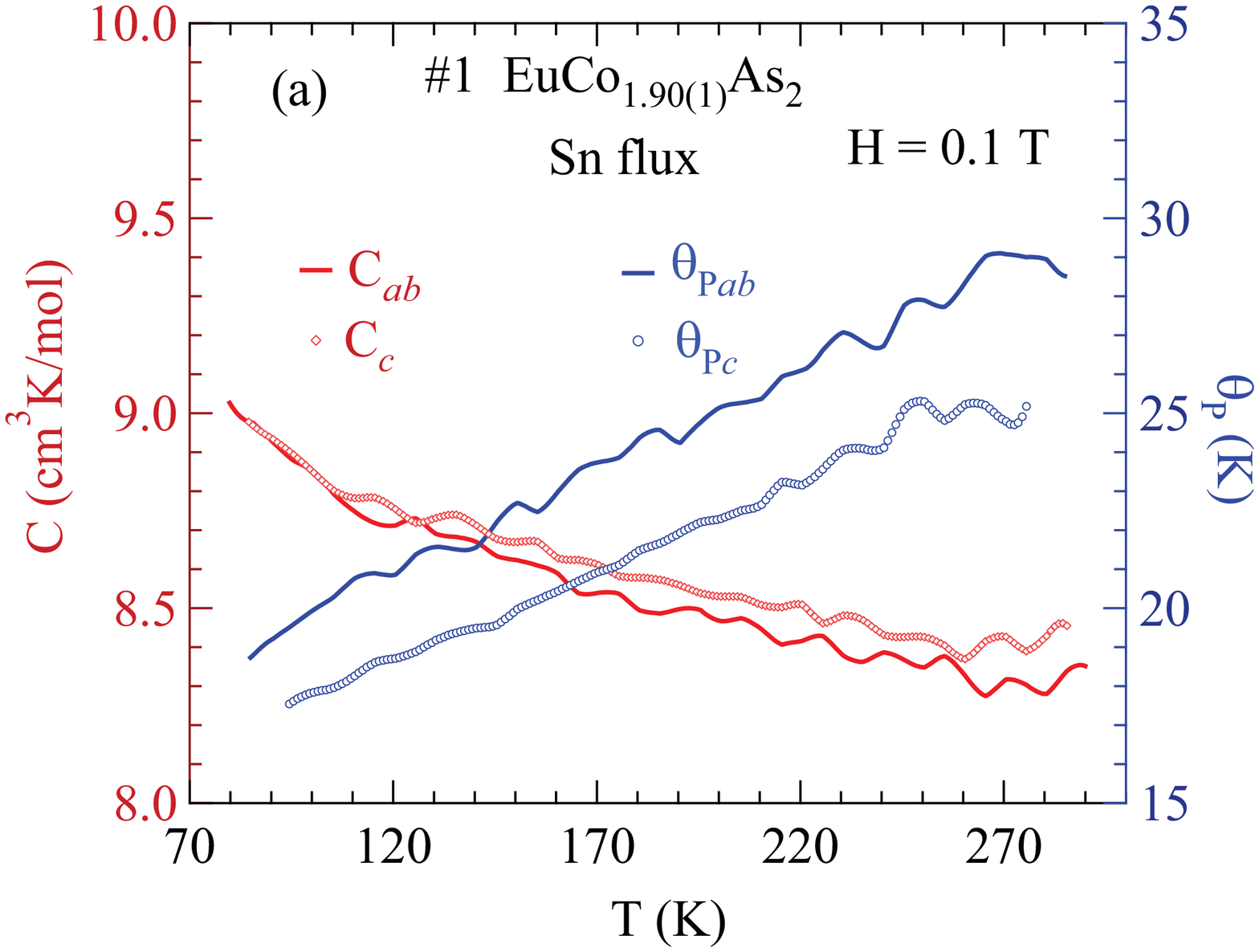}
\includegraphics[width=3in]{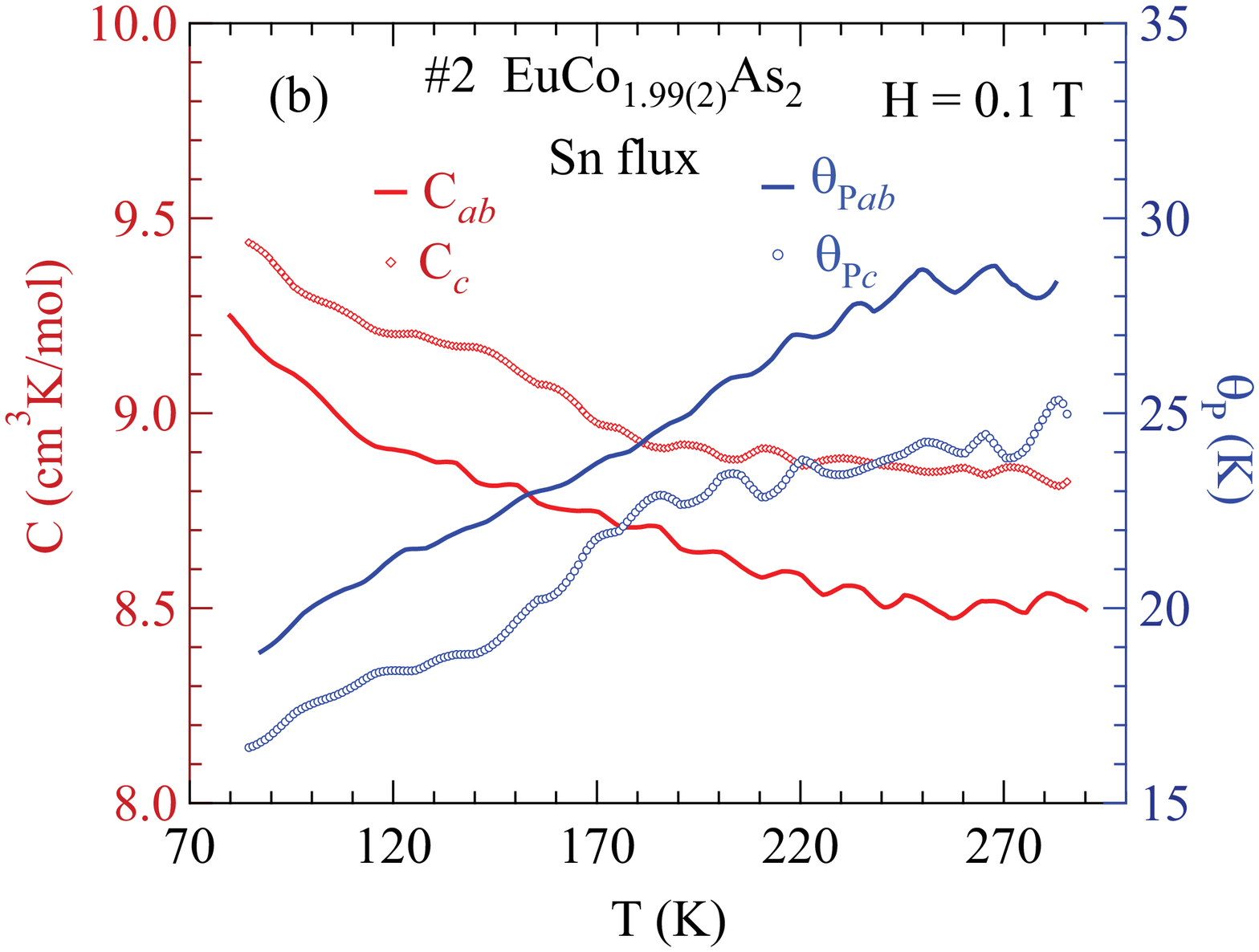}
\includegraphics[width=3in]{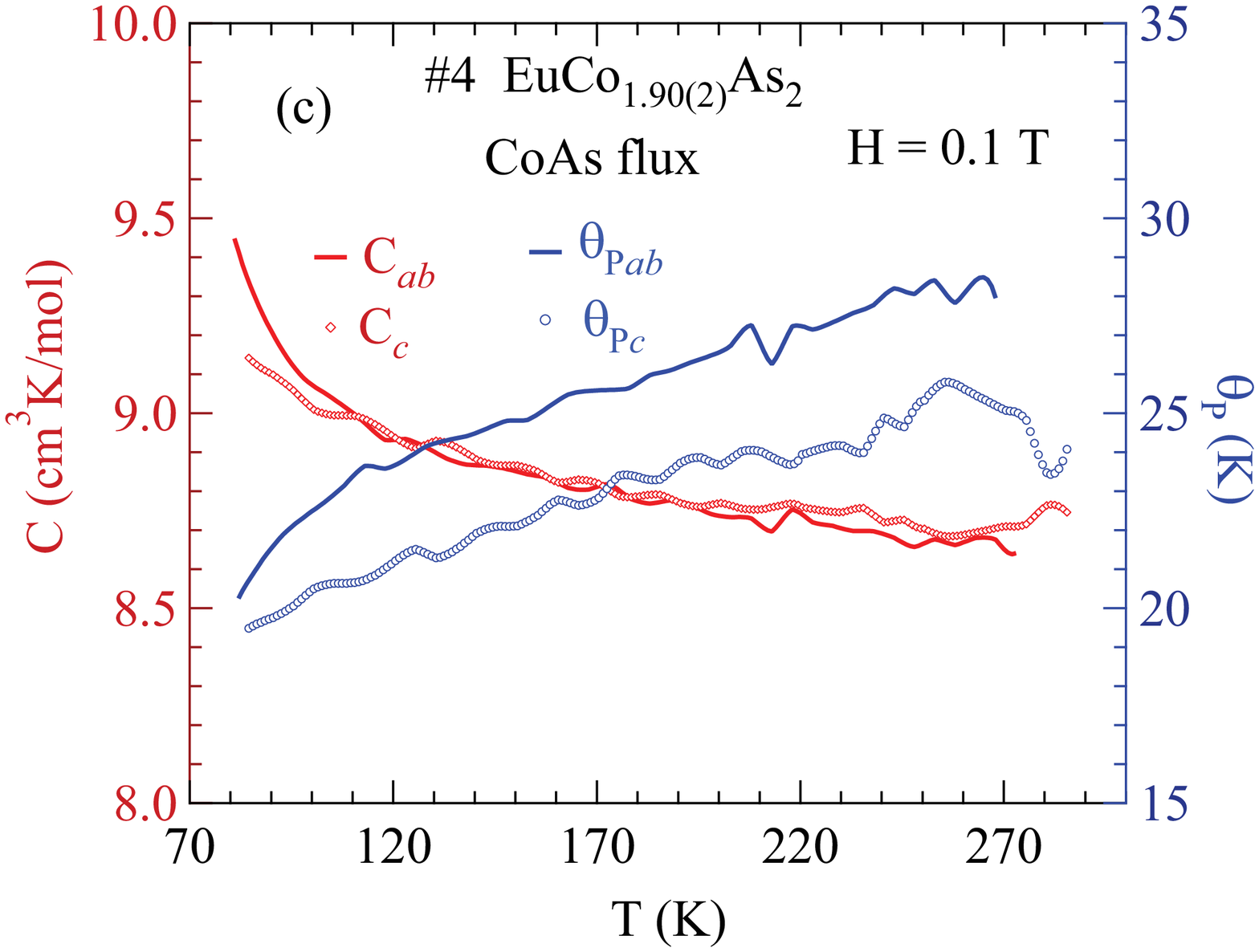}
\caption{Temperature $T$ dependence of the Curie~constant~$C_\alpha$ and Weiss temperature~$\theta_{\rm p\alpha}$ of Sn-flux-grown crystals (a)~\#1~$\rm EuCo_{1.90(1)}As_2$, (b)~\#2~$\rm EuCo_{1.99(2)}As_2$, and (c)~\#4~$\rm EuCo_{1.90(2)}As_2$, derived from Eqs.~(\ref{Eqs:CqFromchichip}).  }
\label{Fig:CC_theta_Snflux}
\end{figure}

Next, we included $\chi_0$ in the fits and the three fitting parameters are listed in Table~\ref{Tab:CuriFit} along with the previous reports for this compound.  Most of the $\chi_0$ values are strongly negative.  The fits are shown as the solid curves in Figs.~\ref{Fig:InvChi_Snflux} and~\ref{Fig:InvChi_CoAsflux}.

Now we obtain an estimate of $\chi_0$ expected for \ecaa.  \ecaa\ is not an ionic compound, so we do not use the ionic values \cite{Selwood1956} for the $\chi^{\rm core}$ contributions.  Instead, we use the atomic core contributions tabulated in Table~2.1 of Ref.~\cite{PMS1977}, which are given per mole of atoms as 
\bse
\bea
\chi^{\rm core}({\rm Eu}) &=& -7.0\times10^{-5}~{\rm cm^3/mol},\\*
\chi^{\rm core}({\rm Co}) &=& -3.1\times10^{-5}~{\rm cm^3/mol},\\*
\chi^{\rm core}({\rm As}) &=& -3.3\times10^{-5}~{\rm cm^3/mol},
\eea
yielding the core susceptility per mole of \ecaa\ as
\be
\chi^{\rm core}({\rm EuCo_2As_2}) = -1.98\times10^{-4}~{\rm cm^3/mol}.
\ee
\ese
Assuming the $g$~factor of the conduction carriers is $g=2$, the Pauli spin susceptibility of the conduction carriers in cgs units  is given by
\be
\chi^{\rm Pauli}{\rm \left[\frac{cm^3}{mol}\right]} = (3.233\times10^{-5}){\cal D}(E_{\rm F}){\rm \left[\frac{states}{eV~f.u.}\right]},
\ee
where f.u.~means the formula unit of \eca\ and the density of states at the Fermi energy ${\cal D}(E_{\rm F})$ is for both spin directions, i.e., taking into account the Zeeman degeneracy of the conduction carriers.  Taking ${\cal D}(E_{\rm F}) \approx 7$~states/eV~f.u.\ obtained from the $C_{\rm p}(T)$ measurements in Table~\ref{Tab:HC} below, one obtains
\be
\chi^{\rm Pauli} \approx 2.3\times10^{-4}~{\rm \frac{cm^3}{mol}}.
\ee
Then taking into account the Landau diamagnetism of the conduction carriers assuming a free-carrier gas gives the $T$-independent contribution to $\chi$ according to Eq.~(\ref{Eq:Chi0Parts}) as
\be
\chi_0 = \chi^{\rm core} + \frac{2}{3}\chi^{\rm Pauli} \approx -4.7\times10^{-5}~{\rm \frac{cm^3}{mol}}.
\ee
This value is much smaller in magnitude than the $\chi_0$ values listed for crystals~\#1, \#2, \#4, and~\#5 in Table~\ref{Tab:CuriFit}, suggesting that these large negative values may instead be reflections of $T$-dependent Curie constants and Weiss temperatures, a possibility examined next.

In order to investigate the possible $T$~dependences of $C_\alpha$ and $\theta_{\rm p\alpha}$, we again set $\chi_0=0$.  We obtained a spline fit to $\chi_\alpha(T)$ from 70 to 300~K, and from that we obtained the temperature derivative $\chi_\alpha^\prime(T)$.  Then one has the two simultaneous equations
\bse
\label{Eqs:CqFromchichip}
\bea
\chi_\alpha(T) &=& \frac{C_\alpha}{T-\theta_{\rm p\alpha}},\\*
\chi_\alpha^\prime(T) &=& -\frac{C_\alpha}{(T-\theta_{\rm p\alpha})^2},
\eea
\ese
from which $C_\alpha$ and~$\theta_{\rm p\alpha}$ were solved for at each~$T$\@.  The results are shown in Fig.~\ref{Fig:CC_theta_Snflux} for Sn-flux-grown crystals \#1~$\rm EuCo_{1.90(1)}As_2$ and \#2~$\rm EuCo_{1.99(4)}As_2$ and for CoAs-flux-grown crystal \#4~$\rm EuCo_{1.90(2)}As_2$.  One sees smooth variations in $C$ and~$\theta_{\rm p}$ versus~$T$ for each crystal, where $C$ increases and $\theta_{\rm p}$ decreases monotonically with decreasing~$T$ for each of the three crystals.  This behavior of $C$ might be expected if the Eu spins polarize the conduction electrons, since the polarization might be expected to increase with decreasing~$T$\@. 

The possibility of conduction-electron polarization due to progressive filling of the $d$~band of the transition metal can lead to a variation of the asphericity of the valence shells of the Eu atoms through a weaker hybridization between the Eu valence states and Co $d$ states. As a result, the itinerant electrons are strongly coupled to the localized moment, leading to an observed effective moment for an $s$-state Eu spin-7/2 given by~\cite{Stewart1972}
\be
\mu_{\rm eff}^{\rm obs} = \mu_{\rm eff}\left[1+\frac{2}{g}\rho_{0}(E_{\rm F})J_{sf}\right].
\ee
Here we take $g=2$, $\rho_0(E_{\rm F})$ is the density of states per atom at the Fermi surface for one spin direction, and $J_{sf}$ is the effective $sf$ exchange interaction due to either direct exchange (positive) or $sf$ mixing (negative). The values of $\rho_0(E_{\rm F}) J_{sf}$ esimated from the effective moments of \eca\ compounds are given in the last column of Table~\ref{Tab:CuriFit}. The positive sign of the quantity suggests that the $sf$ interaction mechanism in these compounds could be due to direct exchange. These interactions are expected to be affected by the change in lattice parameters~$a$ and~$c$, and the overall unit-cell volume $V_{\rm cell}$. Another possible reason for the excess Eu moment is related to the contribution of the non-4$f$ electrons of Eu, which is mainly from on-site 5$d$ electrons. This gives rise to dressing of a bare rare-earth spin with a conduction electron spin cloud which for \eca\ would add a portion of conduction-electron spin magnetization to the free electron moment. These effects are associated with the indirect RKKY (Ruderman-Kittel-Kasuya-Yosida) exchange interaction \cite{Ruderman1954, Kasuya1956, Yosida1957}  and this may affect the $g$~factor. Electron-spin resonance measurements may be useful to confirm or refute the hypothesis that the Curie constant changes with temperature as suggested in Fig.~\ref{Fig:CC_theta_Snflux}.

\begin{table*}
\caption{\label{Tab:CuriFit} Parameters obtained from Modified Curie-Weiss fits of the magnetic susceptibility data between 100 and 300~K for \eca, where $T_{\rm N}$ is the N\'eel temperature, $\chi_0$ is the $T$-independent contribution to the susceptibility, $C_\alpha$ is the molar Curie constant for fields in the $\alpha = ab,\ c$ direction, $\mu_{\rm eff\alpha}$ is the effective moment, $\theta_{\rm p\alpha}$ is the Weiss temperature, $\theta_{\rm p\,ave}$ is the spherical average of $\theta_{\rm p\alpha}$, $\rho_0(E_{\rm F})$ is the density of states at the Fermi energy per atom for one spin direction, and $J_{sf}$ is the exchange interaction between a local $f$-electron atom and the $s$~conduction electrons. For reference, the effective moment for Eu$^{+2}$ with $g=2$ and $S=7/2$ is $\mu_{\rm eff} = g\sqrt{S(S+1)}\mu_{\rm B} = 7.94~\mu_{\rm B}$. The quantity $f_J$ is defined as $f_J = \theta_{\rm p\,ave}/T_{\rm N}$.  PW means present work and N/A means not applicable.}

\begin{ruledtabular}
\begin{tabular}{lclcccccclc}
Compound 											&   Ref.				& Field	& $T_{\rm N}$ 	&$\chi_0$			& $C_\alpha$ 			& $\mu_{\rm eff\alpha}$ & $\theta_{\rm p\alpha}$ & 	$\theta_{\rm p\,ave}$	&	 $\rho_0(E_{\rm F}) J_{sf}$ &	$f_J$ \\
 												& 					& Axis\,$\alpha$	& (K) 		& (${\rm \frac{10^{-3}\,cm^3}{mol}})$ & (${\rm \frac{cm^3\,K}{mol}})$ & ($\mu_{\rm B}$/Eu) & ~(K)  		&	(K)			& ~~(K)				\\
\hline
\#1 EuCo$_{1.90(1)}$As$_2$\footnotemark[1]	& PW 				& H$\parallel ab$ 	& 45.1(8) & $-1.4(2)$	& 8.98(1)	& 8.476(4)	& 19.76(9)	& 19.07	& 0.067	& 0.430	\\
									&					& H$\parallel c$ 	&		& $-1.2(1)$	& 8.970(5)& 8.471(2)	& 17.70(5)	&		& 0.0668	 	\\
\#2 EuCo$_{1.99(2)}$As$_2$\footnotemark[2]	& PW					& H$\parallel ab$ 	& 44.9(5)	& $-0.54(1)$	& 9.214(3)& 8.585(1)	& 20.10(3)	& 19.33	& 0.081	& 0.441	\\
									& 					& H$\parallel c$ 	&		& $-1.2(3)$	& 9.38(1) & 8.662(4)	& 17.8(1)		& 		& 0.09		\\
\#3 EuCo$_{1.92(4)}$As$_2$\footnotemark[3]	& PW					& H$\parallel ab$ 	& 40.8(7)	& 0.07(3)		& 9.23(1) & 8.593(5)	& 23.8(1)		& 22.99	& 0.08	& 0.563	\\
									&					& H$\parallel c$ 	&		& 2.75(2)		& 9.005(6)& 8.488(3) 	& 21.38(6)	&		& 0.07		\\
\#4 EuCo$_{1.90(2)}$As$_2$\footnotemark[4]	& PW					& H$\parallel ab$ 	& 40.6(7)	& $-0.87(1)$	& 9.062(5)& 8.514(2)	& 23.33(4)	& 22.54	& 0.072	& 0.555	\\
									&					& H$\parallel c$ 	&		& $-0.68(1)$	& 9.028(5)& 8.498(2) 	& 20.97(5)	&		& 0.07		\\
\#5 EuCo$_{1.92(1)}$As$_2$\footnotemark[4]	& PW					& H$\parallel ab$ 	& 40.3(5)	& $-1.33(6)$	& 9.15(2)	& 8.556(9)	& 24.9(2)		& 24.23	& 0.077	& 0.601	\\
									&					& H$\parallel c$ 	&		& $-1.45(3)$	& 9.48(1)	& 8.708(4) 	& 22.9(1)		&		& 0.097		\\
EuCo$_2$As$_2$\footnotemark[4]			& \cite{Ballinger2012}	& H$\parallel ab$	& 38.5 	& 2.12		& 8.45	& 8.22		& 28.7		& 27.2	& 0.035	& 0.706	\\
									&					& H$\parallel c$	&		& $-1.52$		& 8.68	& 8.33		& 25.7		&		& 0.049		\\
EuCo$_2$As$_2$\footnotemark[5]$^,$\footnotemark[6]& \cite{Tan2016}	& H$\parallel ab$	& 47 	& 			& 7.65(1) & 7.82(1)		& 20.5(1)		& 20.65	& 0.081	& 0.44	\\
									&					& H$\parallel c$	& 		& 			& 8.39(1) & 8.19(1) 	& 20.8(3)		&		& 0.0025		\\
EuCo$_2$As$_2$\footnotemark[7]			& \cite{Raffius1993} 	&  N/A			& 47(2) 	& 			&		& 7.4(1)		& 18(4)		& 18		& 		& 0.38	\\

\end{tabular}

\end{ruledtabular}
\footnotetext[1]{Grown in Sn flux}
\footnotetext[2]{Grown in Sn flux with H$_2$-treated Co powder}
\footnotetext[3]{Grown in CoAs flux with H$_2$-treated Co powder}
\footnotetext[4]{Grown in CoAs flux}
\footnotetext[5]{Grown in Bi flux}
\footnotetext[6]{The data were sent to us by the authors and we fitted them by $\chi=C/(T-\theta)$ from 100 to 300~K}
\footnotetext[7]{Poycrystalline sample}

\end{table*}

\subsection{\label{Sec:MH} High-Field Magnetization}

\begin{figure}
\includegraphics[width=3.in]{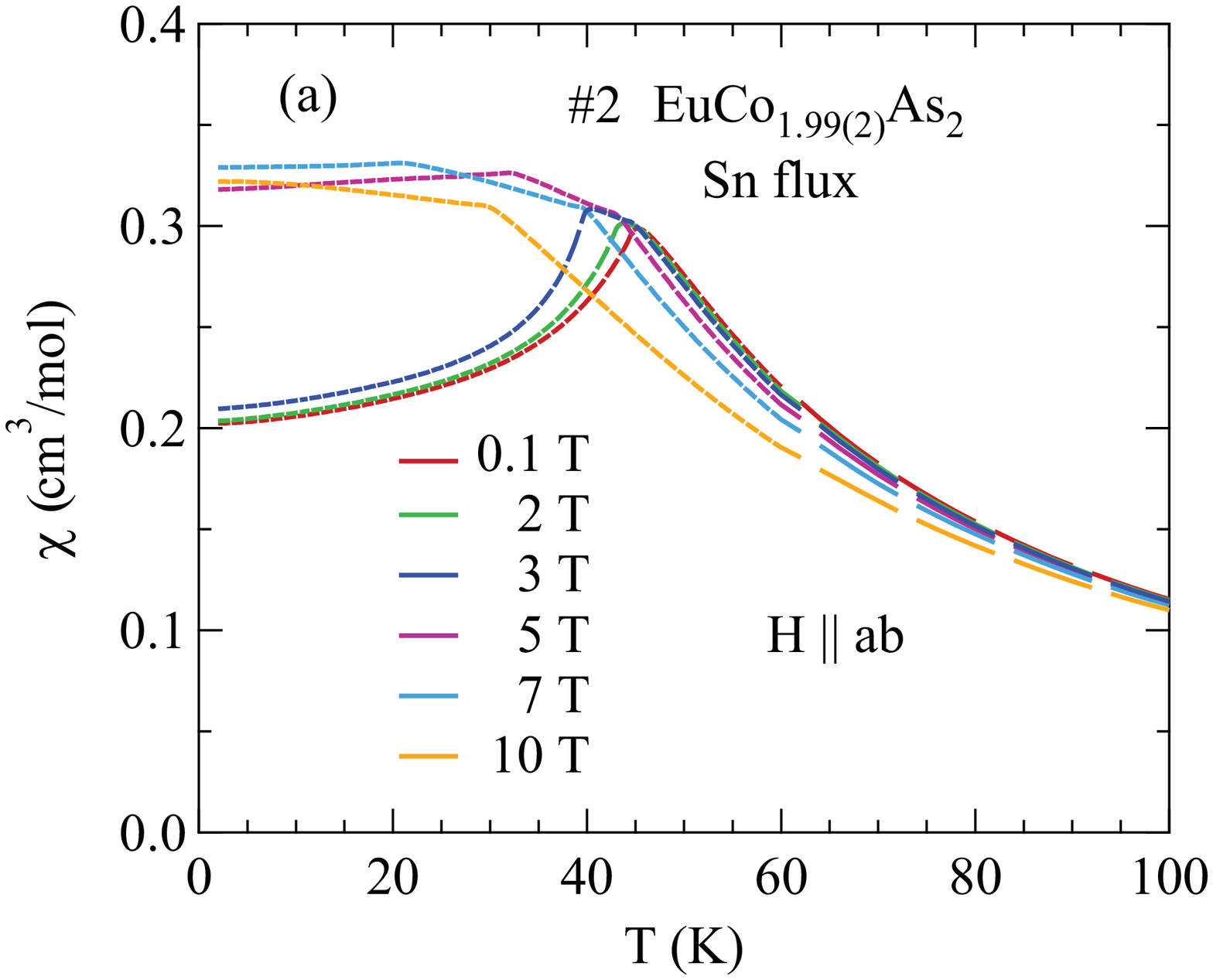}
\includegraphics[width=3.in]{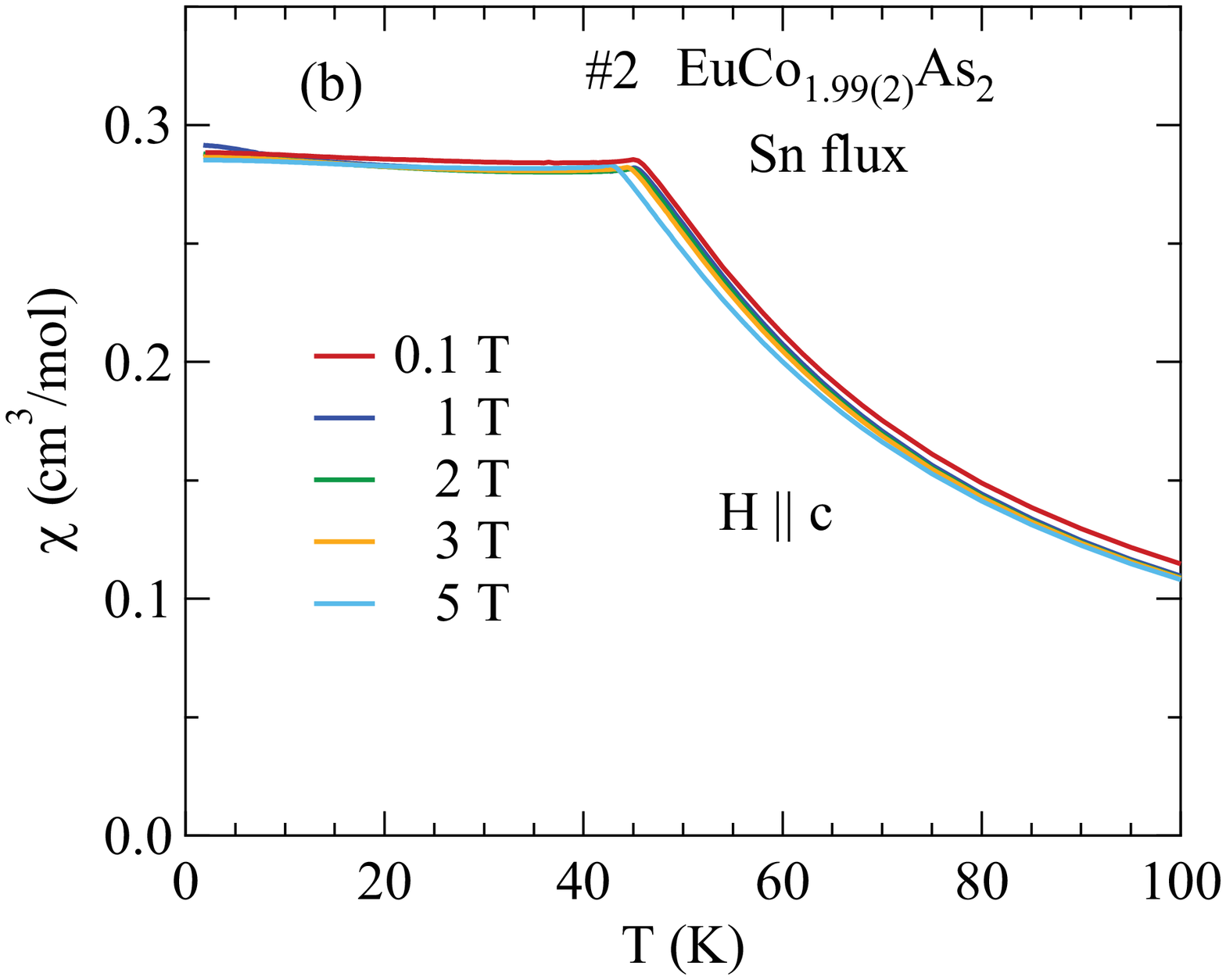}
\caption{Zero-field-cooled (ZFC) magnetic susceptibility $\chi \equiv M/H$ of Sn-flux-grown crystal \#2$~\rm EuCo_{1.99(2)}As_2$ as a function of temperature~$T$ for various magnetic fields~$H$ applied (a) in the $ab$~plane ($\chi_{ab},\ H\parallel ab$) and (b) along the $c$~axis ($\chi_c, H\parallel c$).}
\label{Fig:MT_Snflux}
\end{figure}

\begin{figure}
\includegraphics[width=3.in]{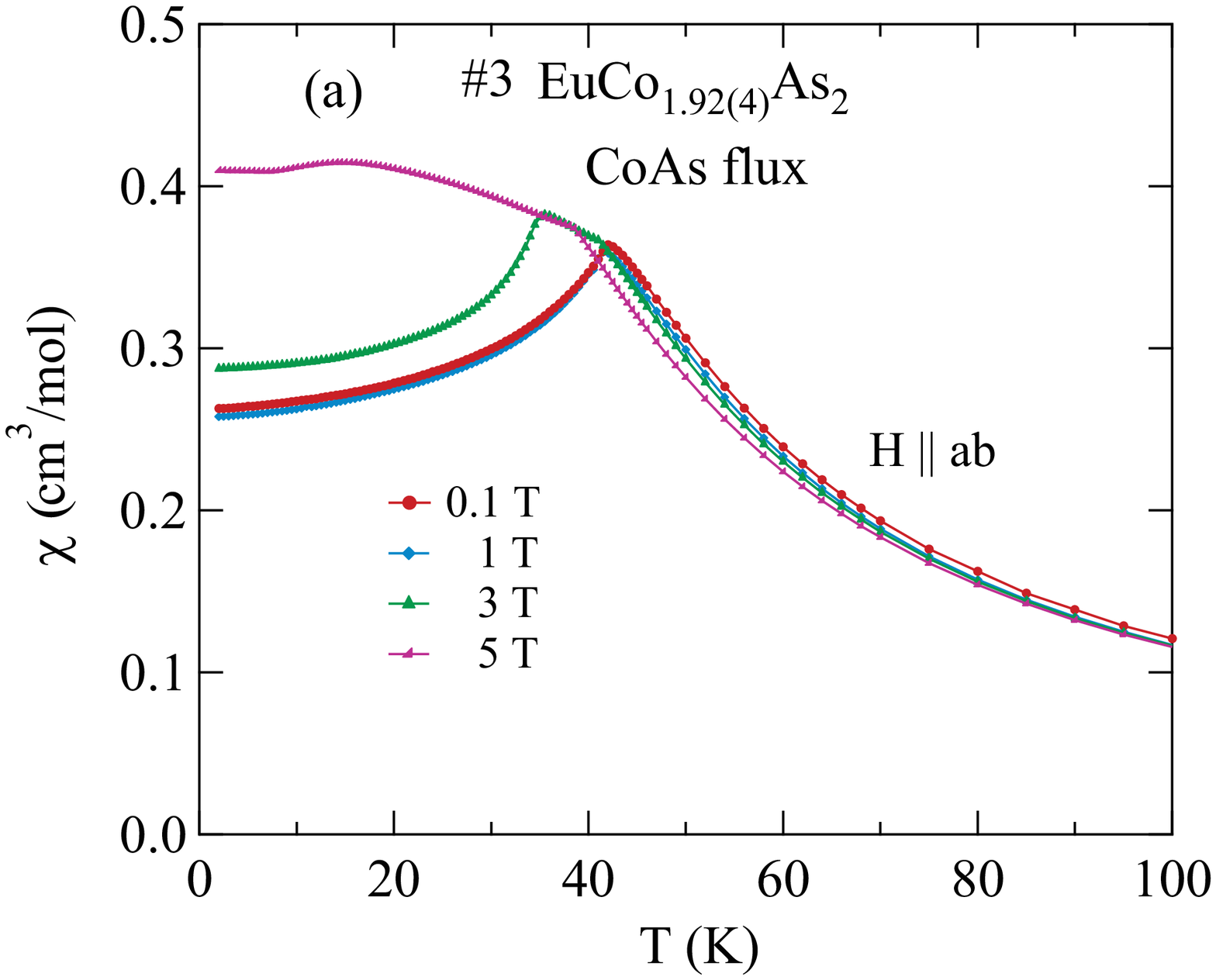}
\includegraphics[width=3.in]{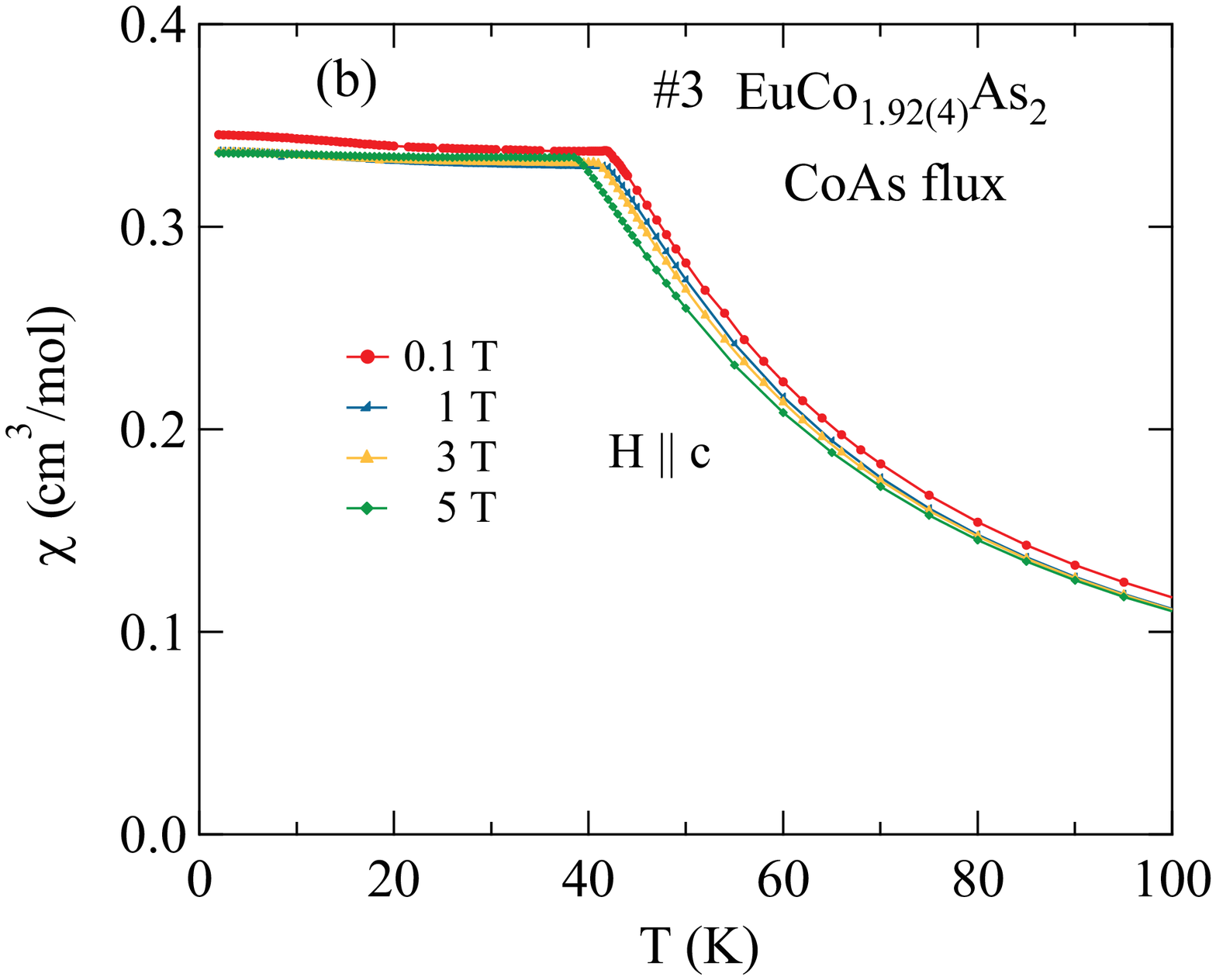}
\caption{Same as Fig.~\ref{Fig:MT_Snflux} except that the crystal measured is CoAs-flux-grown crystal \#3~$\rm EuCo_{1.92(4)}As_2$.}
\label{Fig:MT_CoAsflux}
\end{figure}

The $T$- and $H$-dependent magnetic susceptibility $\chi(T,H)$ was measured for one of the two crystals from each of the Sn-flux and CoAs-flux crystal growths. Figures~\ref{Fig:MT_Snflux} and~\ref{Fig:MT_CoAsflux} show  $\chi(T)$ of Sn-flux-grown crystal~\#2$~\rm EuCo_{1.99(2)}As_2$ and CoAs-flux-grown crystal \#3~$\rm EuCo_{1.92(4)}As_2$, respectively, for various values of $H$ applied in the $ab$ plane ($\chi_{ab}$, $H~||~ab$) and along the $c$ axis ($\chi_{c}$, $H~||~c$) for $2~{\rm K}\leq T\leq 100$~K\@. As shown in Figs.~\ref{Fig:MT_Snflux}(a) and~\ref{Fig:MT_CoAsflux}(a), the lowest-$T$ data reveal a \mbox{metamagnetic} (MM) transition for $H~||~ab$ between $H=3$~T and~5~T\@.  In addition, breaks in slope of $\chi(T)$ at each field are observed, signifying the $H$-dependent $T_{\rm N}$ which decreases with increasing~$H$ as expected for an AFM\@.  Figures~\ref{Fig:MT_Snflux}(b) and~\ref{Fig:MT_CoAsflux}(b) show that $T_{\rm N}$ is much less sensitive to~$H\parallel c$ than to $H\parallel ab$ as seen in Figs.~\ref{Fig:MT_Snflux}(a) and~\ref{Fig:MT_CoAsflux}(a).

\begin{figure}
\includegraphics[width=3.in]{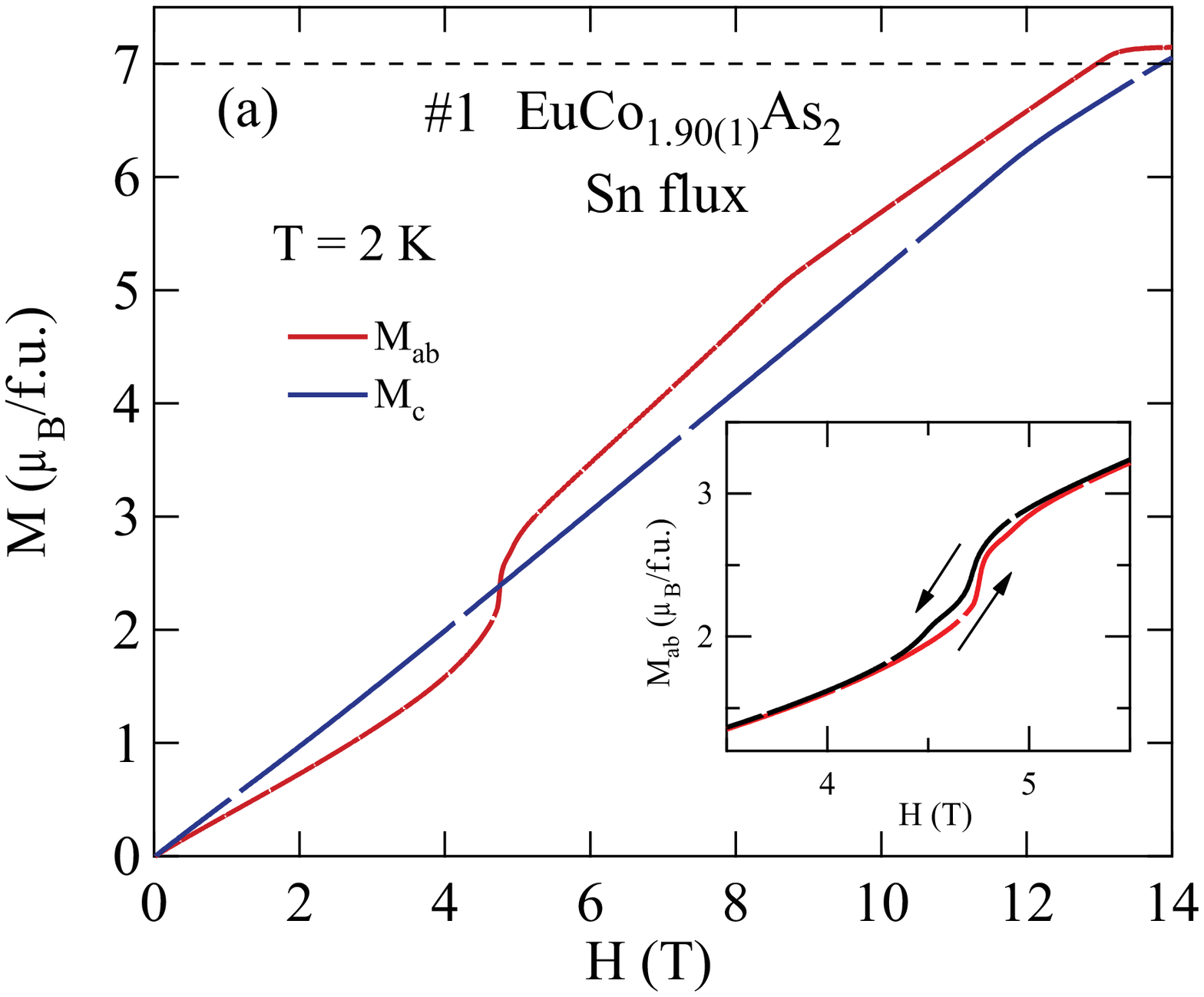}\vspace{0.1in}
\includegraphics[width=3.in]{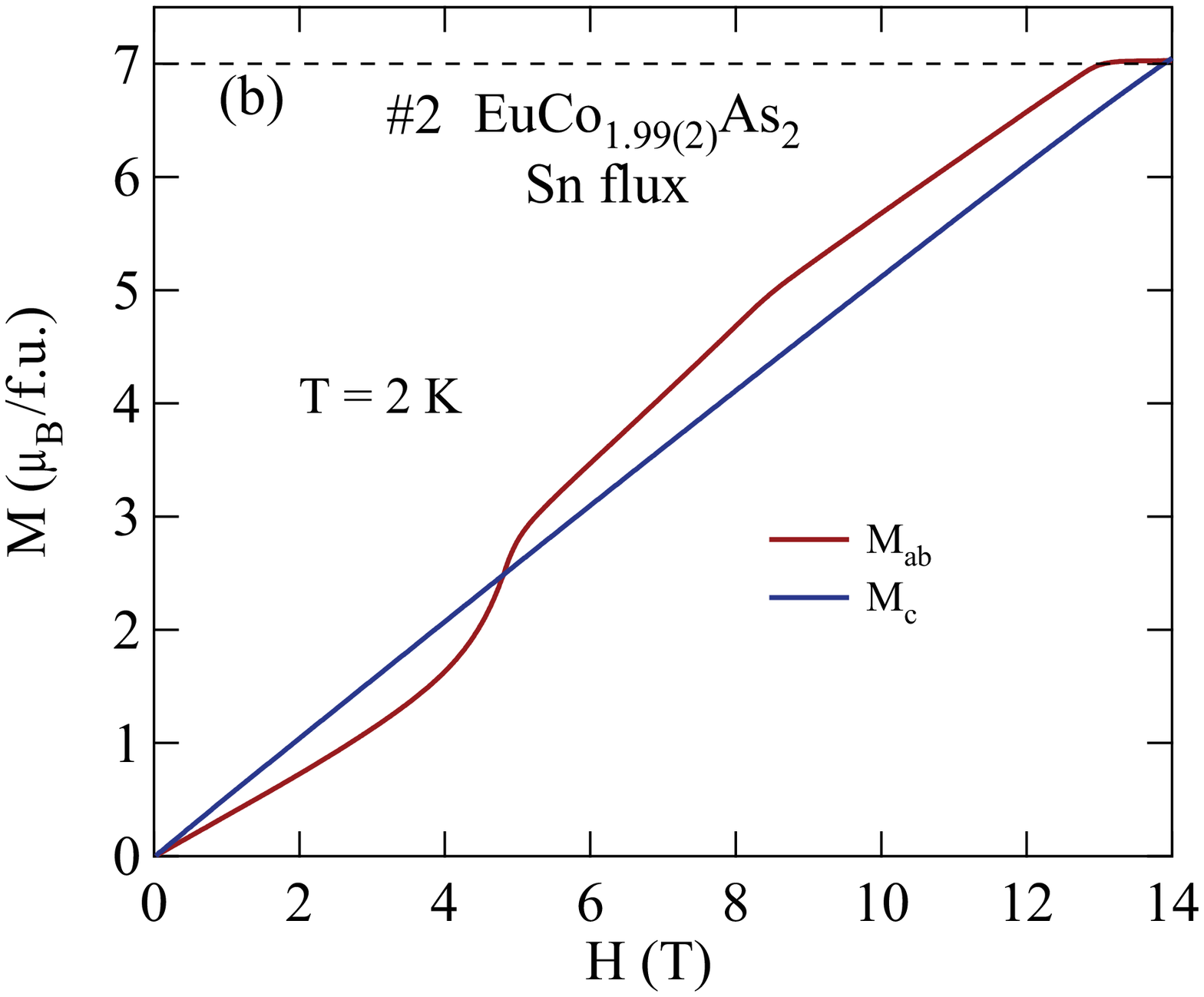}
\caption{Isothermal magnetization $M$ of Sn-flux-grown crystals (a)~\#1~EuCo$_{1.90(1)}$As$_2$ and (b)~\#2~EuCo$_{1.99}$As$_2$ as a function of applied magnetic field $H$ measured at 2~K for $H$ applied in the $ab$~plane ($M_{ab},~H\parallel ab$) and along the $c$~axis ($M_c,~H\parallel c$).}
\label{Fig:MH_Snflux_2K}
\end{figure}

\begin{figure}
\includegraphics[width=3.in]{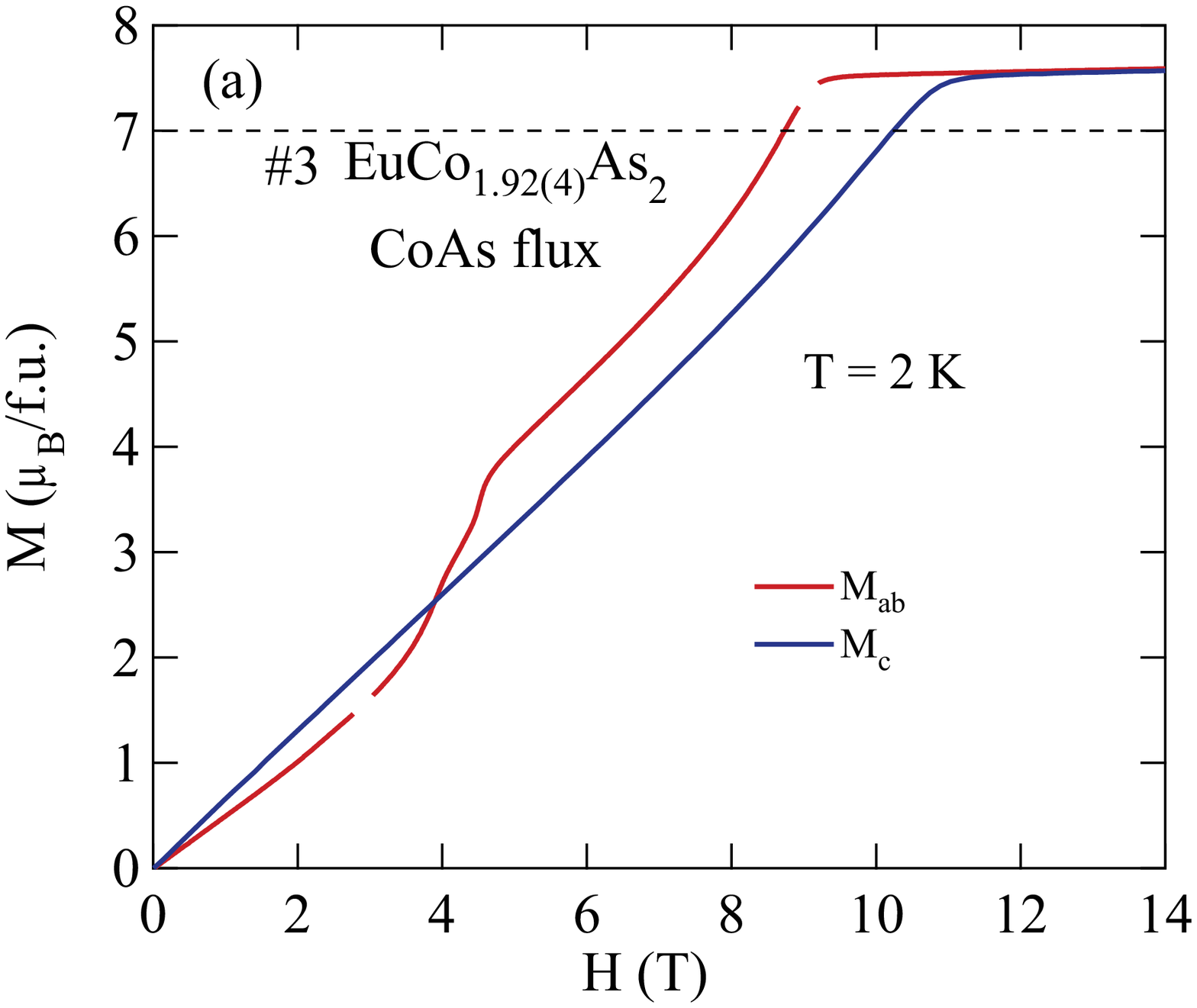}
\includegraphics[width=3.in]{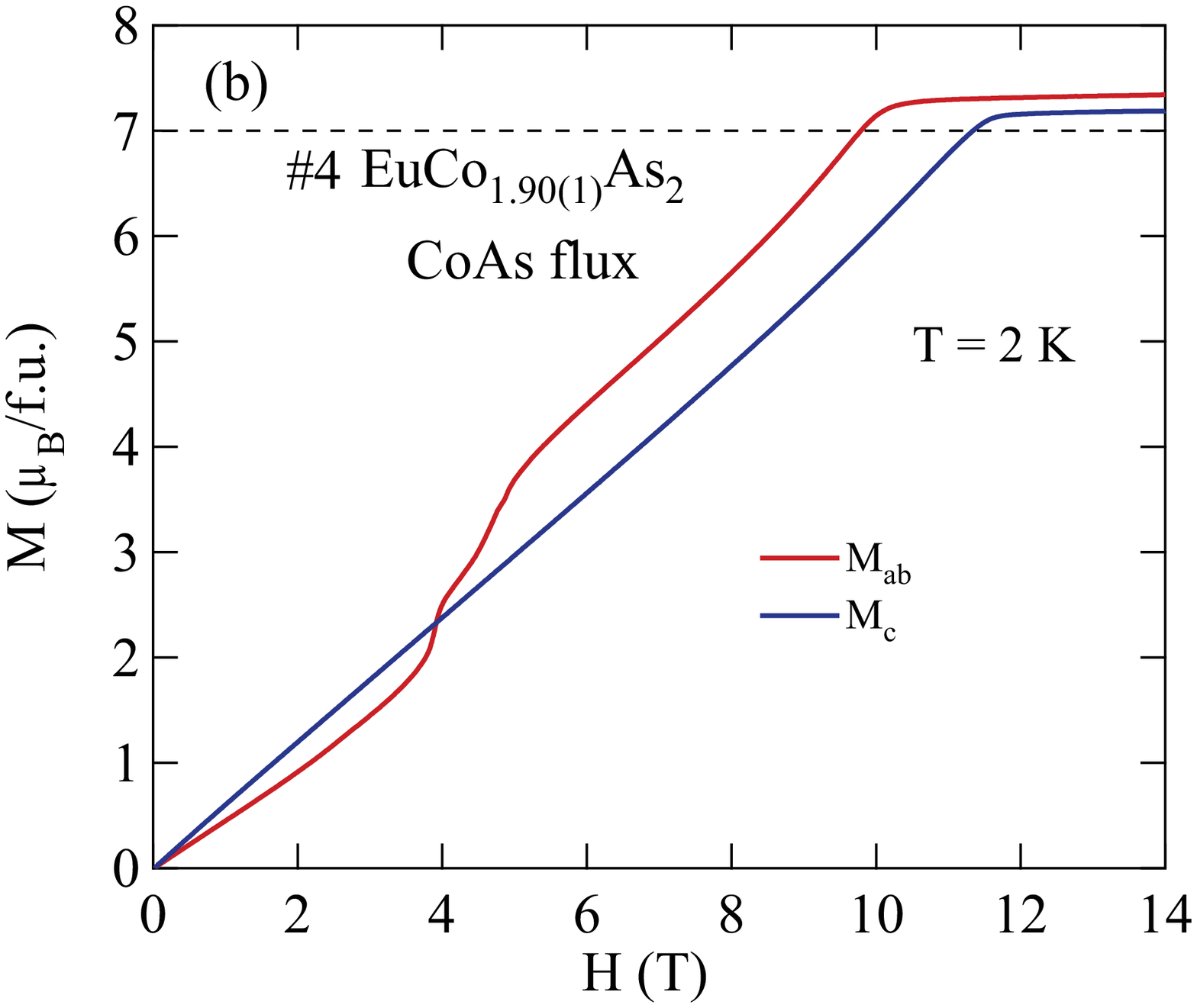}
\caption{Isothermal magnetization $M$ of CoAs-flux-grown crystals (a)~\#3~$\rm EuCo_{1.92(4)}As_2$ and (b)~\#4~$\rm EuCo_{1.90(2)}As_2$ as a function of applied magnetic field $H$at 2~K for $H$ applied in the $ab$~plane ($M_{ab},~H\parallel ab$) and along the $c$~axis ($M_c,~H\parallel c$).}
\label{Fig:MH_CoAsflux_2K}
\end{figure}

Figures~\ref{Fig:MH_Snflux_2K} and~\ref{Fig:MH_CoAsflux_2K} show $M(H)$ isotherms at $T=2$~K with $H$ applied in the $ab$ plane ($M_{ab}, H\parallel ab$) and along the $c$~axis ($M_c, H\parallel c$) obtained for the Sn-flux-grown crystals \#1~$\rm EuCo_{1.90(1)}As_2$ and \#2~$\rm EuCo_{1.99(2)}As_2$ (Fig.~\ref{Fig:MH_Snflux_2K}), and for the CoAs-flux-grown crystals \#3~$\rm EuCo_{1.92(4)}As_2$ and \#4~$\rm EuCo_{1.90(2)}As_2$ (Fig.~\ref{Fig:MH_CoAsflux_2K}). The $M_c(H)$ data are nearly linear in field as predicted at $T \ll T\rm _N$ by MFT for a helix with the applied field along the helix axis, reaching saturation at the perpendicular critical field $H_{\rm c\perp} \sim 10$--15~T, depending on the sample and field direction. 

The $M_{ab}$($H$) isotherms at $T=$ 2~K in Figs.~\ref{Fig:MH_Snflux_2K} and~\ref{Fig:MH_CoAsflux_2K} show what appears to be a field-induced spin-flop (SF) transition at a field $H\rm_{SF}$, with a small hysteresis [see inset of Fig.~\ref{Fig:MH_Snflux_2K}(a)]. The magnetic moment attains its saturation moment $\mu_{\rm sat}$ at the critical field $H\rm_c$ which separates the AFM from the paramagnetic (PM) phases. An additional transition of unknown origin at a field $H_{\rm MM}$ is also seen, with $H_{\rm SF} < H_{\rm MM} < H_{\rm c}$. 

\begin{figure}
\includegraphics[width=3.in]{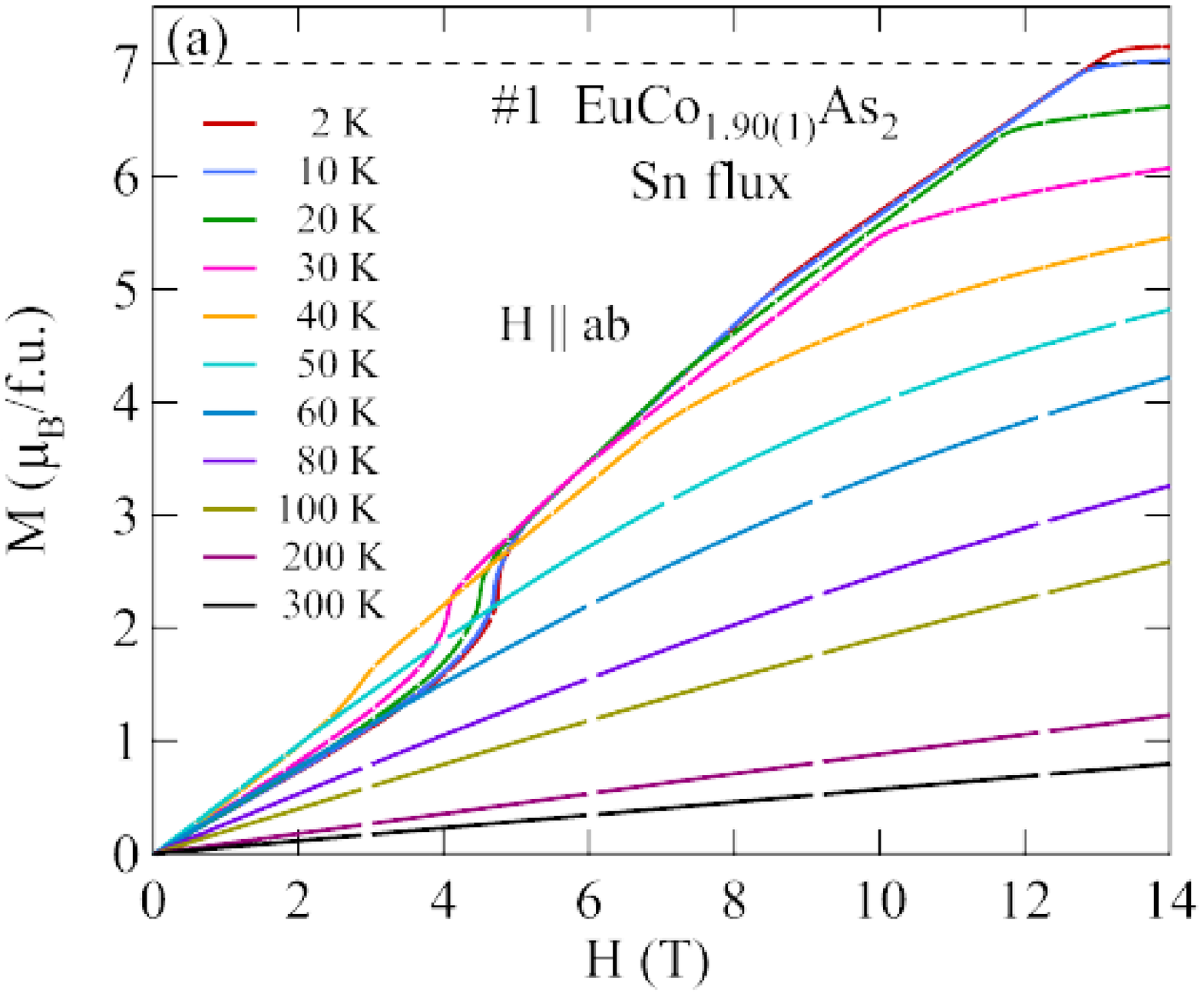}\vspace{0.1in}
\includegraphics[width=3.in]{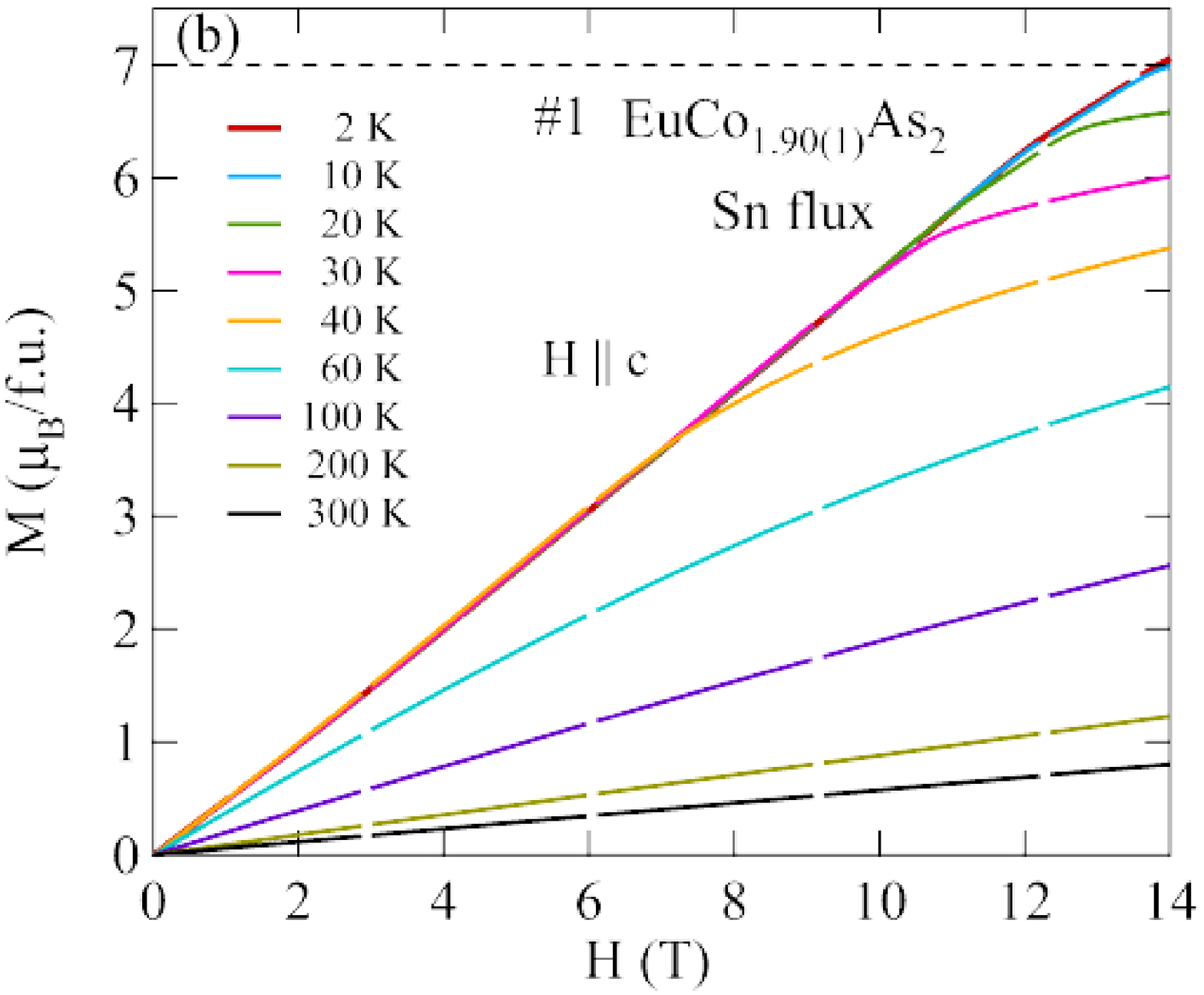}
\caption{Isothermal magnetization $M$ of Sn-flux-grown crystal \#1~EuCo$_{1.90(1)}$As$_2$ as a function of magnetic field $H$ at the indicated temperatures for $H$ applied (a)~in the $ab$~plane ($M_{ab},~H\parallel ab$) and (b)~along the $c$~axis ($M_c,~H\parallel c$).}
\label{Fig:MH_Snflux1}
\end{figure}

\begin{figure}
\includegraphics[width=3.in]{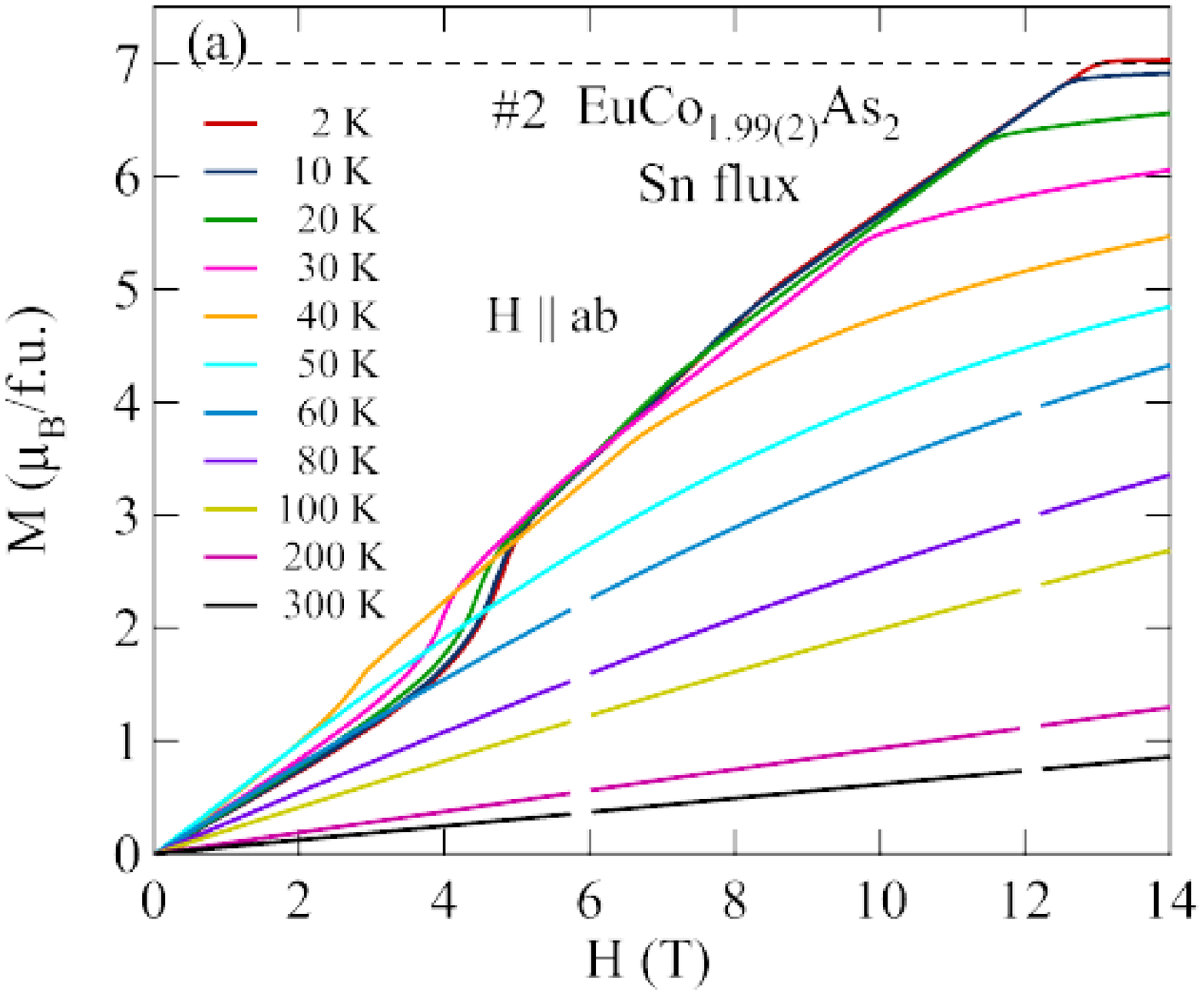}\vspace{0.1in}
\includegraphics[width=3.in]{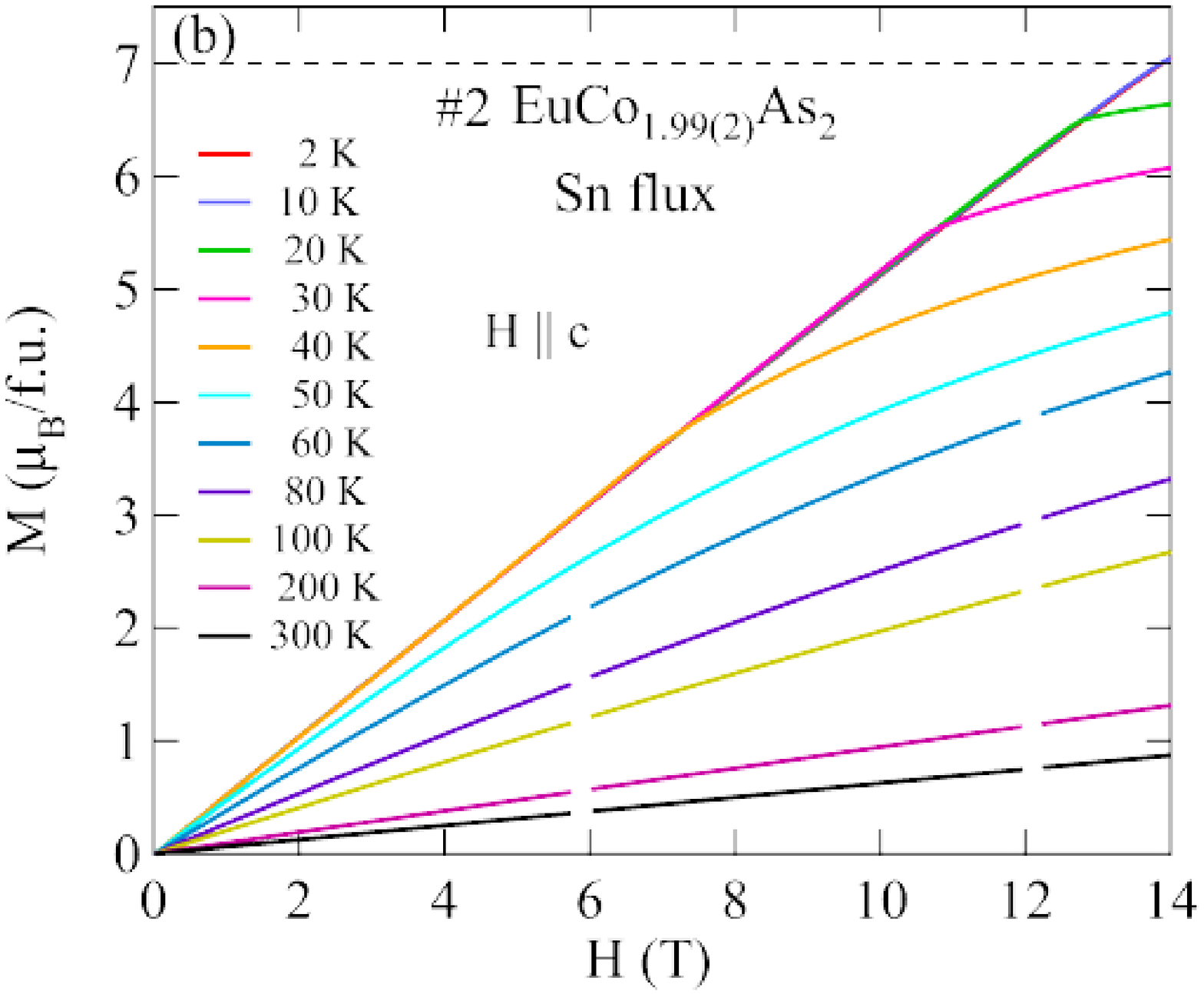}
\caption{Isothermal magnetization $M$ of Sn-flux-grown crystal \#2~EuCo$_{1.99(2)}$As$_2$ as a function of magnetic field $H$ at the indicated temperatures for $H$ applied (a)~in the $ab$~plane ($M_{ab},~H\parallel ab$) and (b) along the $c$~axis ($M_c,~H\parallel c$).}
\label{Fig:MH_Snflux2}
\end{figure}

\begin{figure}
\includegraphics[width=3.in]{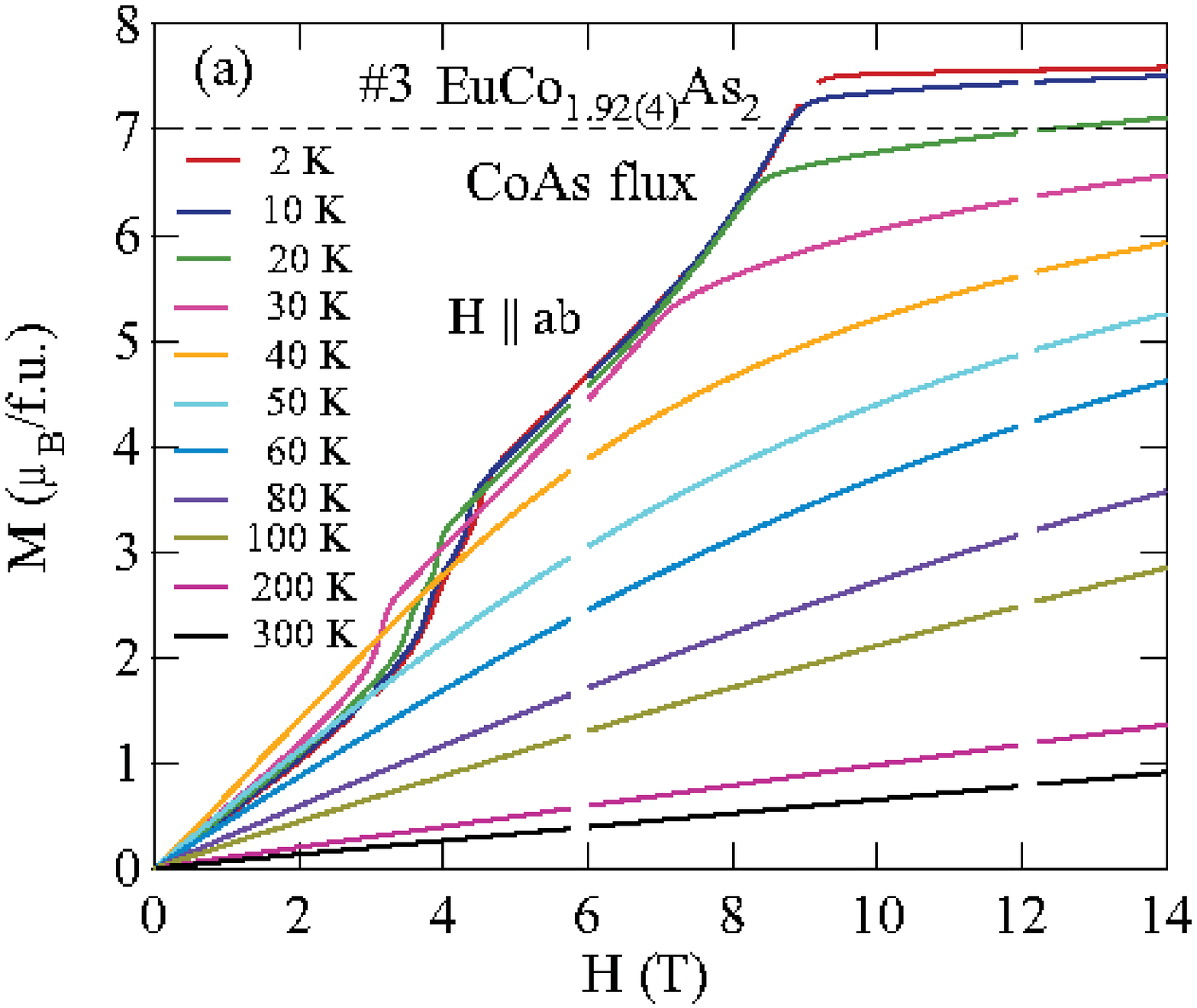}\vspace{0.1in}
\includegraphics[width=3.in]{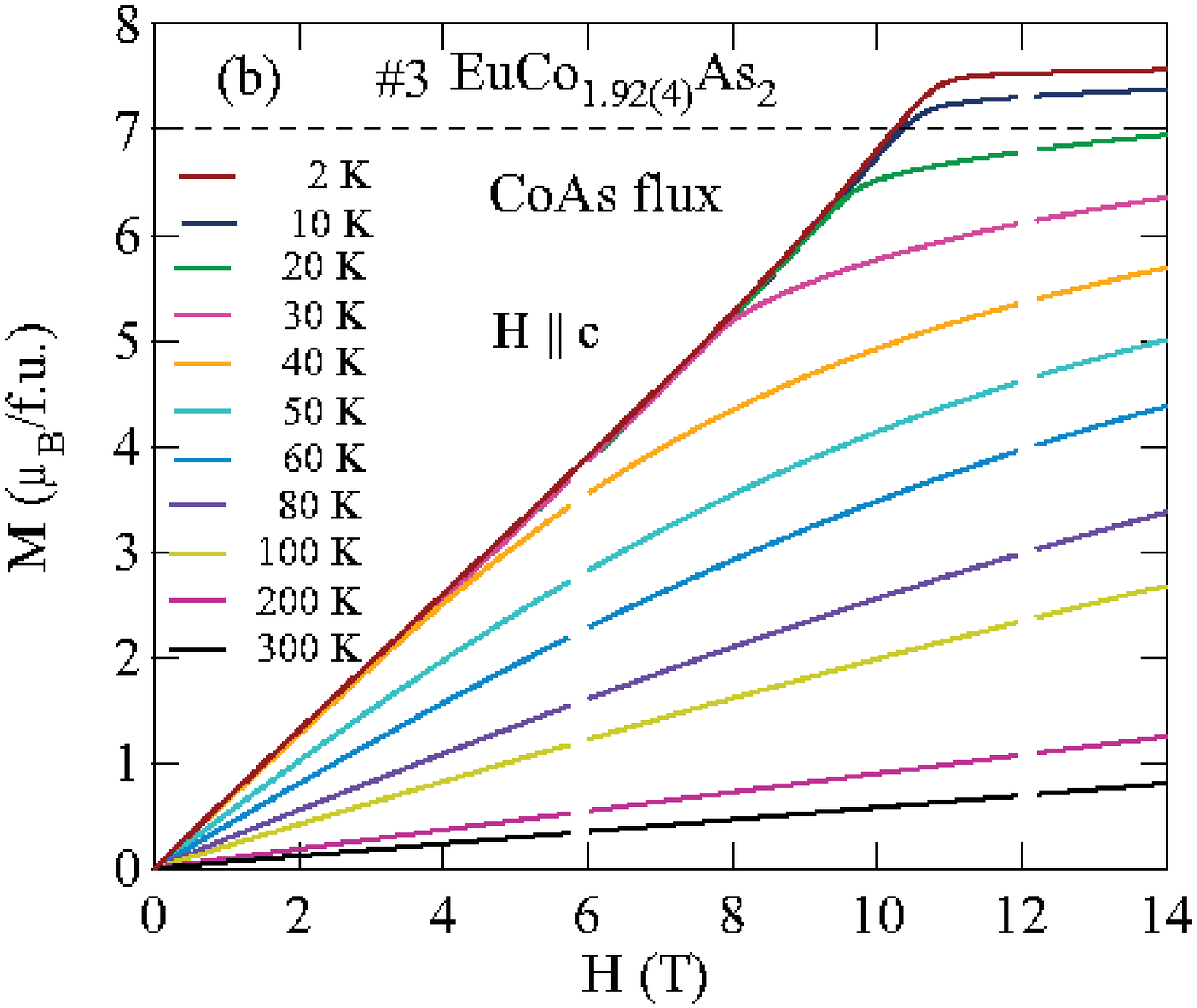}
\caption{(a)~Isothermal magnetization $M$ of CoAs-flux-grown crystal \#3~EuCo$_{1.92(4)}$As$_2$ as a function of magnetic field $H$ applied (a)~in the $ab$~plane ($H \parallel ab$) and (b)~along the $c$~axis ($H \parallel c$) at the indicated temperatures. }
\label{Fig:MH_CoAsflux1}
\end{figure}

\begin{figure}
\includegraphics[width=3.in]{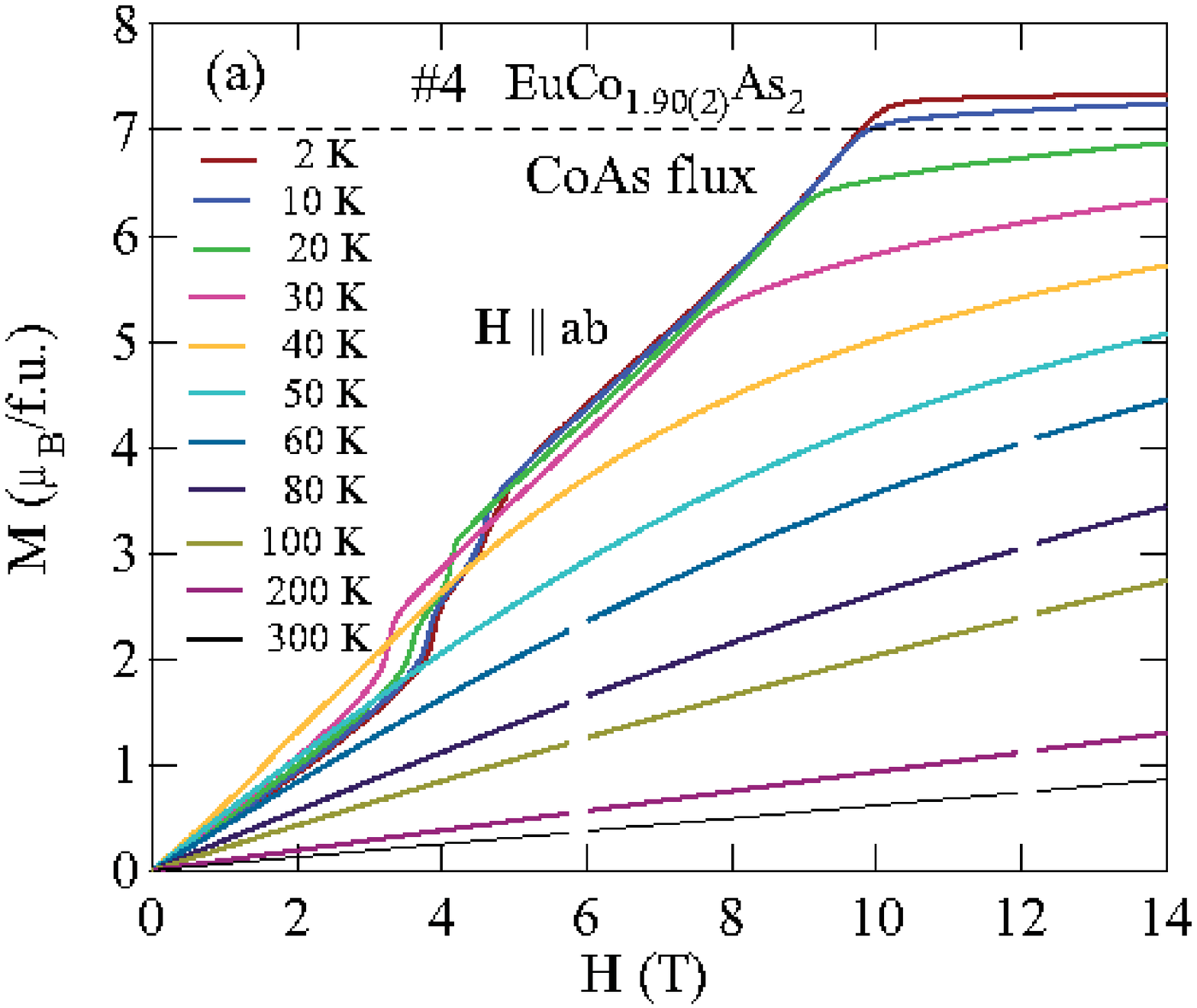}\vspace{0.1in}
\includegraphics[width=3.in]{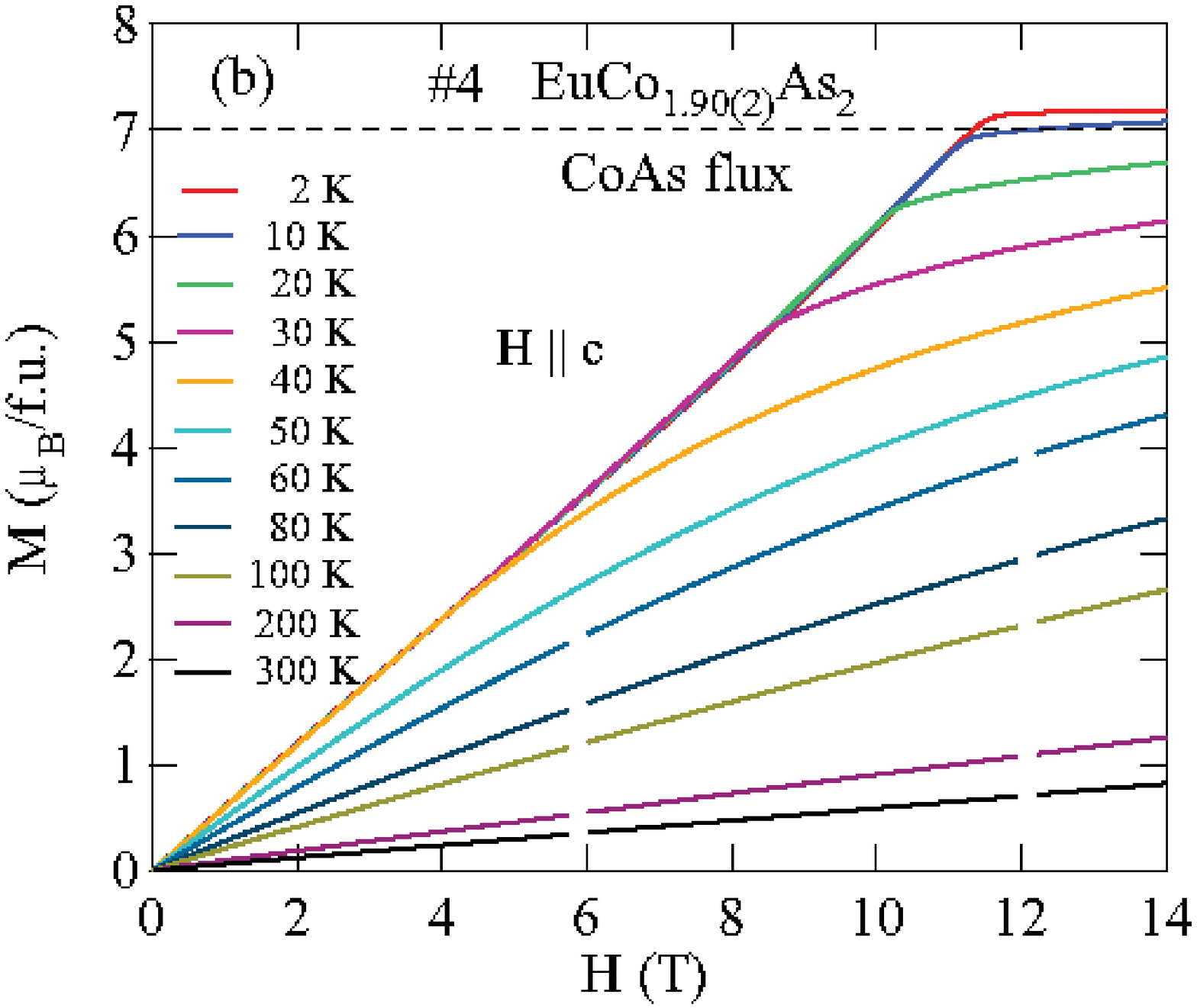}
\caption{(a)~Isothermal magnetization $M$ of CoAs-flux-grown crystal \#4~EuCo$_{1.90(2)}$As$_2$ as a function of magnetic field $H$ applied (a) in the $ab$~plane ($H \parallel ab$) and (b) along the $c$~axis ($H \parallel c$) at the indicated temperatures.}
\label{Fig:MH_CoAsflux2}
\end{figure}

\begin{figure}
\includegraphics[width=1.7in]{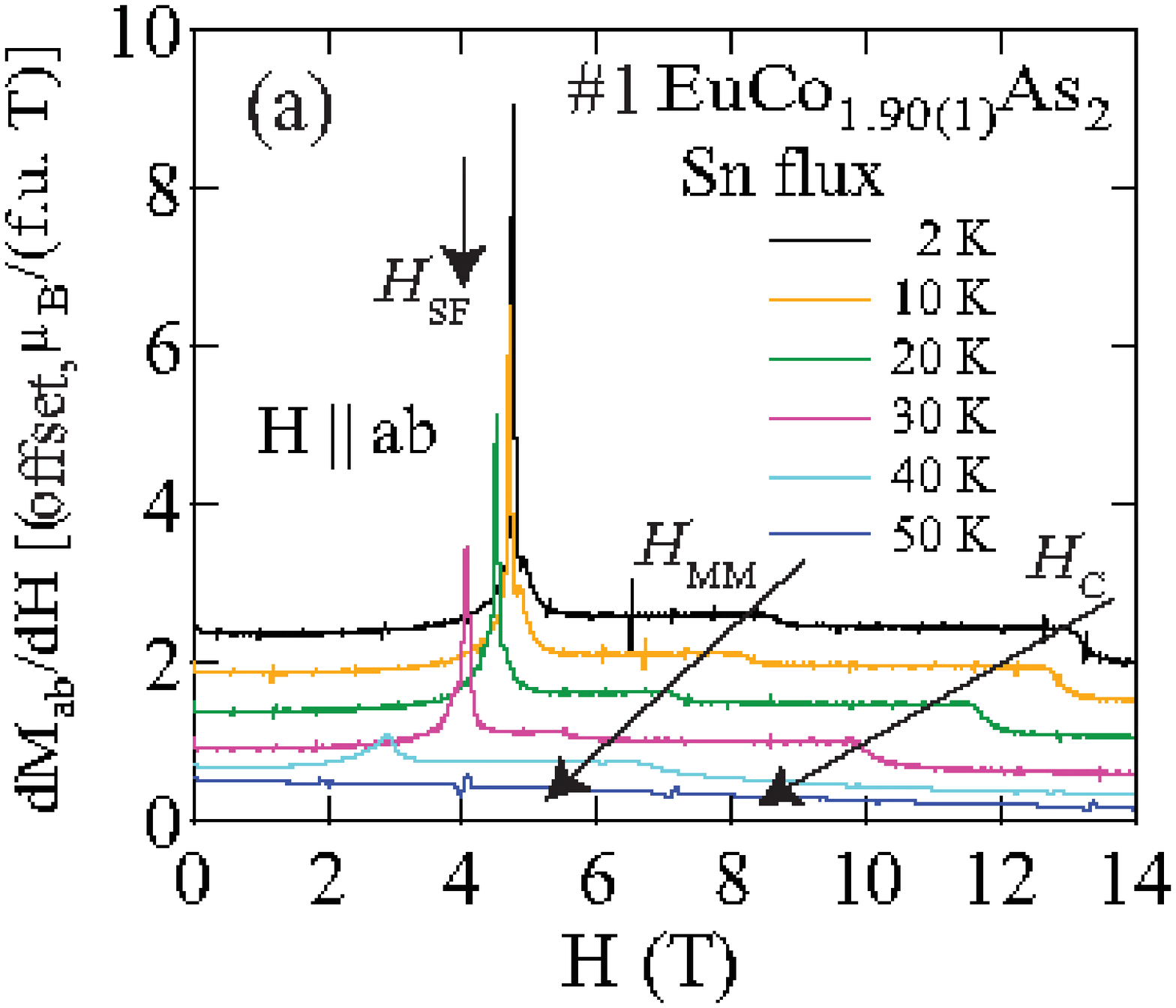}\includegraphics[width=1.7in]{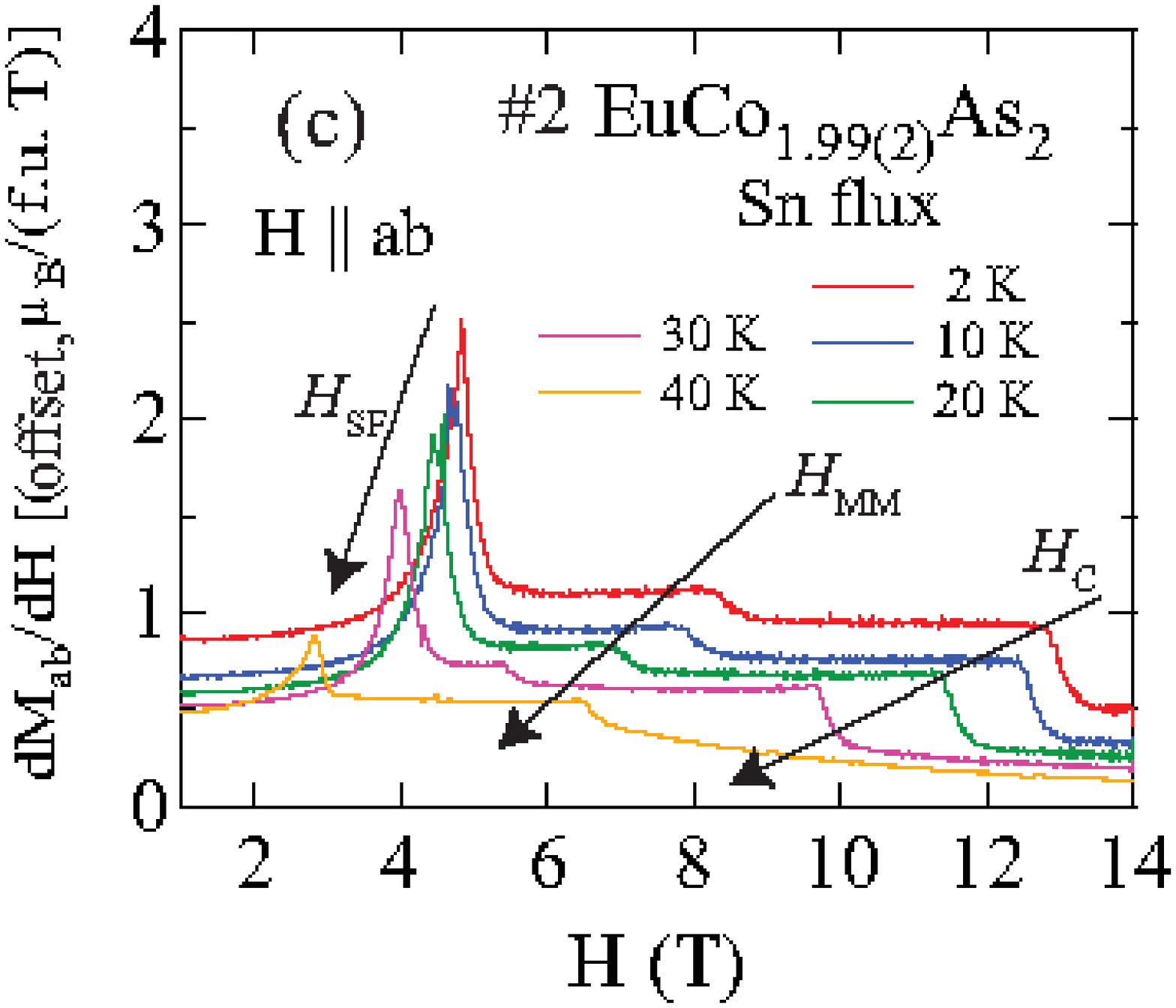}
\includegraphics[width=1.7in]{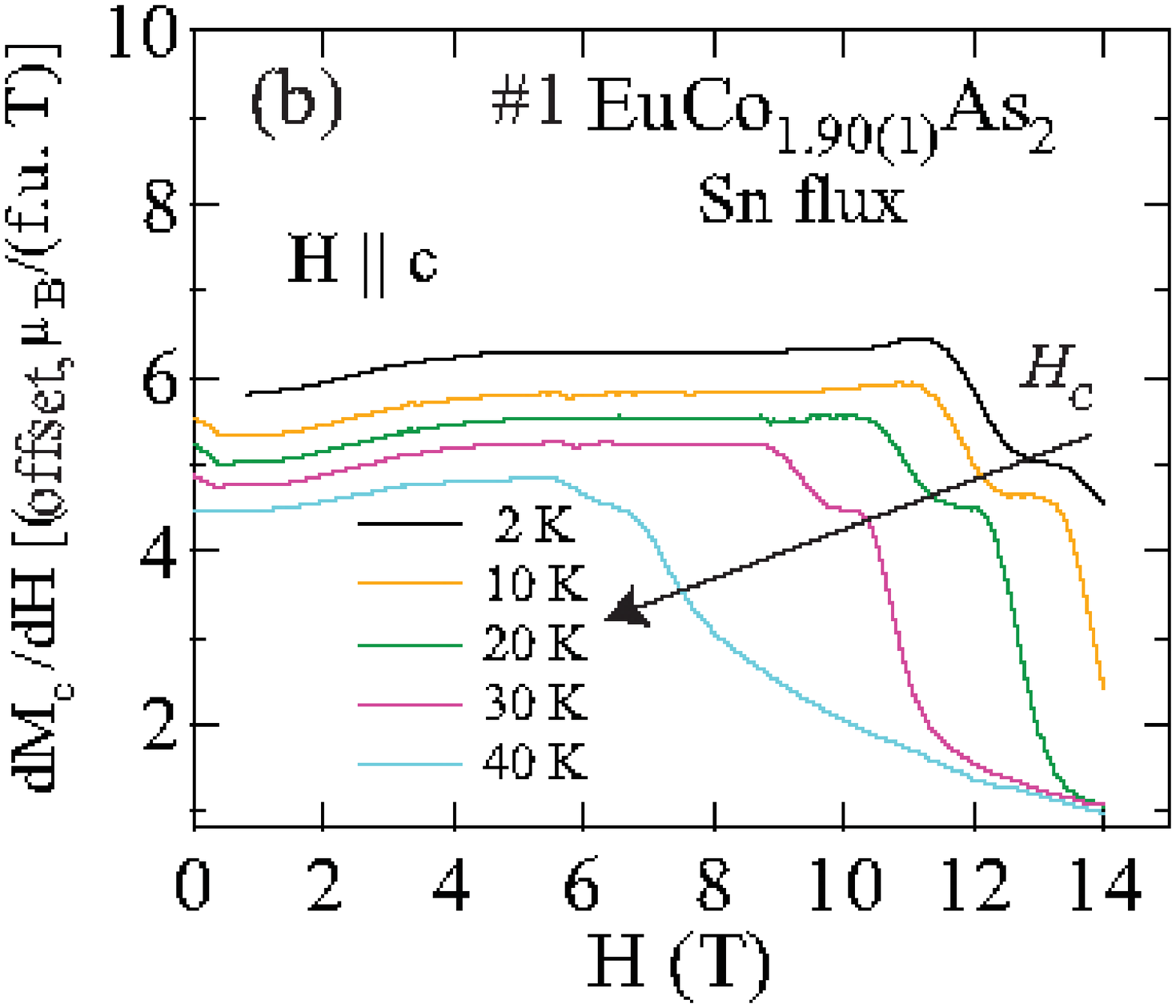}\includegraphics[width=1.7in]{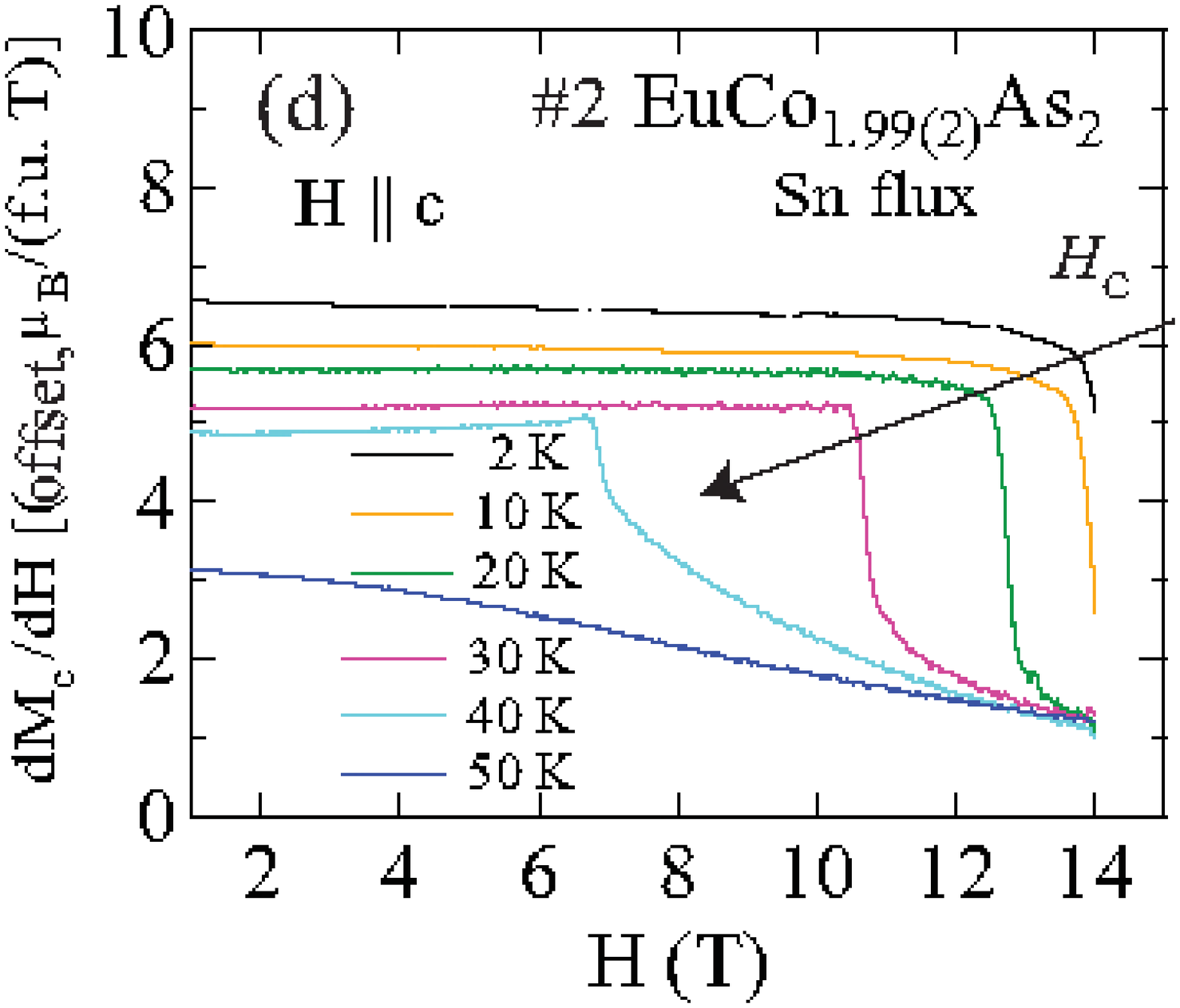}
\caption{Derivative $dM/dH$ versus $H$ for Sn-flux-grown crystals \#1~EuCo$_{1.90(1)}$As$_2$ [(a) $H\parallel ab$, (b) $~H\parallel c$] and \#2~EuCo$_{1.99(2)}$As$_2$ [(c)~$H\parallel ab$, (d)~$H\parallel c$] for several temperatures~$T$ as indicated.}
\label{Fig:dMH_Snflux}
\end{figure}

\begin{figure}
\includegraphics[width=1.7in]{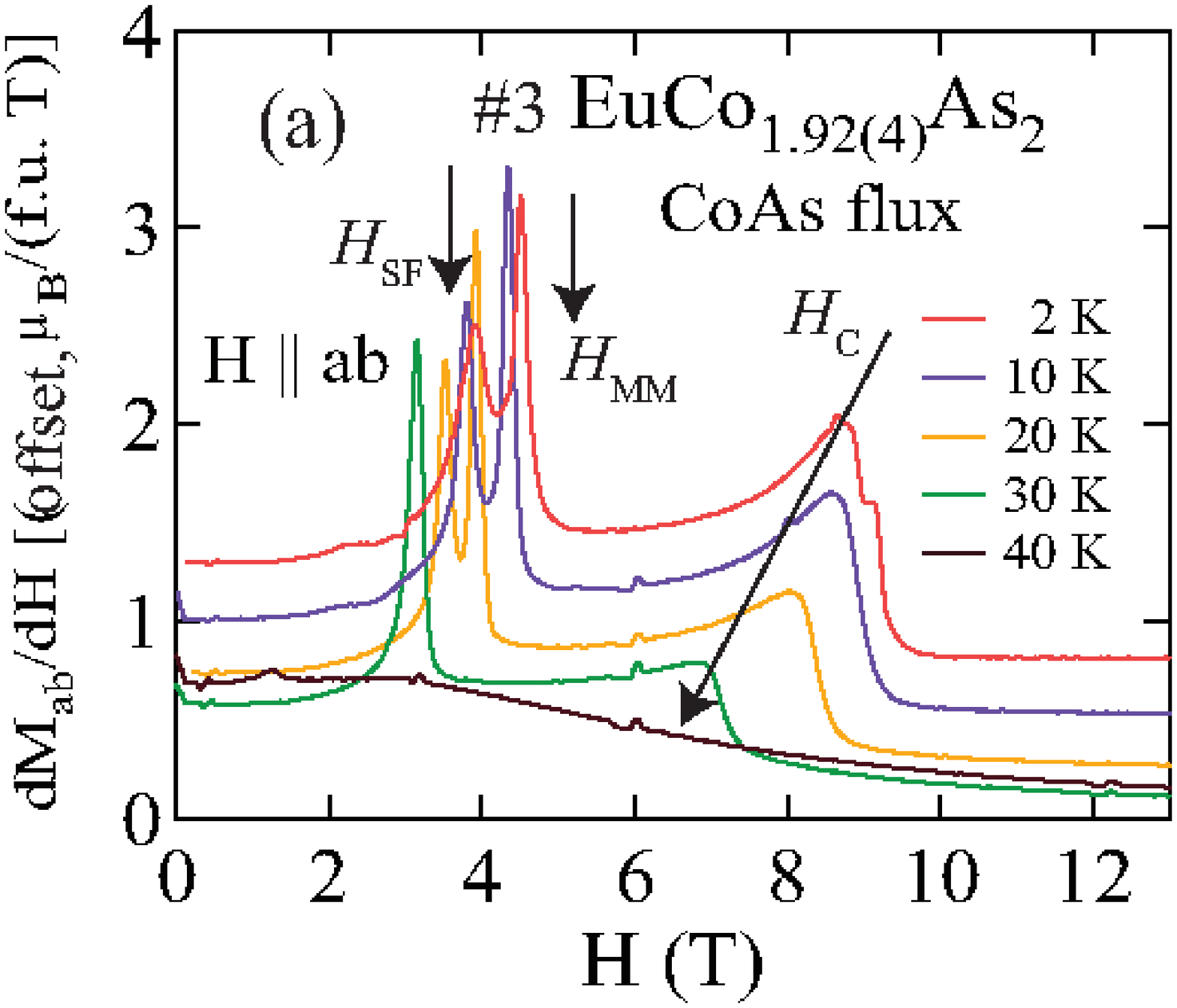}\includegraphics[width=1.7in]{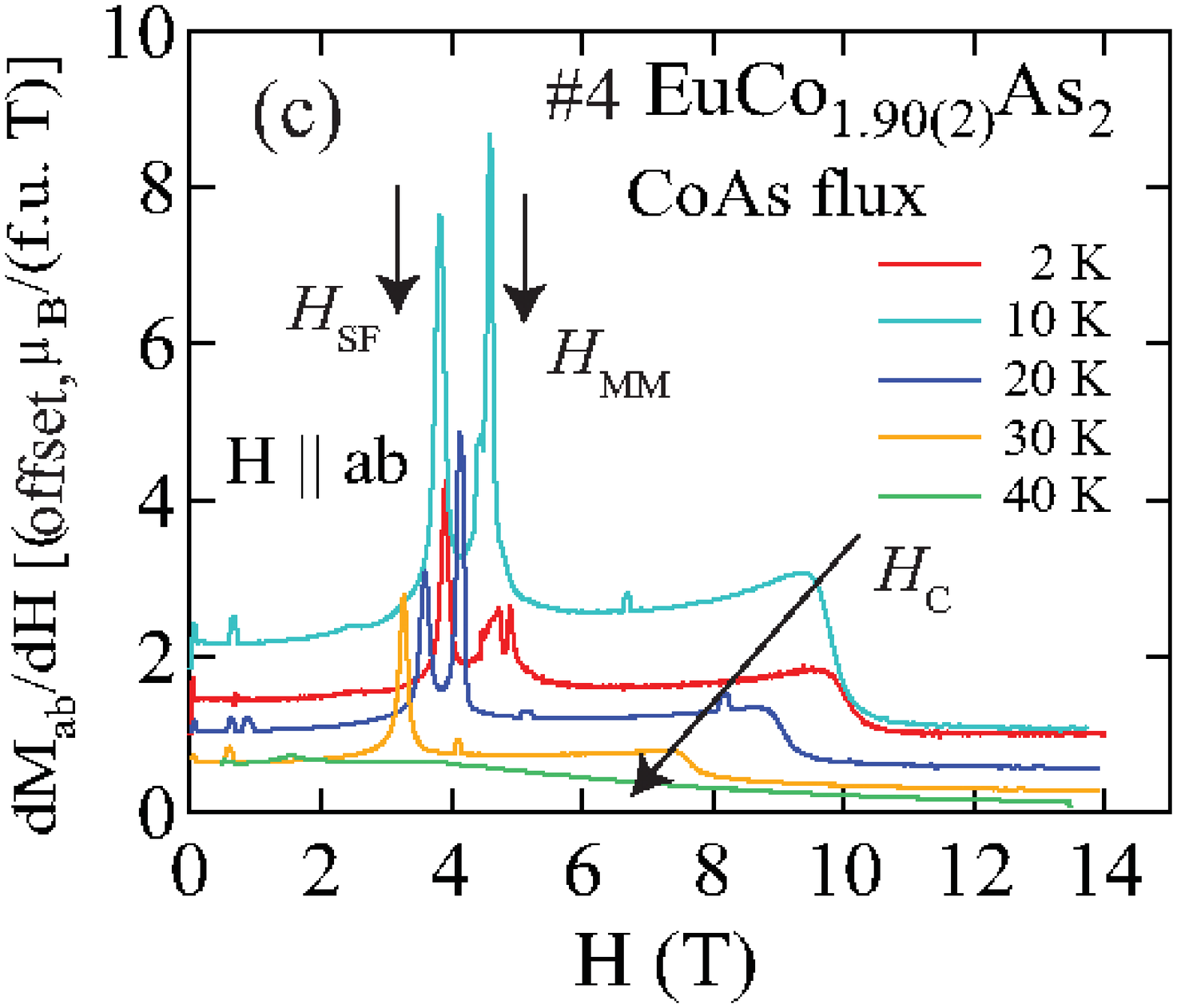}
\includegraphics[width=1.7in]{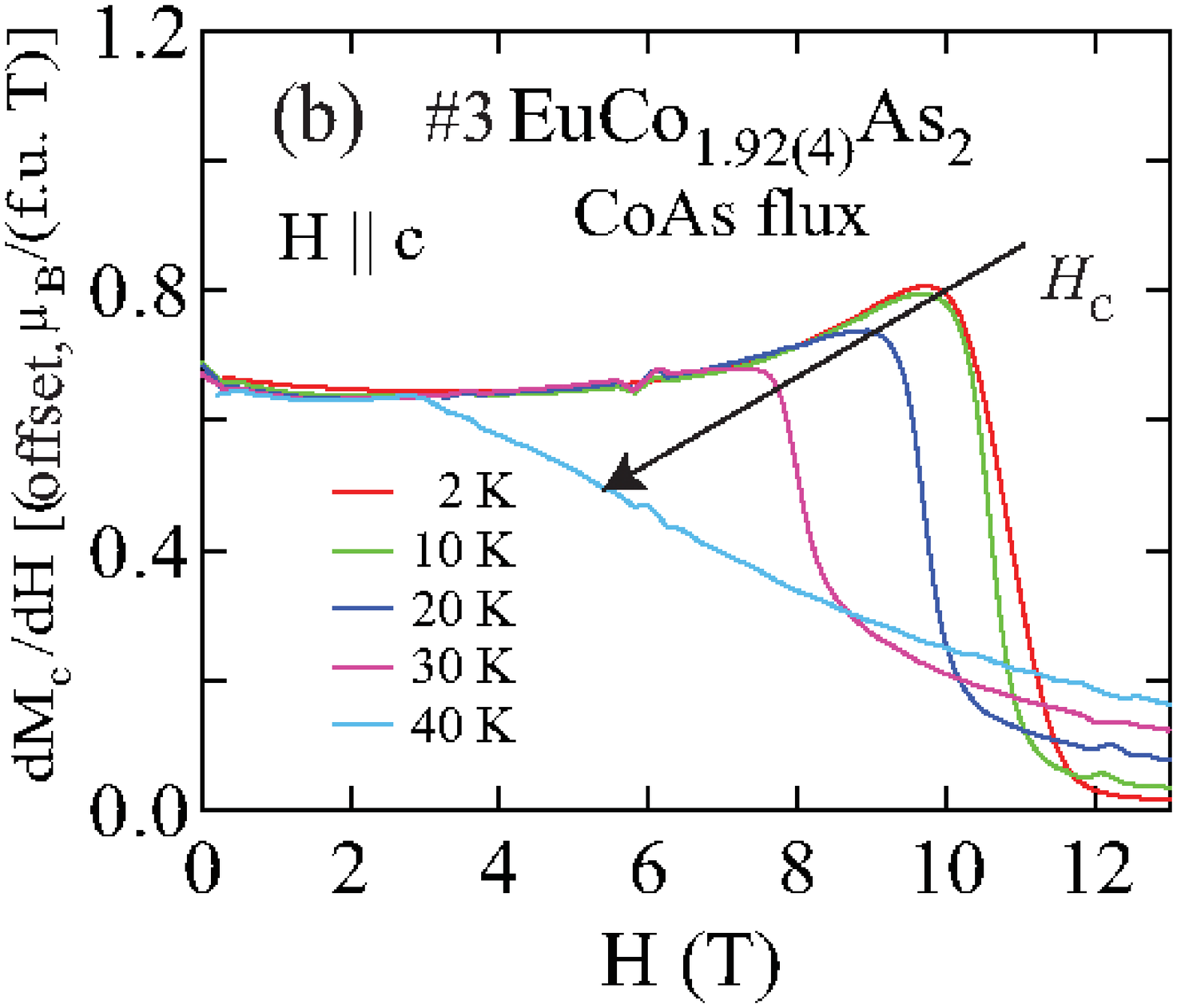}\includegraphics[width=1.7in]{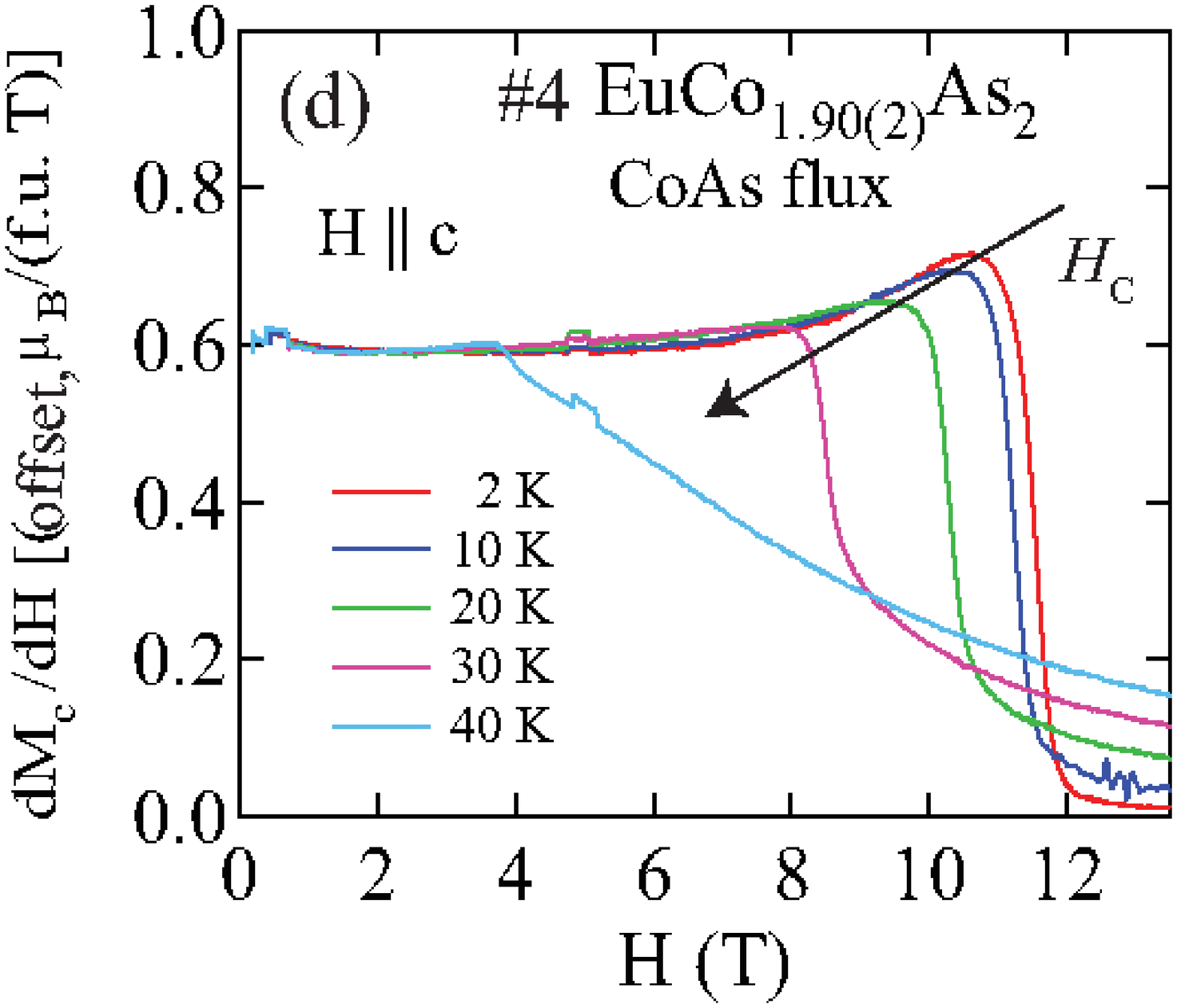}
\caption{Derivative $dM/dH$ versus $H$ of CoAs-flux-grown crystals \#3~EuCo$_{1.92}$As$_2$ [(a) $H~||~ab$, (b) $H~||~c$] and \#4~EuCo$_{1.90}$As$_2$ [(c) $H~||~ab$, (d) $H~||~c$] for several temperatures as indicated.}
\label{Fig:dMH_CoAsflux}
\end{figure}

\begin{table*}
\caption{\label{Tab:MH} Spin-flop transition field $H_{\rm SF}$, metamagnetic transition field $H_{\rm MM}$, critical field $H\rm_c$, and saturation moment $\mu\rm_{sat}$ at $T=2$~K of \ecaa\ single crystals determined from isothermal $M(H)$ data for fields $H\parallel ab$ and $H\parallel c$. }

\begin{ruledtabular}
\begin{tabular}{cccccc}
Crystal								& Field			& $H\rm_{SF}$	&  $H\rm_{MM}$  &  $H\rm_c$  	& $\mu_{\rm_{sat}}$		\\
Designation							& Direction		&     (T)    	&	(T)		&     (T)     & ($\mu\rm_B$/Eu)	\\
\hline
\#1 EuCo$_{1.90(1)}$As$_2$\footnotemark[1]	& $H\parallel ab$ 	& 4.75		& 8.46		& 13.04 		& 7.15	\\
									& $H\parallel c$ 	& 			&			& 13.32		& 7.05	\\
\#2 EuCo$_{1.99(2)}$As$_2$\footnotemark[2]	& $H\parallel ab$ 	& 4.8		& 8.2		& 12.8 		& 7.03	\\
									& $H\parallel c$ 	&			&			& 13.7		& 7.05 	\\
\#3 EuCo$_{1.92(4)}$As$_2$\footnotemark[3]	& $H\parallel ab$ 	& 3.9		& 4.5		& 8.78 		& 7.59	\\
									& $H\parallel c$ 	&			&			& 9.9		& 7.57	\\
\#4 EuCo$_{1.90(2)}$As$_2$\footnotemark[4]	& $H\parallel ab$ 	& 3.8		& 4.6		& 9.5  		& 7.34	\\
									& $H\parallel c$ 	&			&			& 10.86		& 7.19	\\
\#5 EuCo$_{1.92(1)}$As$_2$\footnotemark[4]	& $H\parallel ab$ 	& 3.86		& 4.47		& 8.75  		& 7.50	\\
									& $H\parallel c$ 	&			&			& 9.96		& 7.58	\\
\end{tabular}
\end{ruledtabular}
\footnotetext[1]{Grown in Sn flux}
\footnotetext[2]{Grown in Sn flux with H$_2$-treated Co powder}
\footnotetext[3]{Grown in CoAs flux with H$_2$-treated Co powder}
\footnotetext[4]{Grown in CoAs flux}

\end{table*}

The detailed $M$($H$) isotherms at many temperatures from 2~K to 300~K of Sn-flux-grown crystals \#1~EuCo$_{1.90(1)}$As$_2$ and \#2~EuCo$_{1.99(2)}$As$_2$ are shown in Figs.~\ref{Fig:MH_Snflux1} and \ref{Fig:MH_Snflux2}, respectively, and those of CoAs-flux-grown crystals \#3~EuCo$_{1.92(4)}$As$_2$ and \#4~EuCo$_{1.90(2)}$As$_2$ are shown in Figs.~\ref{Fig:MH_CoAsflux1} and \ref{Fig:MH_CoAsflux2}, respectively, where parts~(a) and~(b) of each of the four figures are for $H\parallel ab\ (M_{ab})$ and  $H\parallel c\ (M_c)$, respectively.  For the Sn-grown crystals, $M_c(H)$ data in Figs.~\ref{Fig:MH_Snflux1}(b) and \ref{Fig:MH_Snflux2}(b) show a negative curvature between 40 and 60~K, but a proportional behaviour of $M_c$($H$) is eventually observed at higher temperature ($T>80$~K). On the other hand, $M_{ab}$($H$) in Figs.~\ref{Fig:MH_Snflux1}(a) and \ref{Fig:MH_Snflux2}(a) show clear spin flop and metamagnetic transitions at $H\rm_{SF}$ and $H\rm _{MM}$, respectively, for $T\ll 40~$K\@. These SF and MM transitions shift to lower field with increasing temperature. As shown in Figs.~\ref{Fig:MH_CoAsflux1} and \ref{Fig:MH_CoAsflux2}, the CoAs-flux-grown crystals exhibit similar behaviors.

The transition fields $H\rm_{SF}$, $H\rm_{MM}$ and $H\rm_c$ versus temperature are taken to be the fields at which $dM$/$dH$  versus $H$ exhibits a peak or a discontinuity (shown in Fig.~\ref{Fig:dMH_Snflux} for Sn-flux-grown crystals and Fig.~\ref{Fig:dMH_CoAsflux} for CoAs-flux-grown crystals). The results are listed in Table~\ref{Tab:MH}. One sees that $H_{\rm c\parallel}$ is different from $H_{\rm c \perp}$ and the saturation moments of these crystals are larger than the theoretical Eu$^{+2}$ value $\mu{\rm_{sat}} = gS\mu_{\rm B}$/Eu = 7 $\mu_{\rm B}$/Eu, where $g=2$ and $S=7/2$. As seen later in Sec.~\ref{Sec:ElecStruct}, this enhancement is due to $d$-electron spin polarization by the ordered Eu spins.

\subsection{\label{Sec:MDIs} Influence of Anisotropy on the Magnetic Properties}

From the above magnetic susceptibility and magnetization data, it is clear that magnetic anisotropy has an important influence on the results.  For example, without anisotropy the spin-flop phase for fields in the $ab$~plane would be the stable phase for all fields less than $H_{\rm c}$.  Here the anisotropy must give rise to an easy $ab$~plane (XY anisotropy) because the helix axis is $c$~axis and the moments are ferromagnetically-aligned within a given $ab$~plane.

Here we estimate the strength of the anisotropy in terms of a generic classical anisotropy field.  The formulas used here are derived in Ref.~\cite{Johnston2017b}.  From the value of the anisotropy field parameter~$h_{\rm A1}$ to be defined below, we estimate the influence of the anisotropy on the N\'eel temperature that would occur in the absence of anisotropy.

The definitions and predictions for this type of anisotropy in the presence of Heisenberg exchange interactions are given in Ref.~\cite{Johnston2017b} for systems comprised of identical crystallographically-equivalent spins as applies to the Eu sublattice in \eca.  The XY anisotropy field ${\bf H}_{{\rm A}i}$ seen by given moment~$\vec{\mu}_i$ making an angle $\phi_i$ with the positive $x$~axis ($a$~axis here, where the $z$~axis is the $c$~axis) is given by an amplitude $H_{{\rm A0}i}$ times the projection of the moment onto the $xy$ plane, i.e.,
\be
{\bf H}_{{\rm A}i} = H_{{\rm A0}i} \sin\theta_i(\cos\phi_i\,\hat{\bf i} - \sin\phi_i\,\hat{\bf j}).
\ee 
The amplitude is expressed in terms of a more fundamental anisotropy field $H_{\rm A1}$ as
\be
H_{{\rm A0}i}(T) = \frac{3H_{\rm A1}}{S+1}\bar{\mu}_i(T),
\ee
where the reduced ordered and/or field-induced moment~$\bar{\mu}_i$ is
\be
\bar{\mu}_i(T) \equiv \frac{\mu_i(T)}{\mu_{\rm sat}} = \frac{\mu_i(T)}{g\mu_{\rm B}S},
\ee
where $\mu_i(T)$ is the $T$-dependent magnitude of $\vec{\mu}_i$.  Finally, $H_{\rm A1}$ is expressed in reduced form $h_{\rm A1}$ as
\be
h_{\rm A1} = \frac{g\mu_{\rm B}H_{\rm A1}}{k_{\rm B}T_{{\rm N}J}},
\ee
where $T_{{\rm N}J}$ is the value that the N\'eel temperature would have been due to Heisenberg exchange interactions alone (in the absence of anisotropy).  Another parameter of the theory is
\be
f_J = \frac{\theta_{{\rm p}J}}{T_{{\rm N}J}},
\ee
where $\theta_{{\rm p}J}$ is the Weiss temperature in the Curie-Weiss law due to exchange interactions alone.

The N\'eel temperature $T_{\rm N}$ in $H=0$ in the presence of both exchange and anisotropy fields is increased in the presence of the XY anisotropy field, as expected, according to the linear relation
\be
T_{\rm N} = T_{{\rm N}J}(1+h_{\rm A1}).
\label{Eq:TNTNJ}
\ee
The anisotropic Weiss temperatures in the Curie-Weiss law for the paramagnetic susceptibility with XY anisotropy are
\bse
\bea
\theta_{{\rm p}z} &=& \theta_{{\rm p}J},\\*
\theta_{{\rm p}xy} &=& \theta_{{\rm p}J} + T_{{\rm N}J}h_{\rm A1},\\*
\theta_{{\rm p}xy} - \theta_{{\rm p}z} &=& T_{{\rm N}J}h_{\rm A1} = \frac{T_{\rm N}h_{\rm A1}}{1+h_{\rm A1}},\\*
\frac{\theta_{{\rm p}xy} - \theta_{{\rm p}z}}{T_{\rm N}} &=& \frac{h_{\rm A1}}{1+h_{\rm A1}},\label{Eq:hA1Fromtheta}
\eea
\ese
where we used Eq.~(\ref{Eq:TNTNJ}) to obtain the third equality.  This allows one to easily determine the parameter $h_{\rm A1}$.  Usually the ratio on the left side of Eq.~(\ref{Eq:hA1Fromtheta}) is small, so one can instead use
\be
\frac{\theta_{{\rm p}xy} - \theta_{{\rm p}z}}{T_{\rm N}} \approx h_{\rm A1}\qquad(h_{\rm A1}\ll 1),\label{Eq:hA1Fromtheta2}
\ee
which is equivalent to the approximation $T_{\rm N}\approx T_{{\rm N}J}$.  Using the $T_{\rm N}$ and $\theta_{{\rm p}ab} - \theta_{{\rm p}c}$ values in Table~\ref{Tab:CuriFit}, one obtains
\be
h_{\rm A1}\approx 0.05\ {\rm for\ EuCo}_{2-y}{\rm As}_2.
\ee
Thus the XY anisotropy increases the N\'eel temperature, also $\theta_{{\rm p}ab}$ by about 5\%, or about 2~K for \eca.

\subsection{Fit of $\chi_{ab}(T\leq T_{\rm N})$ by Molecular Field Theory}

In order to fit the low-field $ab$-plane susceptibility $\chi_{ab}(T\leq T_{\rm N})$ by the unified MFT for Heisenberg AFMs in Refs.~\cite{Johnston2015} and~\cite{Johnston2012}, we assume that the Curie constant~$C_\alpha$ and Weiss temperature~$\theta_{\rm p\alpha}$ $(\alpha = ab$ or~$c$) in the PM state at $T\geq T_{\rm N}$ are independent of $T$ with the values given in Table~\ref{Tab:CuriFit}.  We first remove the contributions of the $T$-independent susceptibility $\chi_0$ and of anisotropy in the PM state to obtain the $\chi_{J\alpha}(T\geq T_{\rm N})$ that would have arisen from exchange interactions alone.

\begin{figure}
\includegraphics[width=3.in]{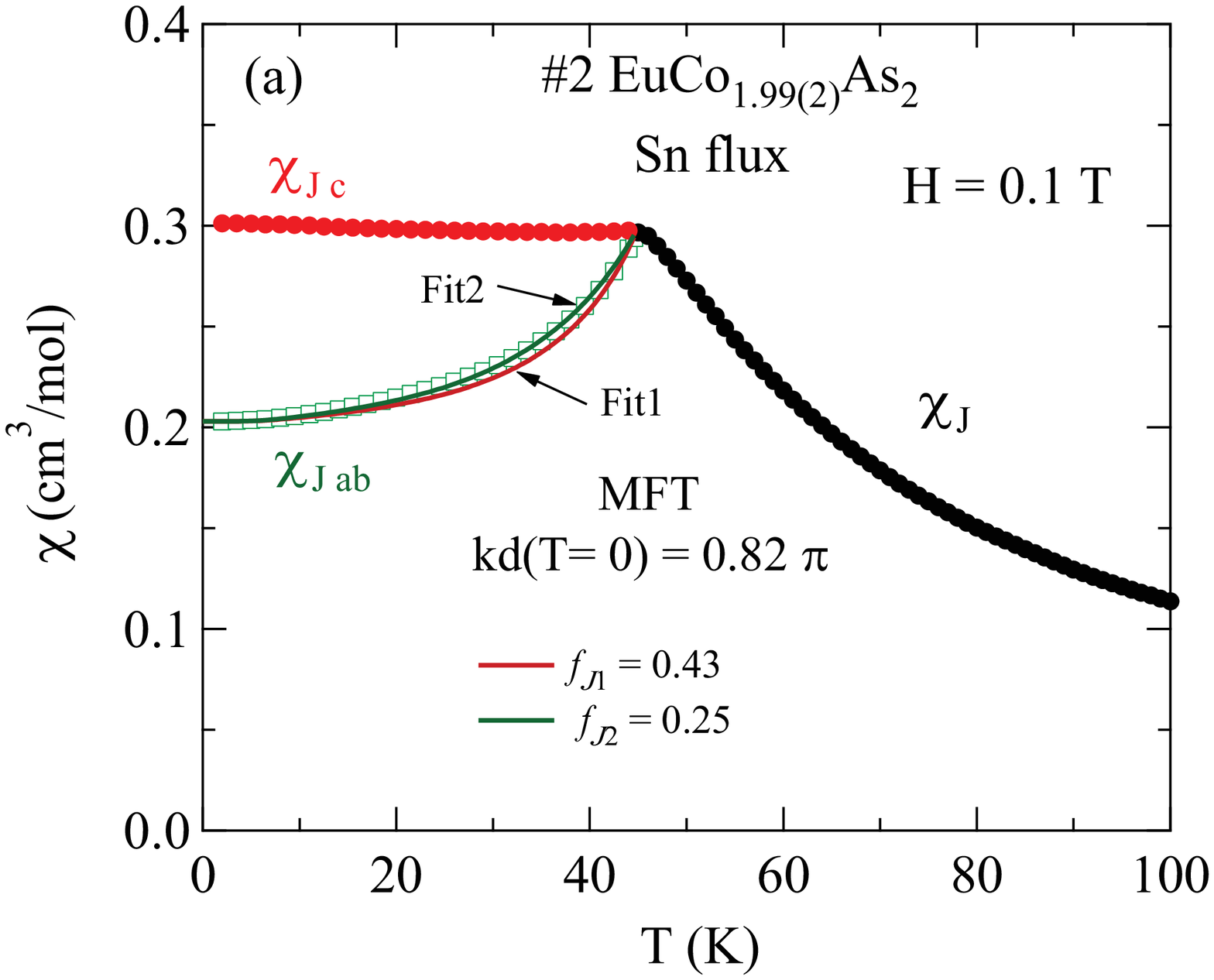}
\includegraphics[width=3.in]{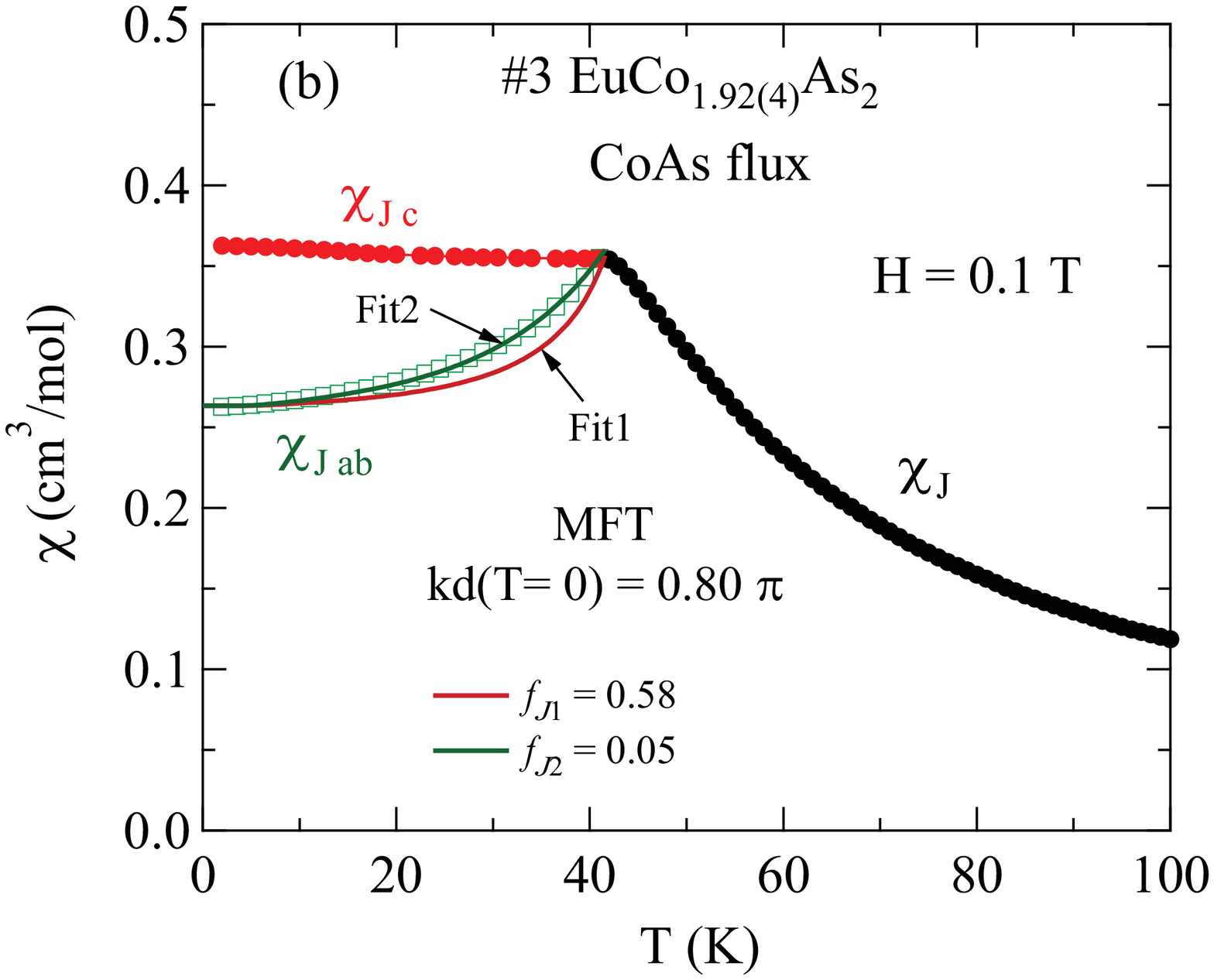}
\caption {$\chi_J(T)$ versus $T$ for $H\parallel ab$ and $H\parallel c$ in $H=0.1$~T for (a)~Sn-flux-grown crystal \#2~EuCo$_{1.99(2)}$As$_2$ and (b)~CoAs-flux-grown crystal \#3~EuCo$_{1.92(4)}$As$_2$.  The fits of $\chi_{Jab}(T)$ for $T\leq T_{\rm N}$ by the MFT prediction for a helix in Eqs.~(\ref{Eqs:Chixy}) are shown as the solid curves. }
\label{Fig:ChiJ_MFT}
\end{figure}

The $T$-independent susceptibility $\chi_{0\alpha}$ is taken into account at all temperatures according to
\be
\chi_{\alpha}^*(T) = \chi_{\alpha}(T) - \chi_{0\alpha},
\ee
where $\chi_{\alpha}(T)$ is the measured susceptibility and the $\chi_{0\alpha}$ values are given in Table~\ref{Tab:CuriFit}.  We assume that the anisotropy in the PM state arises from sources such as magnetic dipole interactions and/or single-ion quantum uniaxial $DS_z^2$ anistropy, for which the magnetic susceptibility tensor is traceless in the PM state \cite{Johnston2016, Johnston2017}.  Then one obtains the Heisenberg susceptibility~$\chi_J$ in the PM state given by
\be
\chi_{J}(T\geq T_{\rm N}) = \frac{1}{3}\Big[2\chi_{ab}^*(T) + \chi_c^*(T)\Big],
\ee
as shown in Fig.~\ref{Fig:ChiJ_MFT} for one each of the Sn-flux-grown and CoAs-flux grown crystals.  As found above in Sec.~\ref{Sec:MDIs}, the anisotropy increases $T_{\rm N}$ by about 5\% and this small change will henceforth be ignored.

Within MFT, for $T\leq T_{\rm N}$ the perpendicular susceptibility $\chi_{Jc}$ is predicted to be independent of $T$, in good agreement with the data in Fig.~\ref{Fig:ChiJ_MFT}.  The normalized $\chi_{Jab}(T \leq T_{\rm N})/\chi_J(T_{\rm N})$ for a helical Heisenberg AFM is given by \cite{Johnston2012,Johnston2015}
\bse
\label{Eqs:Chixy}
\begin{equation}
\frac{\chi_{Jab}(T \leq T_{\rm N})}{\chi_J(T_{\rm N})}=  \frac{(1+\tau^*+2f_J+4B^*)(1-f_J)/2}{(\tau^*+B^*)(1+B^*)-(f_J+B^*)^2},
\label{eq:Chi_plane}
\end{equation}
where
\begin{equation}
B^*=  2(1-f_J)\cos(kd)\,[1+\cos(kd)] - f_J,
\label{eq:Bstar}
\end{equation}
\be
t =\frac{T}{T_{\rm N}},\quad \tau^*(t) = \frac{(S+1)t}{3B'_S(y_0)}, \quad y_0 = \frac{3\bar{\mu}_0}{(S+1)t}, 
\ee
\ese
the ordered moment versus $T$ in $H=0$ is denoted by $\mu_0$, the reduced ordered moment $\bar{\mu}_0 = \mu_0/\mu_{\rm sat}$ is determined by numerically solving the self-consistency equation
\be
\bar{\mu}_0 = B_S(y_0),
\label{Eq:barmuSoln}
\ee
$B'_S(y_0) = [dB_S(y)/dy]|_{y=y_0}$ and our definition of the Brillouin function $B_S(y)$ is given in Refs.~\cite{Johnston2015} and~\cite{Johnston2012}.

We fitted the in-plane $\chi_{Jab}(T)$ data in Fig.~\ref{Fig:ChiJ_MFT} by Eqs.~(\ref{Eqs:Chixy}) using \mbox{$S=7/2$} and the indicated $f_J$ values. For $kd(T)$ we used the neutron diffraction value $kd(T=47~{\rm K}) = 0.79\pi$ \cite{Tan2016}. In order to fit the lowest-$T$ data, we used $kd(T=0) = 0.82\pi$ for the Sn-flux-grown crystal and 0.798$\pi$ for the CoAs-flux-grown crystal, calculated from Eqs.~(\ref{Eqs:Chixy}), which are comparable to the experimentally observed value with respect to neutron diffraction studies \cite{Tan2016}. A rough estimated value of $f_J$ is $f_J \sim $(20~K)/(42~K) ~$\sim0.5$. We treated $f_J$ as an adjustable parameter.  The $\chi_{Jab}(T\leq T_{\rm N})$ fits thus obtained are plotted as the solid blue curves in Figs.~\ref{Fig:ChiJ_MFT}(a) and~\ref{Fig:ChiJ_MFT}(b).  Also shown are the $\chi_{Jab}(T\leq T_{\rm N})$ curves using the approximate measured values of $f_J$.  The discrepancy between the two fitted curves in each figure is a measure of the deficiency of MFT in predicting $\chi_{Jab}(T)$, as previously pointed out in Ref.~\cite{Johnston2012}.


\section{\label{Cp} Heat Capacity}

\subsection{Zero-Field Heat Capacity}

\begin{figure}
\includegraphics[width=3.in]{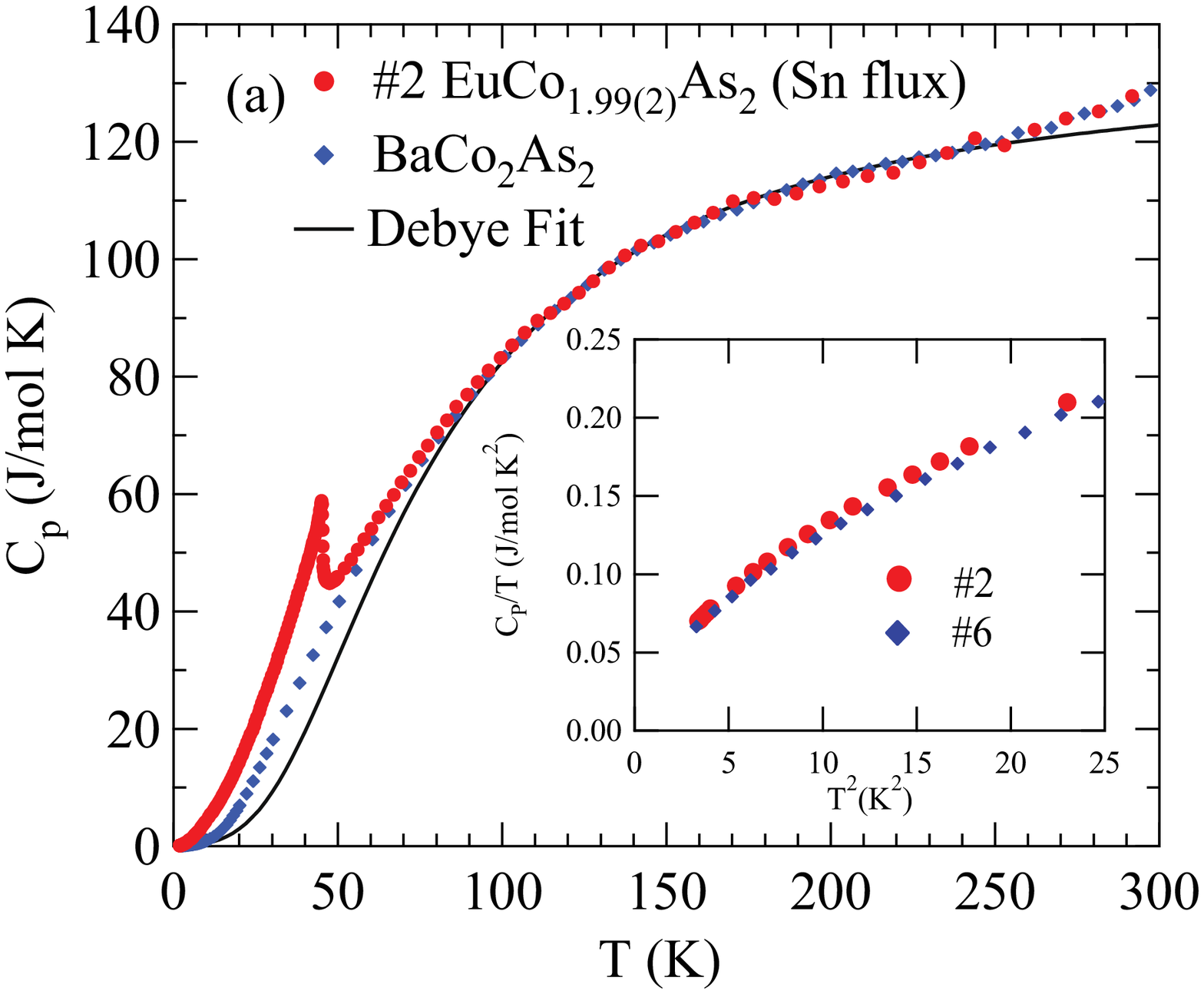}
\includegraphics[width=3.in]{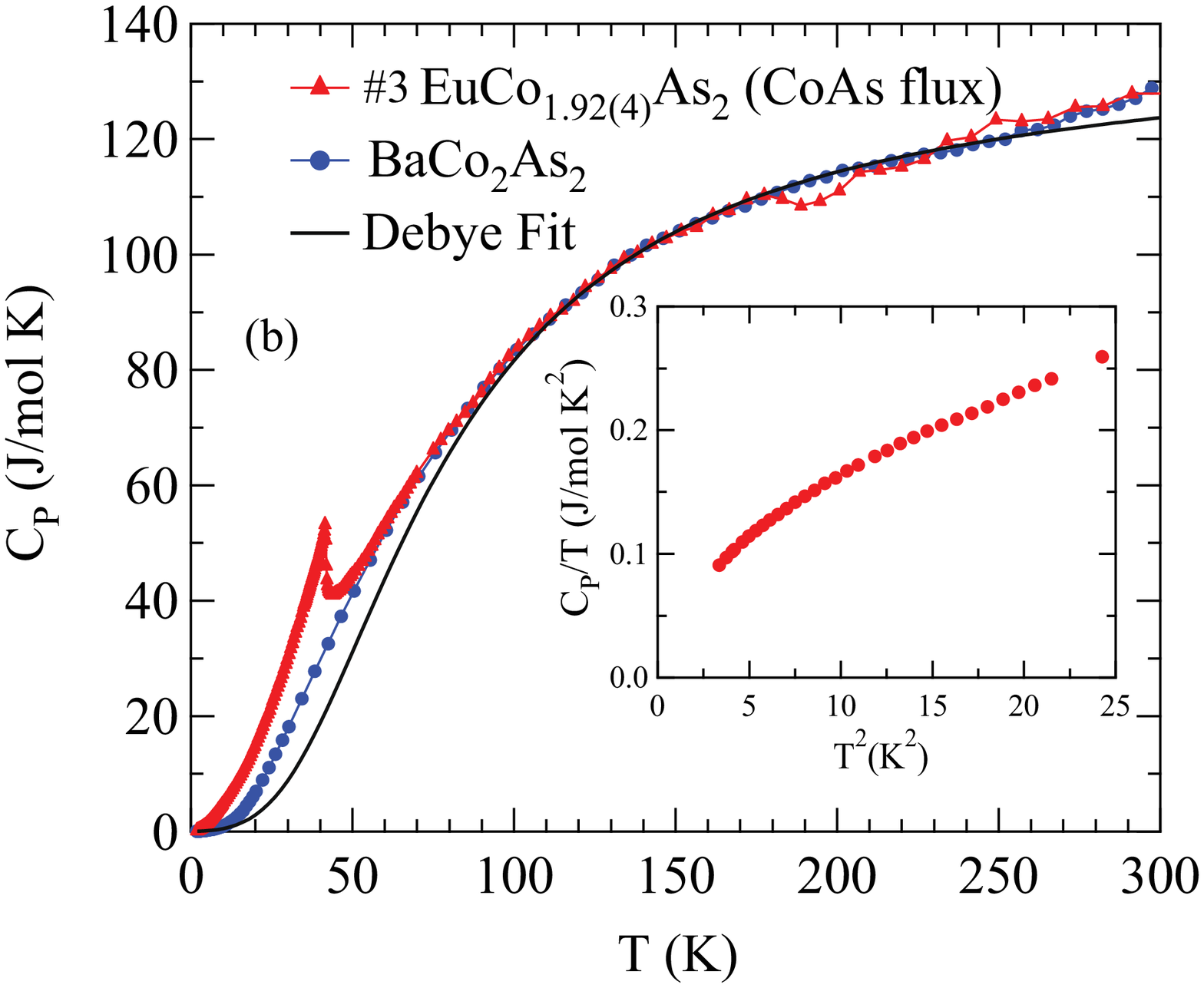}
\includegraphics[width=3.in]{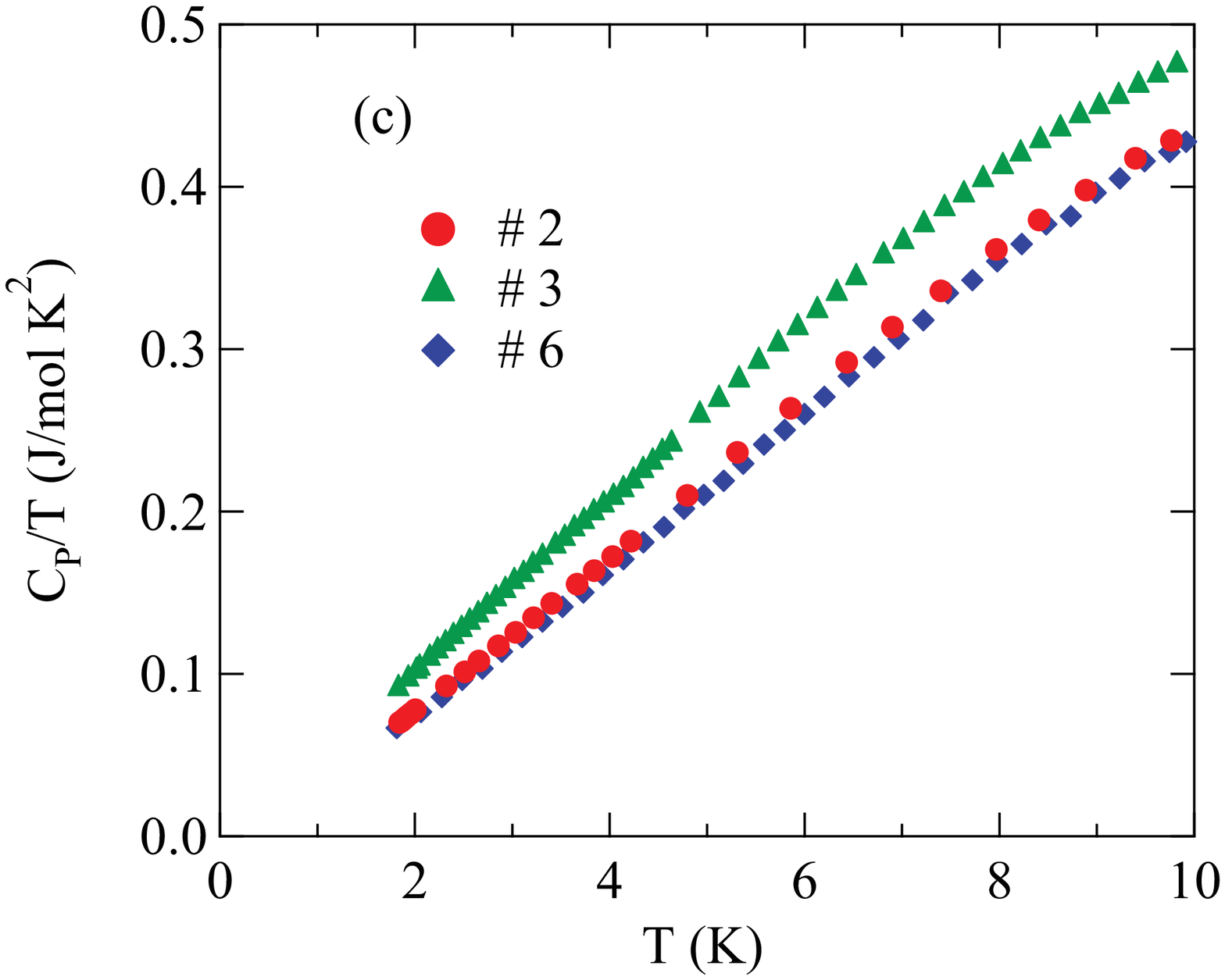}
\caption{Temperature dependence of the heat capacity $C{\rm_p}(T)$ in $H=0$ for (a)~Sn-flux-grown \#2~$\rm EuCo_{1.99(2)}As_2$ and (b)~CoAs-flux-grown \#3~$\rm EuCo_{1.92(4)}As_2$ crystals.  Both panels also show $C{\rm_p}(T)$ of the nonmagnetic reference compound \bca~\cite{Sangeetha2016}. The black curves are Debye lattice heat capacity model fits to the data between 100 and 280--300~K by Eq.~(\ref{Eq:Debye_Fit}). Insets: $C{\rm_p}/T$ versus $T^2$ for the three crystals \#2, \#3, and Sn-flux-grown \#6~EuCo$_{1.94(2)}$As$_2$.  The data do not follow the behavior expected from Eq.~(\ref{Eq:Cp_linearFit}). (c)~Plots of $C{\rm_p}(T)/T$ versus~$T$.}
\label{Fig:HC_0T}
\end{figure}

The heat capacities $C{\rm_p}(T)$ for Sn-flux-grown crystal \#2~EuCo$_{1.99(2)}$As$_2$, CoAs-flux-grown crystal \#3~EuCo$_{1.92(4)}$As$_2$, and the nonmagnetic reference compound \bca~\cite{Sangeetha2016} measured in the temperature range from 1.8 to 300~K are shown in Fig.~\ref{Fig:HC_0T}. The data exhibit a prominent peak at $T_{\rm N} = 45.1(2)$~K and $T_{\rm N} = 40.02(4)$~K for crystals~\#2 and \#3, respectively. Low-temperature $C_{\rm p}/T$ vs $T^2$ plots in the range 1.8 to 5~K for the above two crystals and for Sn-flux-grown crystal \#6~EuCo$_{1.94(2)}$As$_2$ are shown in the insets of Fig.~\ref{Fig:HC_0T}.  The data for all three crystals exhibit negative curvature below $\sim 3$~K and hence cannot be fitted by the conventional expression \cite{Kittel2005}
\bea
\frac{C_{\rm p}(T)}{T} = \gamma + \beta T^2,
\label{Eq:Cp_linearFit}
\eea
where $\gamma$ is the Sommerfeld coefficient associated with the conduction electrons and $\beta$ is the coefficent of the $T^3$ lattice and three-dimensional AFM spin-wave contributions. Below we attempt to find $\gamma$ by fitting the high-$T$ data.  In Table~\ref{Tab:HC} are shown data obtained for similar isostructural compounds.

Shown in Fig.~\ref{Fig:HC_0T}(c) are plots of $C{\rm_p}(T)/T$ versus~$T$ for the three crystals \#2, \#3, and \#6.  One sees that each crystal shows approximately linear behavior over the $T$~range from 3 to 6~K, i.e., that $C_{\rm p}$ has an approximately $T^2$ contribution over this $T$ range.  From preliminary linear spin-wave calculations, this behavior may arise from the temperature-dependent heat capacity of AFM spin waves.

\begin{table}
\caption{\label{Tab:HC} Parameters $\gamma$ and $\beta$ obtained for pnictide compounds isostructural to \eca.  Also listed are the Debye temperatures $\Theta_{\rm D}$ obtained from $\beta$ according to Eq~(\ref{Eq:thetaD}) and the density of states at the Fermi energy $D_\gamma(E_{\rm F})$ obtained from $\gamma$ via Eq.~(\ref{Eq:DOS2}). Values of $\gamma$ and $\Theta_{\rm D}$ for both EuCo$_{1.99(2)}$As$_2$ and EuCo$_{1.92(4)}$As$_2$ are obtained by fitting the $C_{\rm p}(T)$ data between 100 and 280~K in Fig.~\ref{Fig:HC_0T} by the Debye model plus a $\gamma T$ term according to Eq.~(\ref{Eq:Debye_Fit}).}

\begin{ruledtabular}
\begin{tabular}{lcccc}
Crystal            &   $\gamma$  						& $\beta$  						& $\Theta_{\rm D}$ 	& $D_\gamma$($E_{\rm F}$)\\
		&  $\left({\rm \frac{mJ}{mol~K^2}}\right)$	& $\left({\rm \frac{mJ}{mol~K^4}}\right)$	& (K)  		& (${\rm\frac{states}{eV~f.u.}}$)  \\   
\hline
\#2~EuCo$_{1.99(2)}$As$_2$\footnotemark[1]$^,$\footnotemark[3]	& 15(2) 	& 0.33(1) 	& 308(3)		& 6.3(8)	\\
\#3~EuCo$_{1.92(4)}$As$_2$\footnotemark[2]$^,$\footnotemark[3]	& 18(3) 	& 0.31(1) 	& 314(4)		& 7(1) \\
${\rm EuCo_2P_2}$ \cite{Sangeetha2016}	& 23.7(5)	& 2.8(1)	& 151(2)				& 10.0(2)	\\
								& 		& 		& 480(6)\footnotemark[5]	& 	\\
${\rm BaCo_2P_2}$ \cite{Sangeetha2016}	& 37.3(3)	& 0.21(1)	& 359(6)				& 15.8(2)	\\
${\rm SrCo_2P_2}$ \cite{Pandey2013}	& 37.8(1)	& 0.611(7)& 251(1)				& 16.0(3)	\\
${\rm BaCo_2As_2}$\footnotemark[4] \cite{Anand2014b}	& 39.8(1)	& 0.386(4)& 293(2)	& 16.9(1)	\\
${\rm CaCo_{1.86}As_2}$\footnotemark[4] \cite{Anand2014}	& 27(1)	& 1.00(8) & 212(1)	& 11.4(5)	\\

\end{tabular}
\end{ruledtabular}
\footnotetext[1]{Grown in Sn flux with H$_2$-treated Co powder}
\footnotetext[2]{Grown in CoAs flux with H$_2$-treated Co powder}
\footnotetext[3]{From a 100--280~K fit of $C_{\rm p}(T)$ by Eq.~(\ref{Eq:Debye_Fit})}
\footnotetext[4]{Grown in Sn flux}
\footnotetext[5]{From a 200--280~K fit of $C_{\rm p}(T)$ by Eq.~(\ref{Eq:Debye_Fit})}
\end{table}

The $C_{\rm p}(T)$ data for our crystals in the temperature range $120~{\rm K} \leq T \leq 280$~K are analysed using an electronic $\gamma T$ term plus the Debye model for the lattice heat capacity \cite{Kittel2005}
\bea
C_{\rm p}(T) &=& \gamma T + n C_{\rm V\,Debye}(T/\Theta_{\rm D}),\label{Eq:Debye_Fit} \\*
C_{\rm V\,Debye}(T/\Theta_{\rm D}) &=& 9R \left(\frac{T}{\Theta_{\rm D}}\right)^3\int_{0}^{\Theta_{\rm D}/T}\frac{x^4e^x}{(e^x-1)^2} dx,\nonumber
\eea
The representation of the Debye function $C_{\rm V\,Debye}(T/\Theta_{\rm D})$ used here is an accurate analytic Pad\'{e} approximant function of $T/\Theta_{\rm D}$ \cite{Goetsch2012}. The fits to the $C{\rm_p(}T)$ data over the temperature range 100 to 280~K by Eq.~(\ref{Eq:Debye_Fit}) are shown as the black solid curves in Figs.~\ref{Fig:HC_0T}(a) and~\ref{Fig:HC_0T}(b) and the fitted values of $\gamma$ and $\Theta_{\rm D}$ are listed in Table~\ref{Tab:HC}.

The density of conduction carrier states at the Fermi energy $E\rm_F$, $D_\gamma$($E\rm_F$), is obtained from $\gamma$ according to \cite{Kittel2005}
\bse
\bea
D_\gamma(E\rm_F)=\frac{3\gamma}{\pi^2k_B^2},
\label{Eq:DOS1}
\eea
\rm{which gives}
\bea
D_\gamma(E\rm_F)\left[\frac{states}{eV~f.u.}\right] = \frac{1}{2.359}~\gamma\left[\frac{mJ}{mol~ K^2}\right].
\label{Eq:DOS2} 
\eea
\ese
The $D_\gamma$($E\rm_F$) values calculated for \eca\ crystals~\#2 and~\#3 from their $\gamma$ values using Eq.~(\ref{Eq:DOS2}) are listed in Table~\ref{Tab:HC}, where values from the literature for similar compounds \cite{Anand2014, Pandey2013, Anand2014b, Sangeetha2016} are also given.

The Debye temperature is estimated from the value of $\beta$ in Eq.~(\ref{Eq:Cp_linearFit}) from the expression \cite{Kittel2005}
\bea
\Theta_{\rm D} = \left(\frac{12\pi^4nR}{5\beta}\right)^{1/3},
\label{Eq:thetaD}
\eea
where $n$ is the number of atoms per formula unit (\mbox{$n=5-y$} for \eca) and $R$ is the molar gas constant.  The values of $\Theta_{\rm D}$ obtained from the $\beta$ vaues for other compounds \cite{Anand2014, Pandey2013, Anand2014b, Sangeetha2016} are listed for comparison with those for our crystals in Table~\ref{Tab:HC}.

\begin{figure*}
\includegraphics[width=3.in]{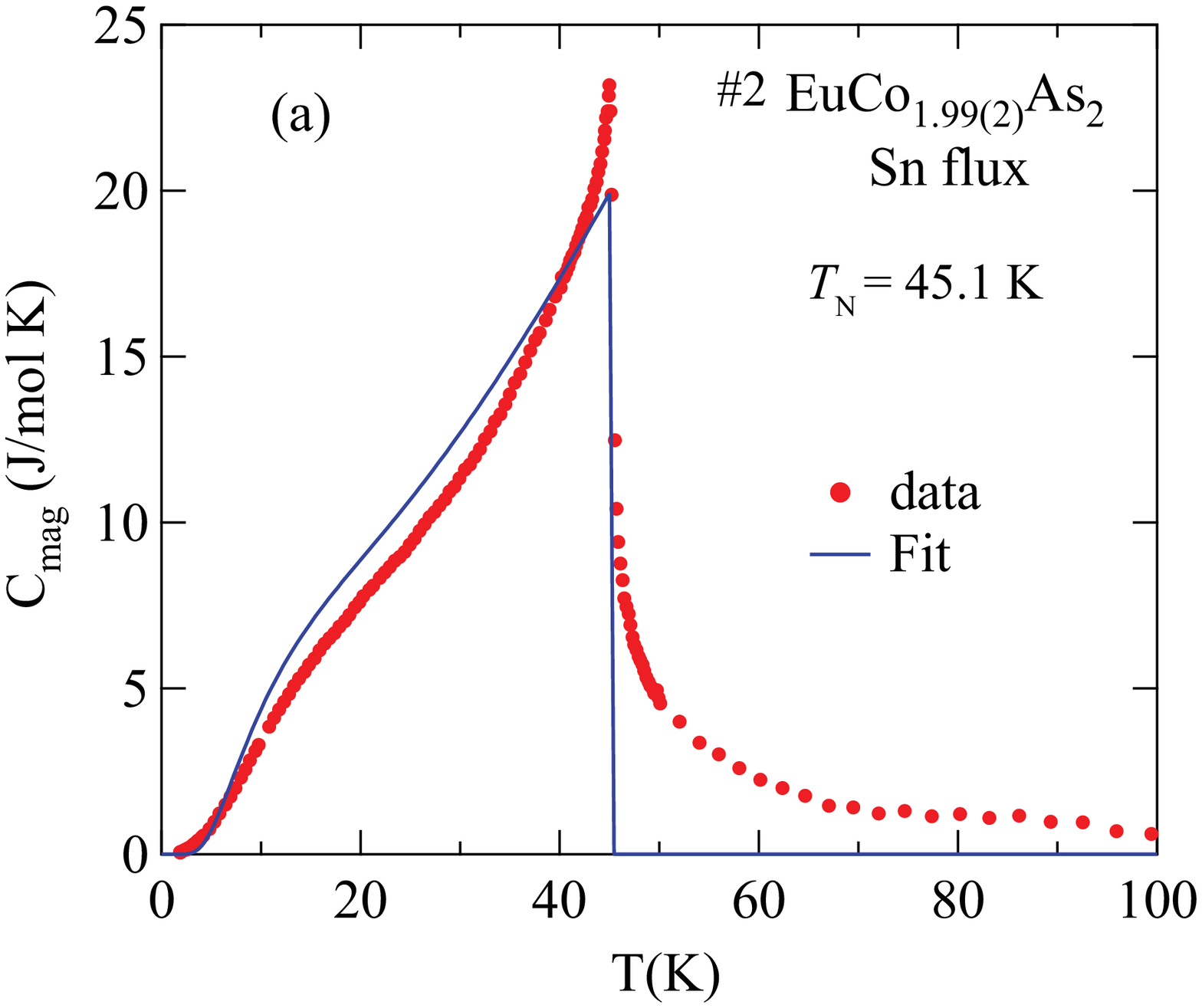}\includegraphics[width=3.in]{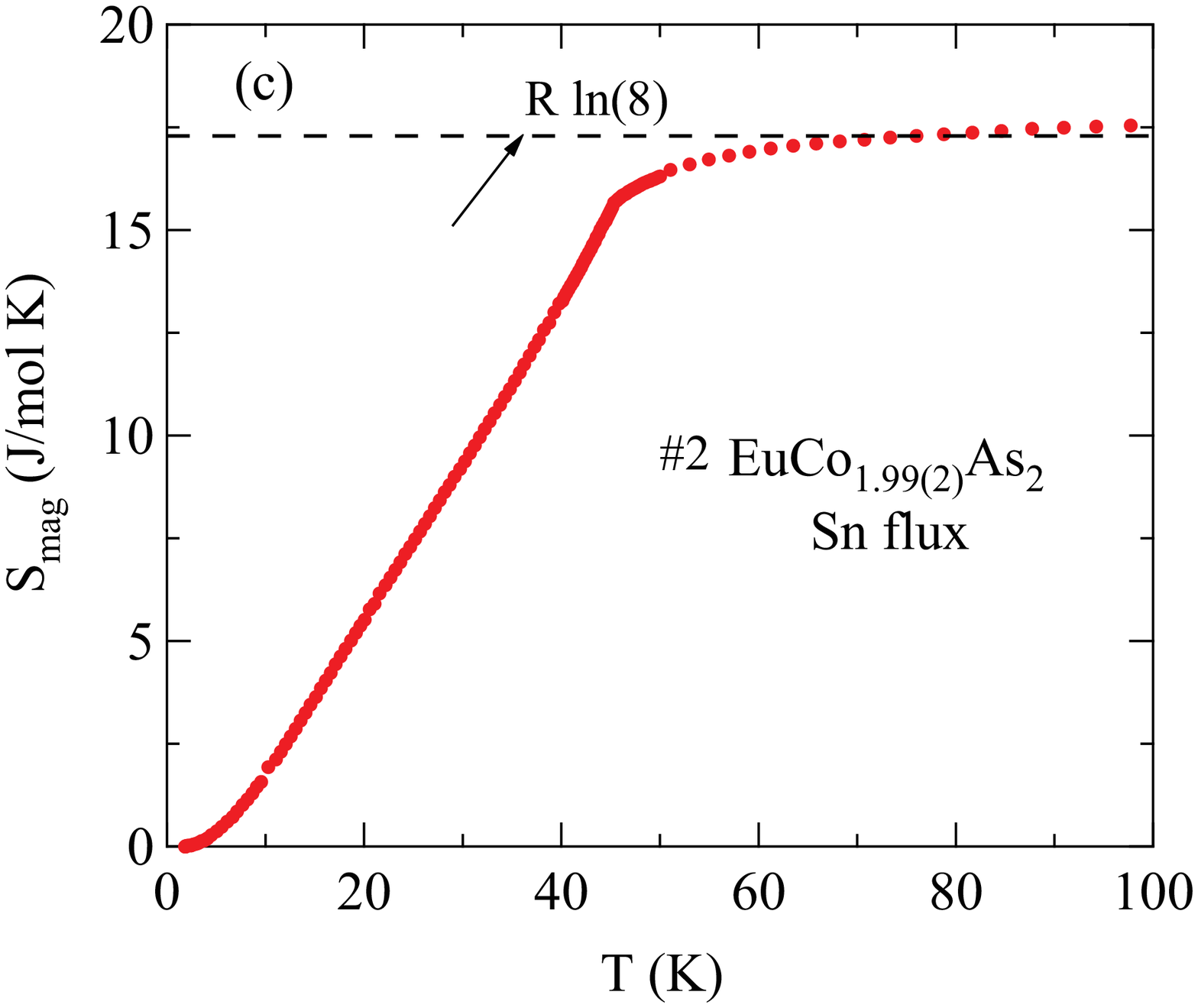}
\includegraphics[width=3.in]{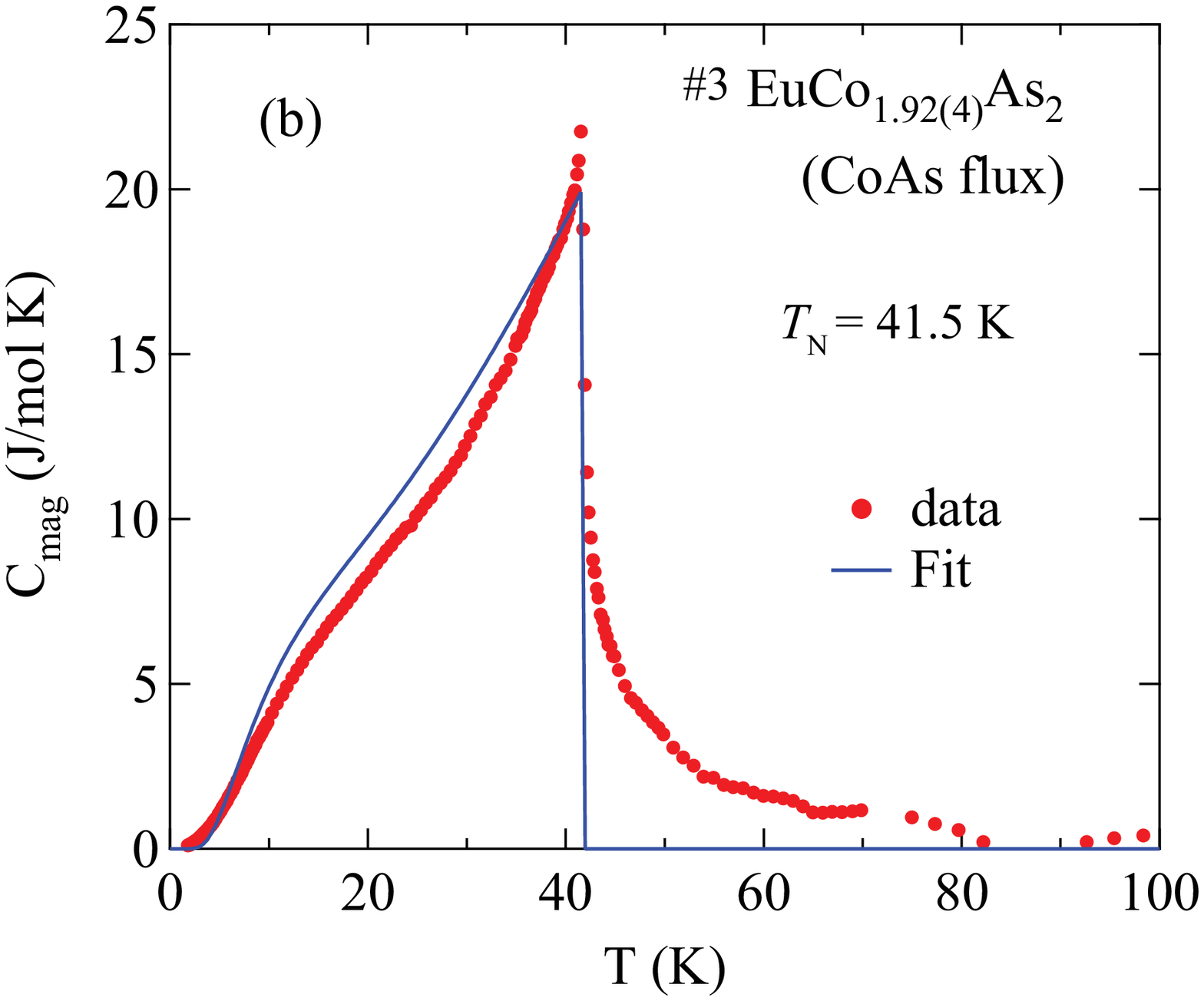}\includegraphics[width=3.in]{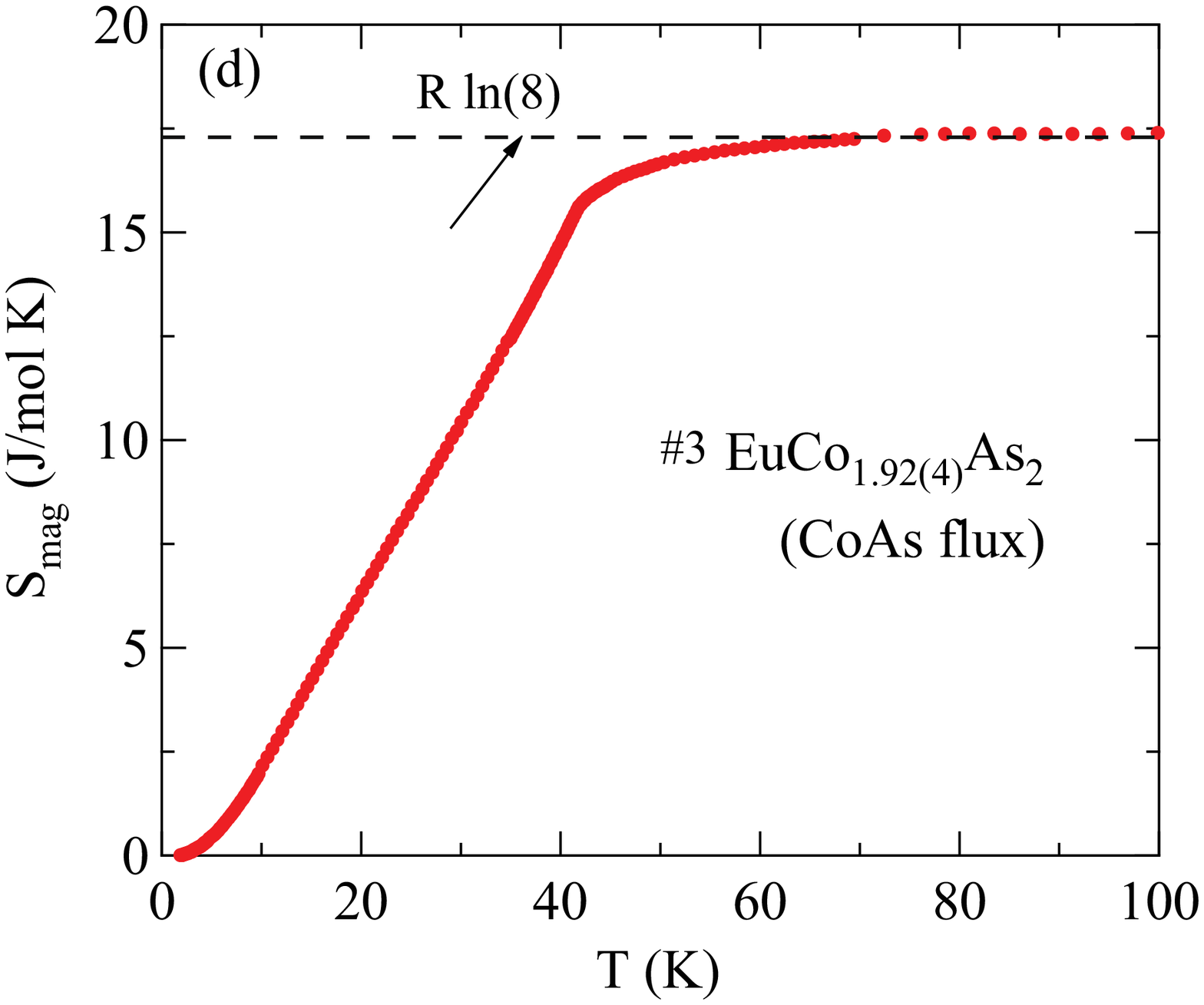}
 \caption{Magnetic contributions $C_{\rm mag}(T)$ and $S_{\rm mag}(T)$ to the heat capacity and entropy, respectively, of  (a,c)~Sn-flux-grown crystal \#2~$\rm EuCo_{1.99(2)}As_2$ and (b,d)~CoAs-flux-grown crystal \#3~$\rm EuCo_{1.92(4)}As_2$. In (c,d), the horizontal dashed line is the theoretical high-$T$ limit $S_{\rm mag} = R\ln(2S+1) = 17.29$~J/mol\,K for Eu$^{+2}$ with $S=7/2$. }
\label{Fig:HC_mag}
\end{figure*}

\begin{figure}
\includegraphics[width=2.75in]{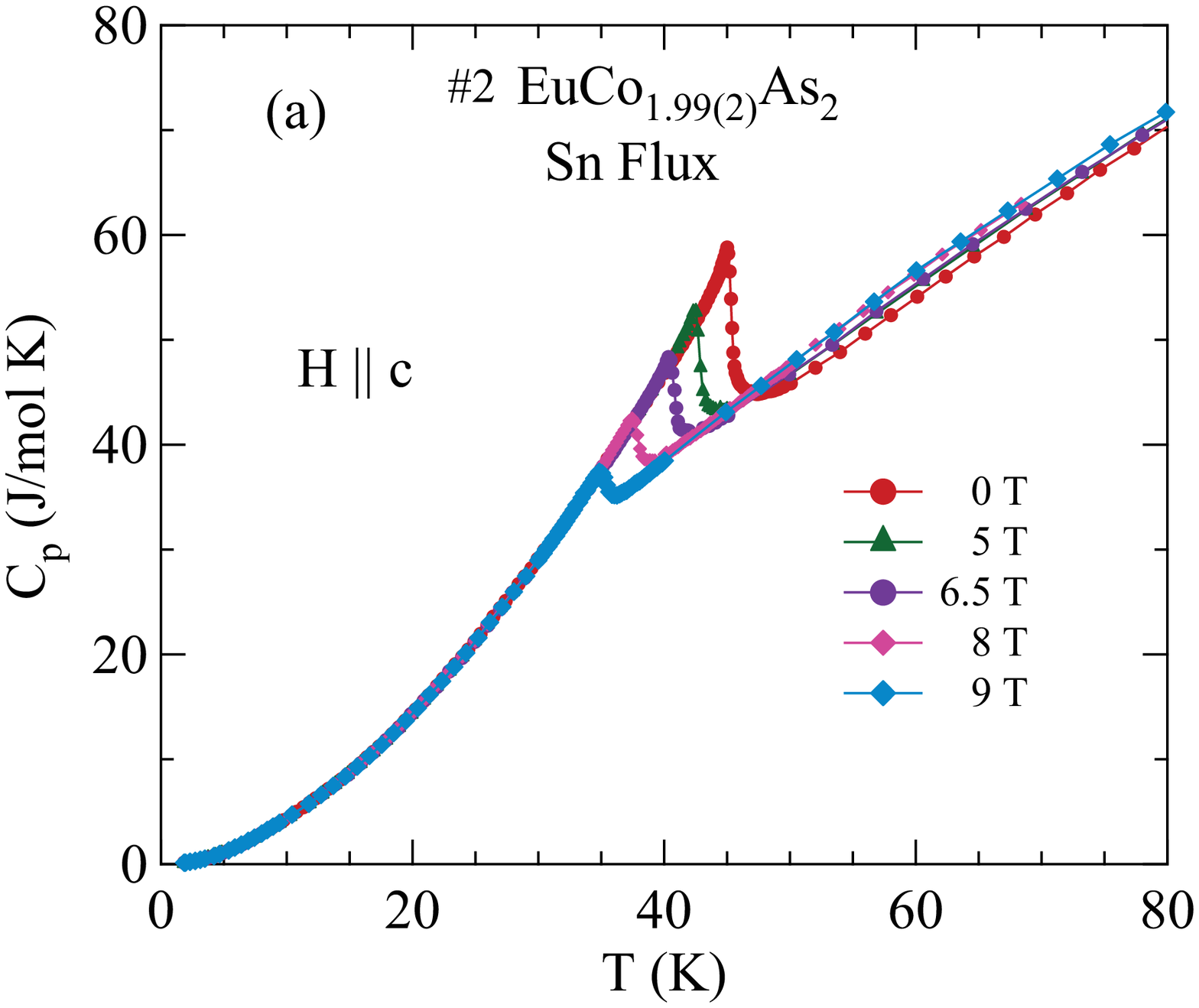}\vspace{0.05in}
\includegraphics[width=2.75in]{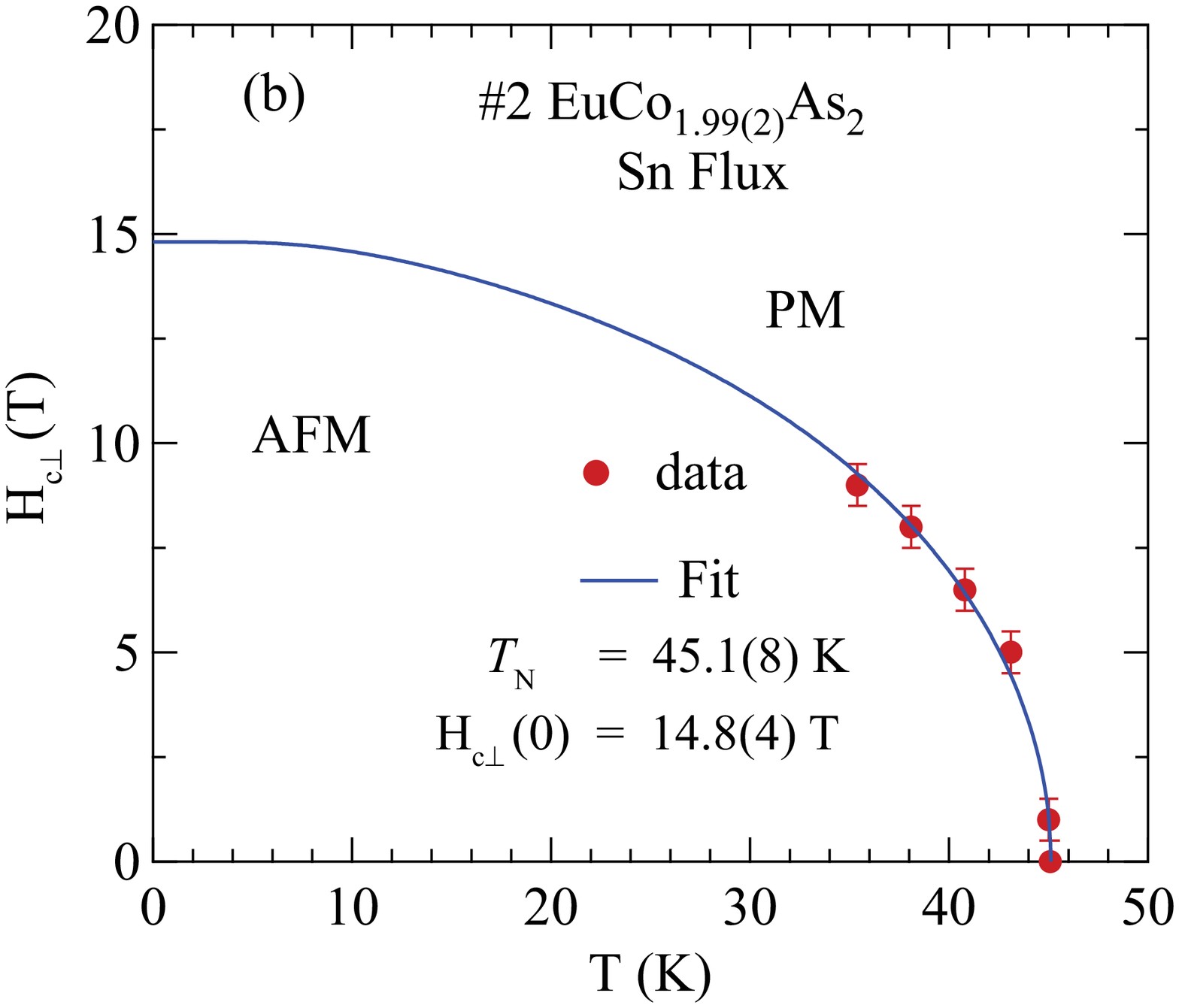}
 \caption{(a)~Heat capacity $C_{\rm p}$ versus temperature~$T$ of Sn-flux-grown crystal \#2~$\rm EuCo_{1.99(2)}As_2$ in various fields $H_\perp \equiv H||c$ as listed. (b)~Magnetic $H_\perp$-$T$ phase diagram constructed from the $C_p$($H, T$) data in~(a). The solid blue curve is a fit of the data points by Eq.~(\ref{Eq:HcFit}). }
\label{Fig:HC_H_Snflux}
\end{figure}

\begin{figure}
\includegraphics[width=2.75in]{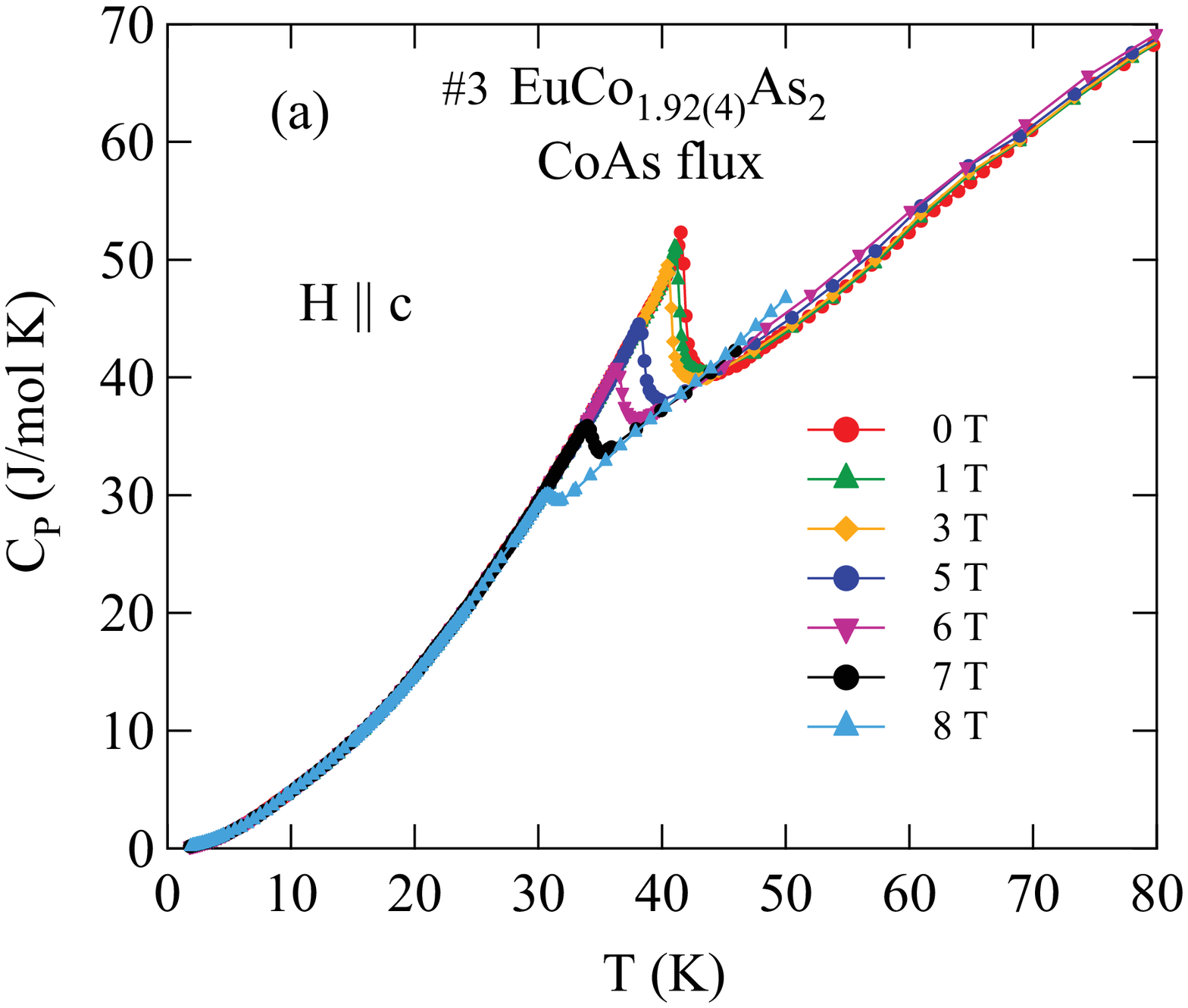}
\includegraphics[width=2.75in]{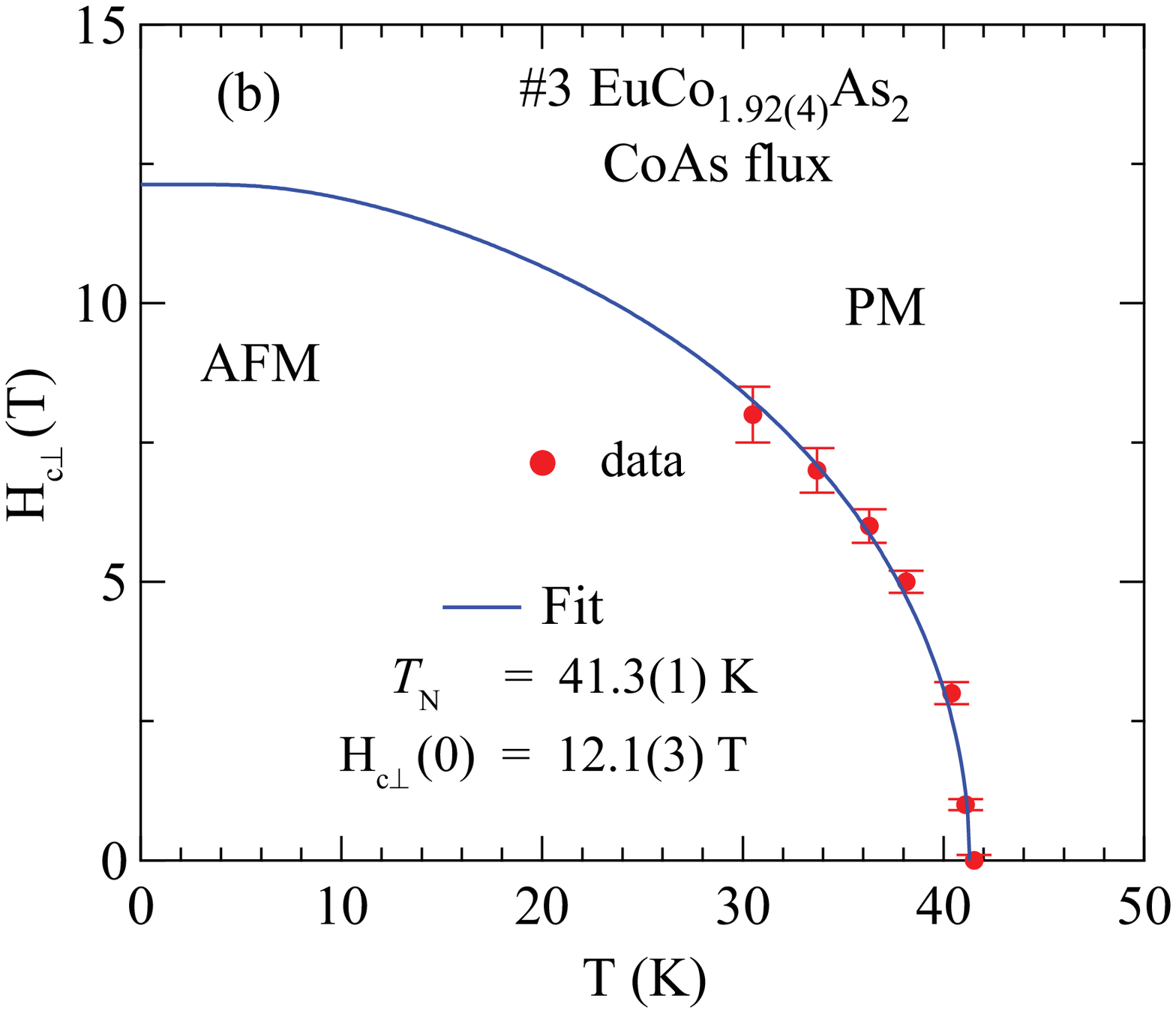}
 \caption{Same as Fig.~\ref{Fig:HC_H_Snflux} except for CoAs-flux-grown crystal \#3~$\rm EuCo_{1.99(2)}As_2$ instead.}
\label{Fig:HC_H_CoAsflux}
\end{figure}

The magnetic contribution $C_{{\rm mag}}(T)$ to $C_{\rm p}(T)$ for the \eca\ crystals is obtained by subtracting $C_{\rm p}(T)$ of the nonmagnetic reference compound \bca\ from those of the \eca\ crystals, as shown in Figs.~\ref{Fig:HC_mag}(a) and \ref{Fig:HC_mag}(b), respectively. Within the Weiss MFT the discontinuity in the magnetic heat capacity at $T_{\rm N}$ for a spin $S=7/2$ system is given by
\bea
\Delta C{\rm_{mag}}=\frac{5RS(S+1)}{1+2S+2S^2} = 20.14~ \rm {J/mol~K}.
\label{Eq:deltaCp}
\eea
The jump in the heat capacity at $T_{\rm N}$ is $\approx23.2$~J/mol~K and 21.74~J/mol~K in the Sn-flux-grown crystal \#2~$\rm EuCo_{1.99(2)}As_2$ and the CoAs-flux-grown crystal \#3~$\rm EuCo_{1.92(4)}As_2$, respectively, which are somewhat larger than the prediction~(\ref{Eq:deltaCp}) of MFT\@. The discrepancy arises from the difference between the observed $\lambda$-shape and the predicted step-shape of $C_{\rm mag}(T)$ at $T_{\rm N}$\@.  The nonzero contribution to $C{\rm_{mag}(}T$) for $T_{\rm N}<T\lesssim~100$~K reflects the presence of dynamic short-range AFM ordering of the Eu spins above $T_{\rm N}$\@. The hump in $C{\rm_{mag}(}T$) below $T_{\rm N}$ at $T\sim 15$~K arises naturally within MFT for large $S$~\cite{Johnston2015}. The solid blue curves in Figs.~\ref{Fig:HC_mag}(a) and~\ref{Fig:HC_mag}(b) represent the MFT predication for $C_{\rm mag}(T)$ calculated for each respective $T_{\rm N}$ and for $S=7/2$ which are in reasonable agreement with the data for each crystal below the respective $T_{\rm N}$.

The magnetic entropy~$S_{\rm mag}(T)$ in $H=0$ is calculated from the $C_{\rm mag}(T)$ data for each crystal according to $S{\rm_{mag}(}T) = \int_{0}^{T}[C{\rm_{mag}(}T)/T]dT$ and the results are shown in Fig.~\ref{Fig:HC_mag}(b) for Sn-flux-grown crystal \#2~$\rm EuCo_{1.99(2)}As_2$ and in Fig.~\ref{Fig:HC_mag}(d) for CoAs-flux-grown crystal \#3~$\rm EuCo_{1.92(4)}As_2$. The horizontal dashed line in each figure is the theoretical high-$T$ limit  $S_{\rm mag} = R\ln(2S+1) = 17.29$~J/mol K for $S=7/2$. For each crystal, the entropy reaches $\approx 90$\% of $R$ln(8) at $T_{\rm N}$ and recovers the full value by $\sim 70$~K\@.

\subsection{High-Field Heat Capacity}

Figures~\ref{Fig:HC_H_Snflux}(a) and~\ref{Fig:HC_H_CoAsflux}(a) show $C{\rm_p(}H, T)$ for Sn-flux-grown crystal $\rm \#2~EuCo_{1.99(2)}As_2$ and CoAs-flux-grown crystal \#3~$\rm  EuCo_{1.92(4)}As_2$, respectively, measured in various applied magnetic fields up to 9~T with $H||c$. Thus the field direction is perpendicular to the $ab$~plane of the ordered moments in $H=0$ which we therefore denote as $H_\perp \equiv H\parallel c$ \cite{Johnston2015}.  It is evident that the AFM transition temperature $T_{\rm N}(H_\perp)$ shifts to lower temperature and that the heat capacity jump at $T_{\rm N}(H_\perp)$  decreases with increasing field, both as predicted from MFT in Ref.~\cite{Johnston2015} for a field parallel to the helix axis. The data in the $H-T$ phase diagrams with $H\parallel c$ in Figs.~\ref{Fig:HC_H_Snflux}(b) and~\ref{Fig:HC_H_CoAsflux}(b) were constructed from the $H_\perp$ dependence of $T_{\rm N}$ obtained from the respective Figs.~\ref{Fig:HC_H_Snflux}(a) and~\ref{Fig:HC_H_CoAsflux}(a).

The MFT prediction for the critical field $H_{\rm c\perp}(T)$ at which the AFM state undergoes a second-order transition to the PM state with increasing field at fixed~$T$ is given by \cite{Johnston2015}
\bse
\be
H_{\rm c\perp}(T) = H_{\rm c\perp}(0)\bar{\mu}_0(T),
\label{Eq:HcFit}
\ee
where the reduced $T$-dependent ordered moment $\bar{\mu}_0(T)$ is obtained by solving Eq.~(\ref{Eq:barmuSoln}) and the zero-temperature critical field is given by
\be
H_{\rm c\perp}(0) = \frac{3k_{\rm B}T_{\rm N}(1-f_J)}{g\mu_{\rm B}(S+1)}.
\ee
In convenient units where $H_{\rm c\perp}(0)$ is expressed in Teslas (1\,T $\equiv 10^4$\,Oe) and taking $g=2$ and $S=7/2$ for Eu$^{+2}$, one has
\be
H_{\rm c\perp}(0)[{\rm T}] = 0.4962(1-f_J)T_{\rm N}[{\rm K}].
\label{Eq:HcPerp02}
\ee
\ese
The values of $T_{\rm N}$ and $f_J$ for the four crystals studied in this paper are given in Table~\ref{Tab:CuriFit}. For Sn-flux-grown crystal \#2~$\rm EuCo_{1.99(2)}As_2$, Eq.~(\ref{Eq:HcPerp02}) gives
\bse
\bea
H_{\rm c\perp}(0) = 12.1~{\rm T},
\label{H0(T)value1}
\eea
and for CoAs-flux-grown crystal \#3~$\rm EuCo_{1.92(4)}As_2$, one obtains
\bea
H_{\rm c\perp}(0)  = 8.9~{\rm T}.\label{H0(T)value2}
\eea
\ese
These values have the same relationship to each other as do the critical fields $H_{\rm c}$ obtained from $M(H,T=2~{\rm K})$ data that are listed in Table~\ref{Tab:MH} for $H\parallel c$.

Using $H_{\rm c\perp}(0)$ as a fitting parameter, we fitted the $H_{\rm c\perp}(T)$ data in Figs.~\ref{Fig:HC_H_Snflux}(b) and~\ref{Fig:HC_H_CoAsflux}(b) by Eq.~(\ref{Eq:HcFit}) and obtained $H_{\rm c\perp}(0) =14.8(4)$~T for crystal~\#2 and $H_{\rm c\perp}(0) =12.1(3)$~T for crystal~\#3.  The fits are shown by the solid blue curves in Figs.~\ref{Fig:HC_H_Snflux}(b) and~\ref{Fig:HC_H_CoAsflux}(b), respectively.


\section{\label{Sec:PhaseDiags} Phase Diagrams in the Field--Temperature Plane}

\begin{figure*}
\includegraphics[width=2.3in]{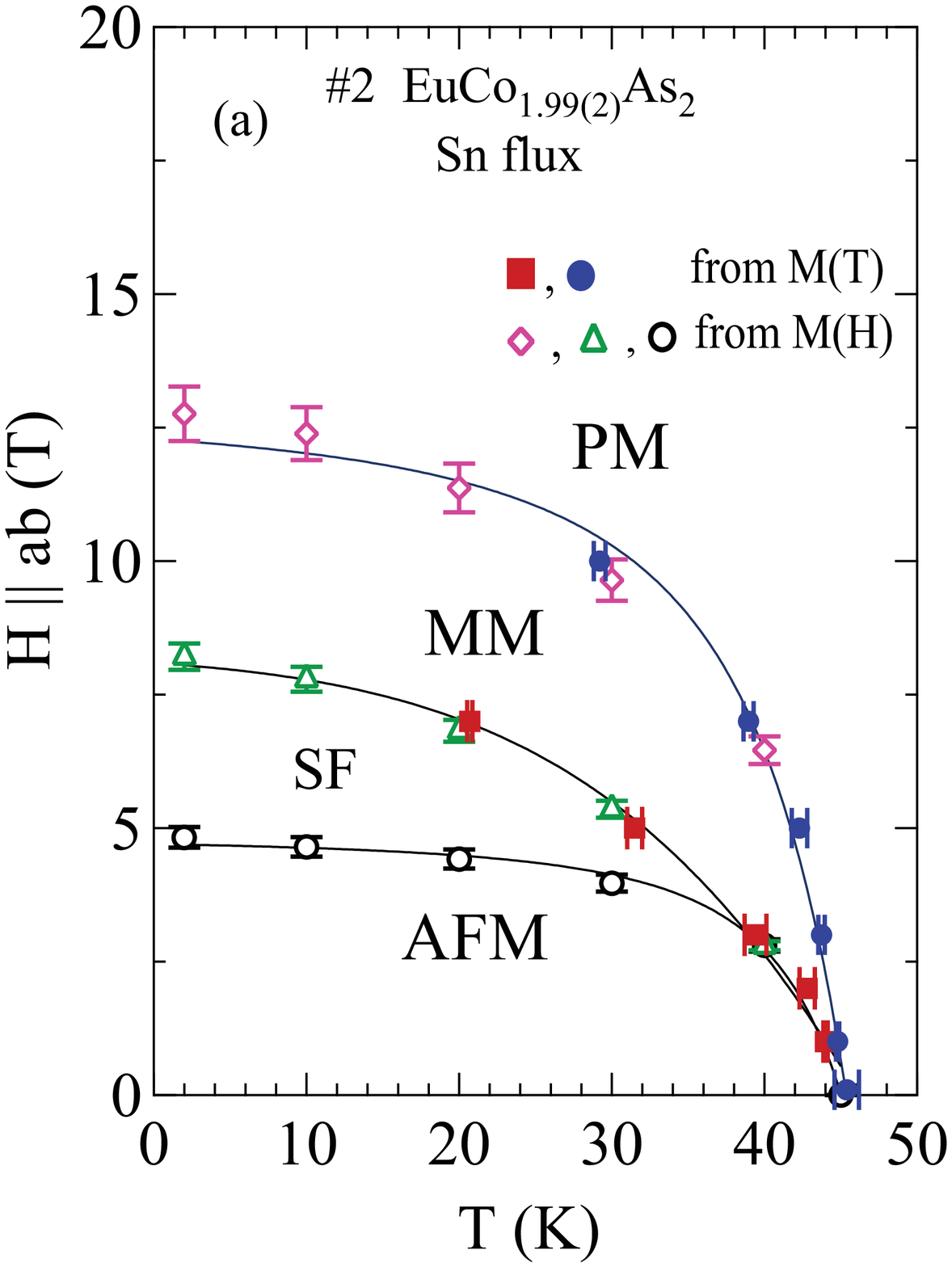}\includegraphics[width=2.2in]{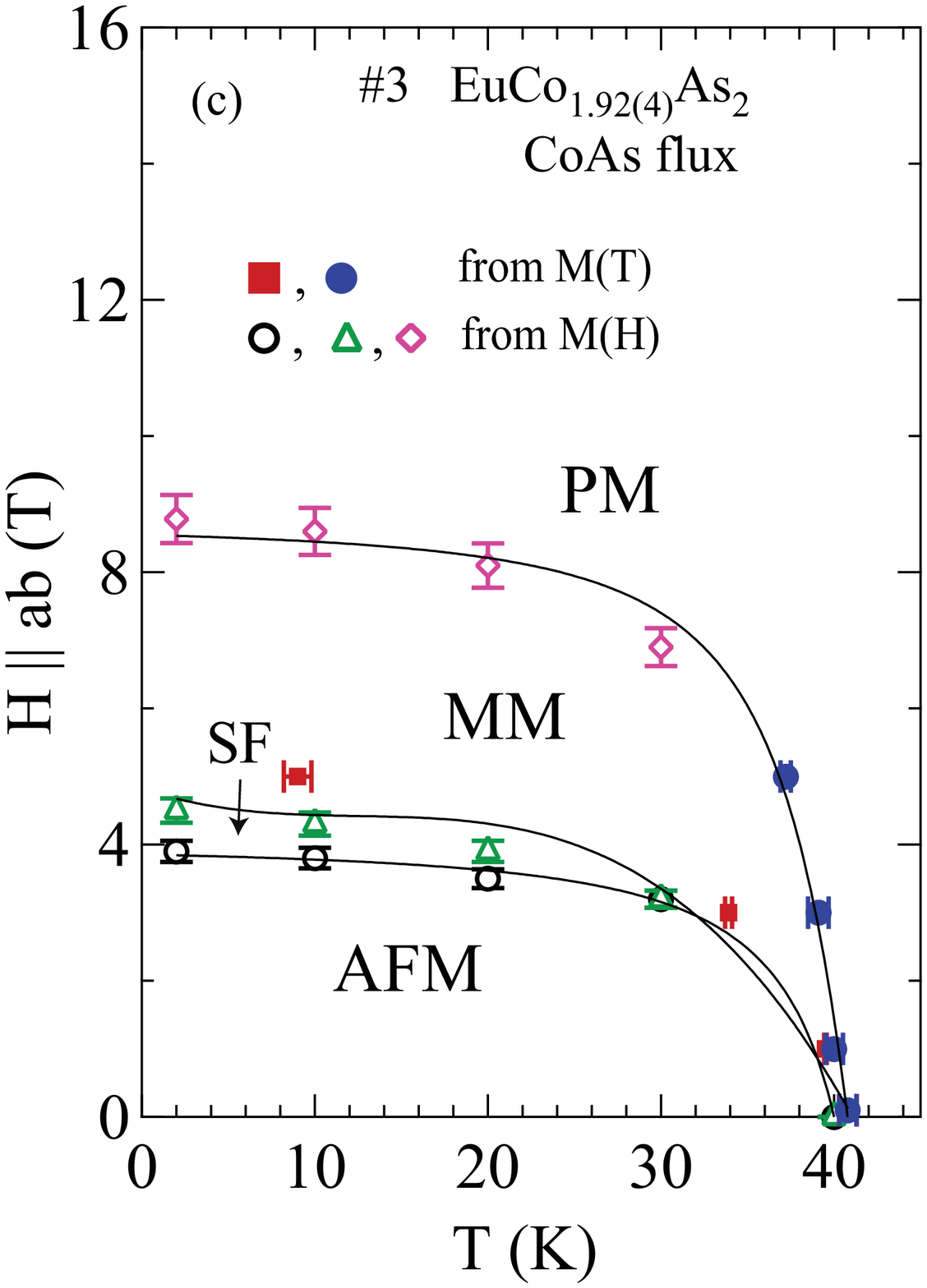}
\includegraphics[width=2.3in]{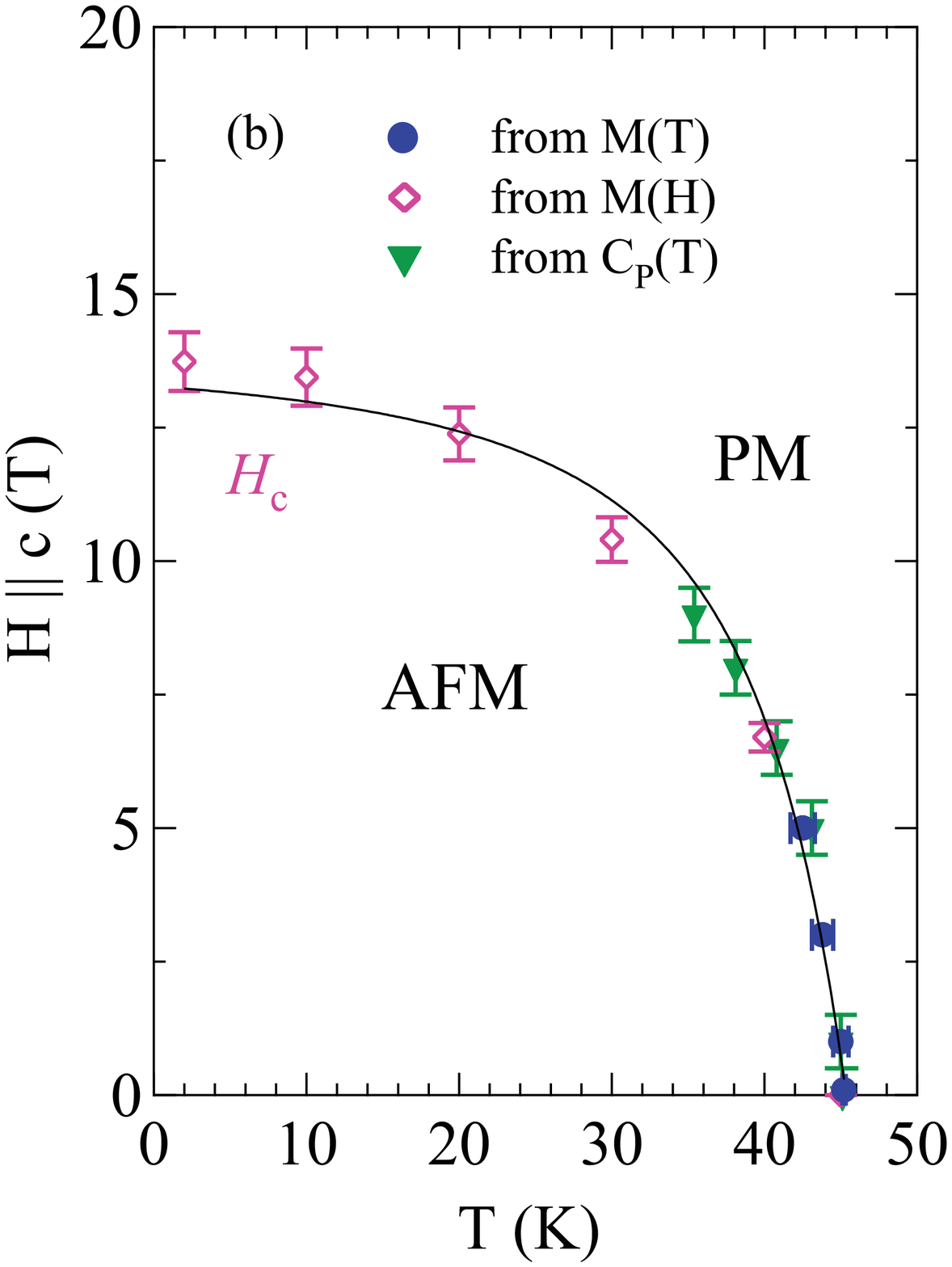}\includegraphics[width=2.2in]{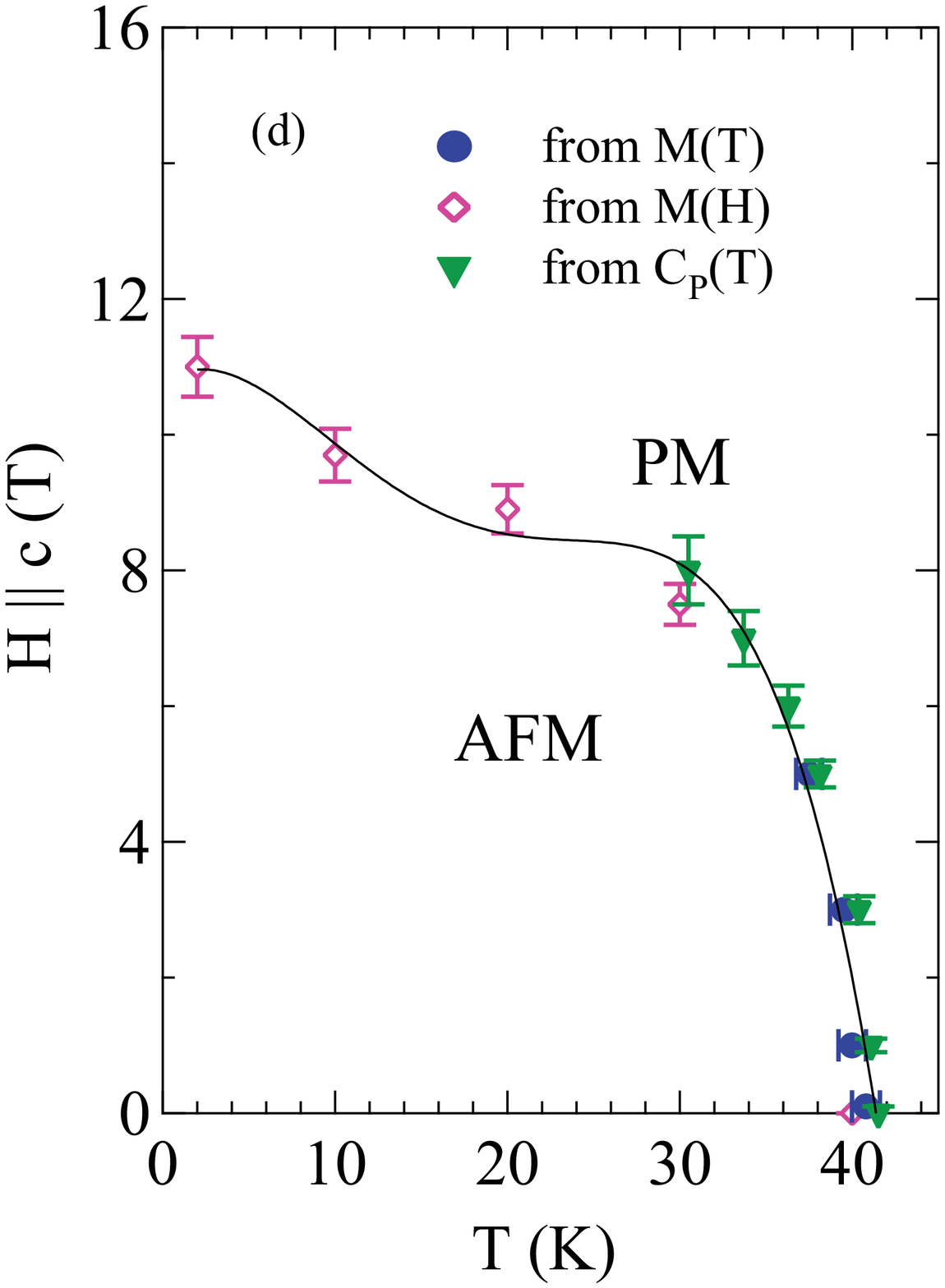}
\caption{The magnetic phase diagrams of (a,b)~Sn-flux-grown crystal \#2~$\rm EuCo_{1.99(2)}As_2$ and (c,d)~CoAs-flux-grown crystal \#3~$\rm EuCo_{1.92(4)}As_2$ for (a,c)~$H\parallel ab$ and (b,d) $H\parallel c$. }
\label{Fig:MT_PD_Snflux}
\end{figure*}

From the transitions observed in Figs.~\ref{Fig:MT_Snflux}--\ref{Fig:dMH_CoAsflux}, \ref{Fig:HC_H_Snflux}, and~\ref{Fig:HC_H_CoAsflux}, the phase diagrams in the $H$--$T$ plane were constructed and are shown for Sn-flux-grown crystal \#2 and CoAs-flux-grown crystal \#3 in Figs.~\ref{Fig:MT_PD_Snflux}(a,b) and~\ref{Fig:MT_PD_Snflux}(c,d), each for both $H\parallel c$ and $H\parallel ab$. For $H~||~c$, the observed phases are the AFM and PM phases, whereas for $H~||~ab$, there are AFM, MM and PM phases.  For $H\parallel c$ in Figs.~\ref{Fig:MT_PD_Snflux}(b,d), the only phase transition line is a second-order transition at the critical field~$H_{\rm c}$ that separates the canted AFM phase from the PM phase. For $H||ab$ in Figs.~\ref{Fig:MT_PD_Snflux}(a,c), there are three phase transition curves: (1)~the first-order spin-flop transition at $H\rm_{SF}$ that separates the canted AFM and SF states; (2)~a second-order intermediate metamagnetic transition at $H\rm_{MM}$ of unknown origin that separates SF and MM phases; and (3)~the second-order critical field transition curve $H_{\rm c}$ that separates the MM and PM states. 

\section{\label{Sec:Res} Electrical Resistivity}

\subsection{Zero-Field Resistivity}

\begin{figure}
\includegraphics[width=3.in]{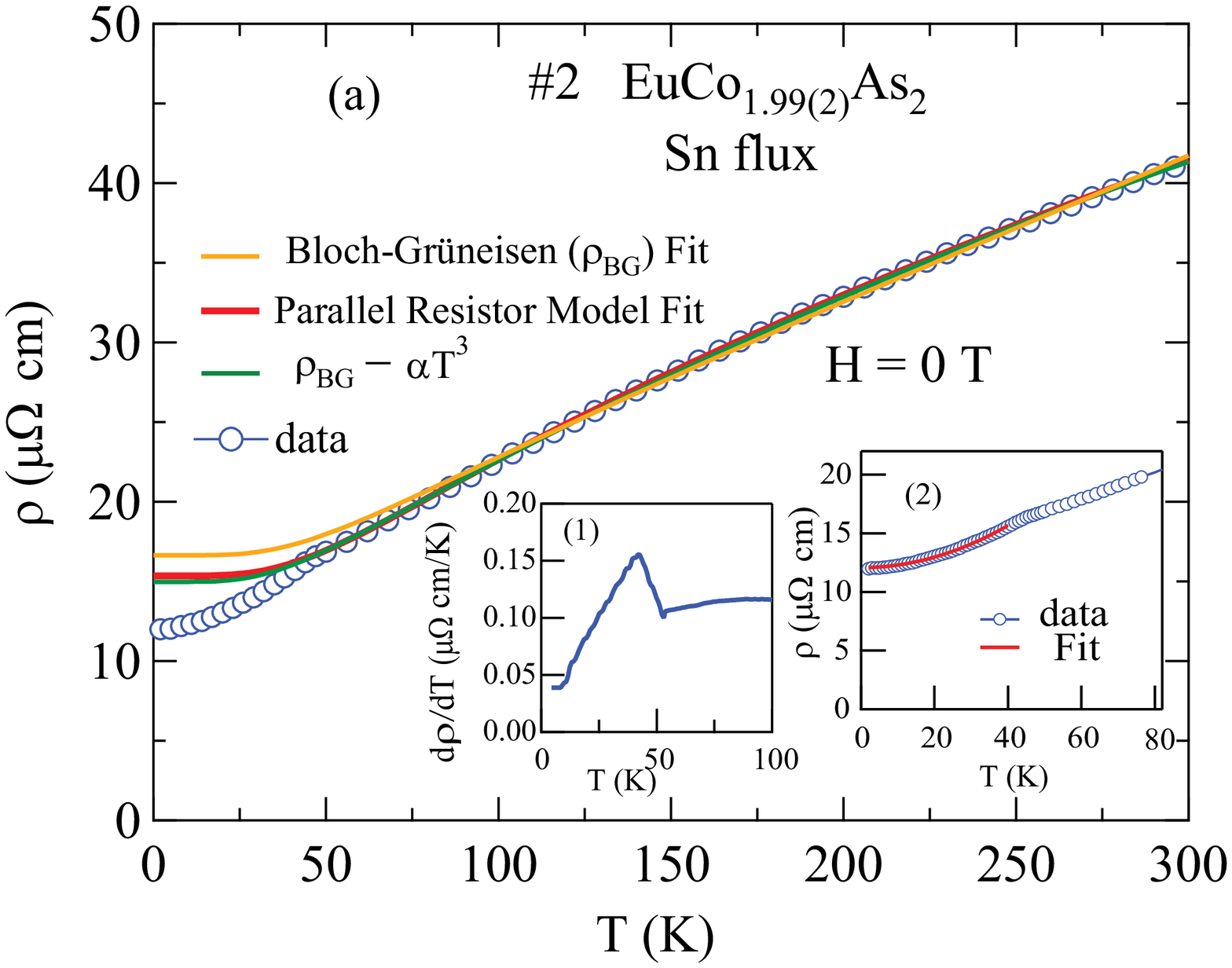}\vspace{0.1in}
\includegraphics[width=3.in]{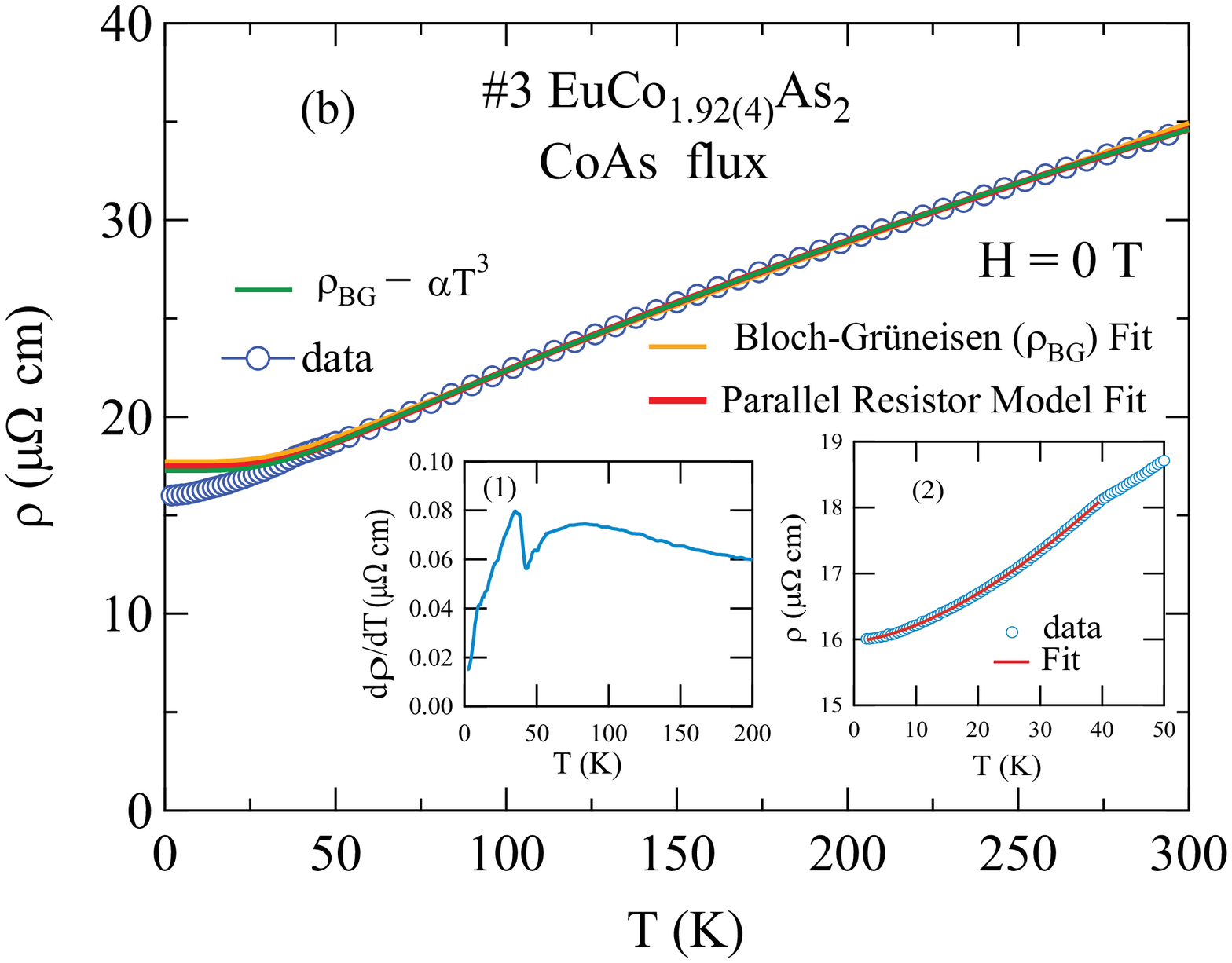}
\caption{In-plane electrical resistivity $\rho$ versus temperature~$T$ of (a)~Sn-flux-grown crystal \#2~$\rm EuCo_{1.99(2)}As_2$ and (b)~CoAs-flux-grown crystal \#3~$\rm EuCo_{1.92(4)}As_2$ as a function of temperature $T$ measured in zero magnetic field. Insets~(1): Temperature derivatives $d\rho/dT$ versus $T$\@. Insets (2): Expanded plots of $\rho(T)$ at low temperatures.  The straight red lines in insets~(2) are fits by $\rho = \rho_0 +AT^2$ over the temperature interval $2~{\rm K} \leq T \leq$ 43~K\@. The fit parameters are listed in Table~\ref{Tab:Rho_T}. The three fits are almost indistinguishable on the scale of the figure.}
\label{Fig_Rho_0T}
\end{figure}

The $ab$-plane electrical resistivity $\rho$ as a function of temperature $T$ from 1.8 to 320~K measured in $H=0$ for Sn-flux-grown crystal~\#2 and CoAs-flux-grown crystal~\#3 are shown in Figs.~\ref{Fig_Rho_0T}(a) and~\ref{Fig_Rho_0T}(b), respectively. The $\rho(T)$ data for both crystals exhibit metallic behavior. For the Sn-flux-grown crystal~\#2, the residual resistivity is $\rho_0 = 12.0~\mu\Omega$~cm at $T= 1.8$~K and the residual resistivity ratio is RRR~$\equiv \rho(320~{\rm K})/\rho(1.8~{\rm K}) = 3.85$. As shown in the inset of Fig.~\ref{Fig_Rho_0T}(a), a slope change in $\rho(T)$ occurs at $T_{\rm N} = 45.0(4)$~K, a value consistent with the $T_{\rm N}$ found from the above $C_{\rm p}(T)$ and $\chi(T)$ measurements on this crystal. 

The $\rho(T)$ for the CoAs-flux-grown crystal~\#3 is shown in Fig.~\ref{Fig_Rho_0T}(b), where $\rho_0 = 16.0~\mu\Omega$~cm at $T= 1.8$~K and RRR~$= 2.16$. The AFM transition is observed at $T_{\rm N} = 40.0(9)$~K, as clearly shown in the plot of $d\rho(T)/dT$ in Fig.~\ref{Fig_Rho_0T}(b) inset~(1), again in agreement with $T_{\rm N}$ found from our $\chi(T)$ and $C_{\rm p}(T)$ data for this crystal.

\begin{table}
\caption{\label{Tab:Rho_T} The parameters obtained from Bloch-Gr\"{u}neisen, Parallel Resistor, and $sd$-Scattering fits obtained using Eqs.~(\ref{Eq:BGM}), (\ref{Eq:PRM}), and~(\ref{Eq:BG_M}), respectively, to $\rho$($T$) data fir \eca\ single crystals in the temperature range \mbox{$50~{\rm K} <  T<~320$~K.}}
\begin{ruledtabular}
\begin{tabular}{lccc}
Crystal: & \underline{\#2 EuCo$_{1.99(2)}$As$_2$\footnotemark[1]} & \underline{\#3 EuCo$_{1.92(4)}$As$_2$\footnotemark[2]} \\
\underline{Fit}\\
\underline{Bloch-Gr\"{u}neisen}\\
$\rho_0$~($\mu\Omega$\,cm) 				& 	16(1)	&	17.7(3)	\\
$F$ ($\mu\Omega$\,cm)					&	21(5)	&	12(1)	\\
$\Theta_{\rm R}$ (K) 					&	257(6)	&	213(3)	\\
\underline{Parallel-Resistor}	\\
$\rho_0$~($\mu\Omega$\,cm) 				&	16.87(4)	&	19.55(4)	\\
$\rho_{\rm max}$~($\mu\Omega$\,cm)			&	168.9(9)	&	164(1)	\\
$F$ ($\mu\Omega$\,cm)					&	32.9(2)	&	18.3(2)	\\
$\Theta_{\rm R}$ (K) 					&	260(2)	&	231(1)	\\
\underline{$sd$-Scattering}	\\
$\rho_0$~($\mu\Omega$\,cm) 				& 	14.7(1)	&	17.38(2)	\\
$F$ ($\mu\Omega$\,cm)					&	20.4(6)	&	12.8(1)	\\
$\Theta_{\rm R}$ (K) 					&	213(6)	&	211(2)	\\
$\alpha$\,($10^{-8}\,\mu\Omega$\,cm/K$^3$)	&	10.8(1)	&	5.14(8)	\\
\end{tabular}
\end{ruledtabular}
\footnotetext[1]{Grown in Sn flux with H$_2$-treated Co powder}
\footnotetext[2]{Grown in CoAs flux with H$_2$-treated Co powder}
\end{table}

The low-$T$ data below $T_{\rm N}$ was fitted well by the quadratic expression $\rho(T)=\rho_0+AT^2$ corresponding to electron-electron scattering, as shown by the solid curve in Fig.~\ref{Fig_Rho_0T}(a) inset~(2) for the Sn-flux-grown crystal and in Fig.~\ref{Fig_Rho_0T}(b) inset(2) for the CoAs-flux-grown crystal, where the fitting parameters are $A = 0.0022(1)~\mu\Omega$~cm/K$^2$ for the Sn-flux-grown crystal, and $A = 0.0065(1)~\mu\Omega$~cm/K$^2$ for the CoAs-flux-grown crystal. 

The $\rho(T)$ above 50~K was fitted by the Bloch-Gr\"{u}neisen (BG) model where the resistivity arises from electron-phonon scattering, given by \cite{Goetsch2012}
\be
\rho{\rm_{BG}}(T)=\rho_0+F\left(\frac{T}{\Theta_{\rm R}}\right)^5\int_{0}^{\Theta_{\rm R}/T}\frac{x^5 dx}{(1-e^{-x})(e^x-1)},\label{Eq:BGM}
\ee
where $F$ is a numerical constant that describes the $T$-independent interaction strength of the conduction electrons with the thermally excited acoustic phonons and contains the ionic mass, Fermi velocity, and other parameters, $x=\frac{\hbar\omega}{2\pi k_{\rm B}T}$, and $\Theta_{\rm R}$ is the resistively-determined Debye temperature \cite{Goetsch2012}. The representation for $\rho{\rm_{BG}}$($T$) used here is an accurate analytic Pad\'e approximant function of $T/\Theta_{\rm R}$ \cite{Goetsch2012}. The fits to the data between 70 and 320~K by Eq.~(\ref{Eq:BGM}) are shown as the yellow curves in the main panels of Figs.~\ref{Fig_Rho_0T}(a) and~\ref{Fig_Rho_0T}(b), and the fitted parameters are listed in Table~\ref{Tab:Rho_T}.

On close examination, the BG model does not provide an optimum fit to the data in Fig.~\ref{Fig_Rho_0T}. A phenomenological model that can describe the negative curvature in $\rho$(T) at high~$T$ is the so-called parallel-resistor model given by \cite{Weismann1977} 
\bea
\frac{1}{\rho(T)} = \frac{1}{\rho_{\rm BG}(T)} + \frac{1}{\rho_{\rm max}},
\label{Eq:PRM}
\eea
where $\rho_{\rm max}$ is the $T$-independent saturation resistivity which is also called the Ioffe-Regel limit \cite{Chakraborty1979}, and $\rho_{\rm BG}(T)$ is the Bloch-Gr\"{u}neisen expression~(\ref{Eq:BGM}).  We fitted the $\rho(T)$ data above $T_{\rm N}$ in the range $50~{\rm K}<T< 320$~K by Eq.~(\ref{Eq:PRM}) as shown by the red curves in Figs.~\ref{Fig_Rho_0T}(a) and~\ref{Fig_Rho_0T}(b). One sees that the data for both crystals are fitted well by the parallel-resistor model and the values of the parameters obtained from the fits are listed in Table~\ref{Tab:Rho_T}.  One sees from the table that the values of $\theta_{\rm R}$ for the two crystals are closer to each other for the parallel-resistor fits compared to the BG fits by themselves and also the fit parameters have higher precision for the parallel-resistor fits.

The negative curvature in the resistivity at the higher temperature that is not fitted by the BG model may be either due to interband scattering or weak additional electron-electron scattering originating from the thermal population of higher-lying energy levels \cite{{Mott_Jones1936},{Mott1936}}. A model that can describe the negative curvature $\rho$($T$) above the ordering temperature is the Bloch-Gr\"{u}neisan-Mott model, given by \cite{Mott1936A}
\bea
\rho{\rm_{BGM}}(T)=\rho{\rm_{BG}}(T)- \alpha T^3,
\label{Eq:BG_M}
\eea
where $\rho{\rm_{BG}}$(T) is the Bloch-Gr\"{u}neisan expression as shown by Eq.(\ref{Eq:BGM}) and $\alpha$ is the $s$-$d$ interband scattering coefficient (Mott coefficient). The fits of the model to the experimental data are shown by the solid green curves in Figs.~\ref{Fig_Rho_0T}(a) and~\ref{Fig_Rho_0T}(b). In this model, when the mean-free path is shorter than on the order of a few atomic spacings, the scattering cross section is no longer linear in~$T$ because under the influence of the lattice vibrations the $s$ electrons may make transitions to the unoccupied or partially-filled $d$~states. As a result, the resistance decreases with increasing temperature and shows negative curvature ($d^2\rho/dT^2<0$). 

\subsection{High-Field Resistivity}

\begin{figure}
\includegraphics[width=3.in]{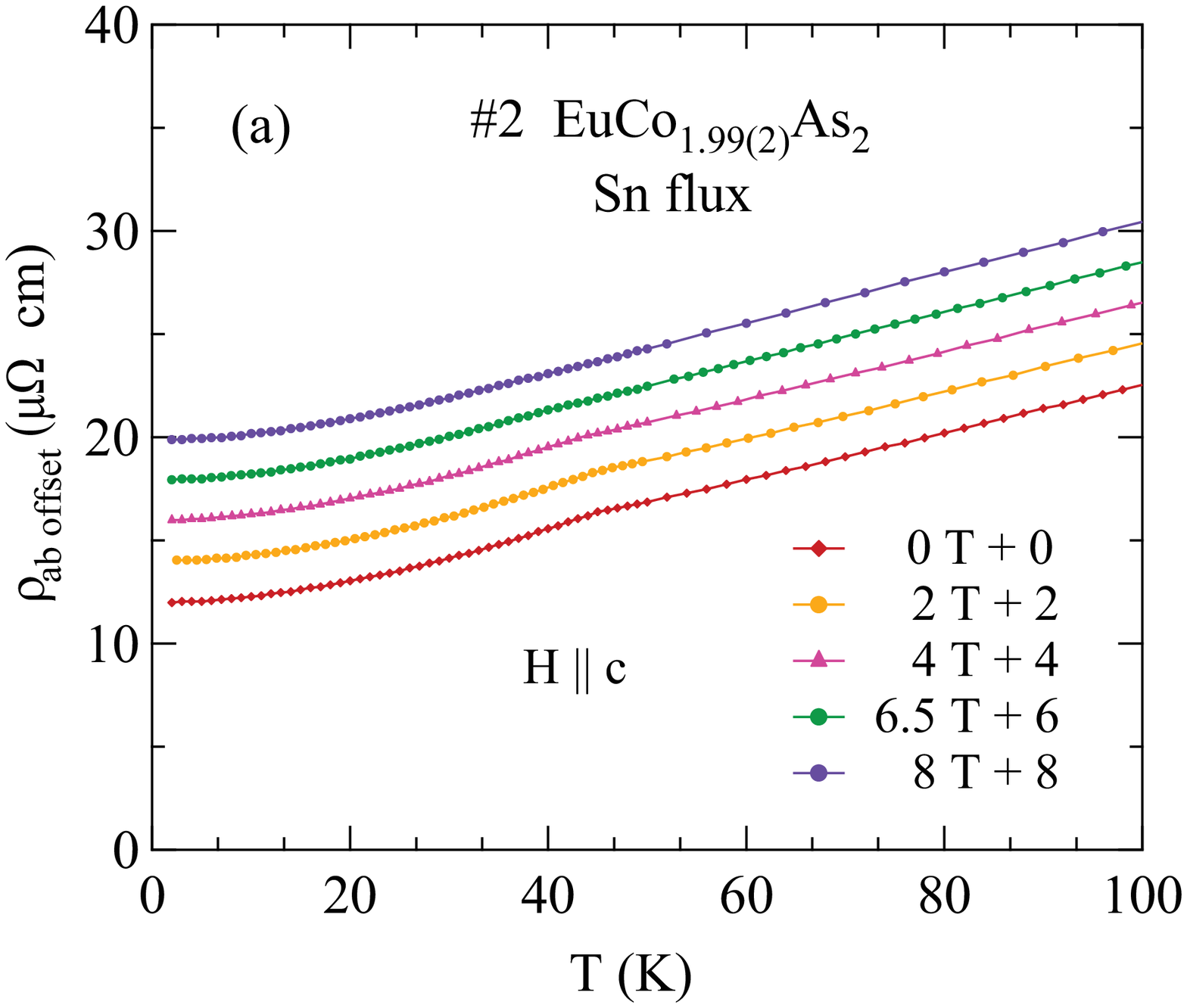}
\includegraphics[width=3.in]{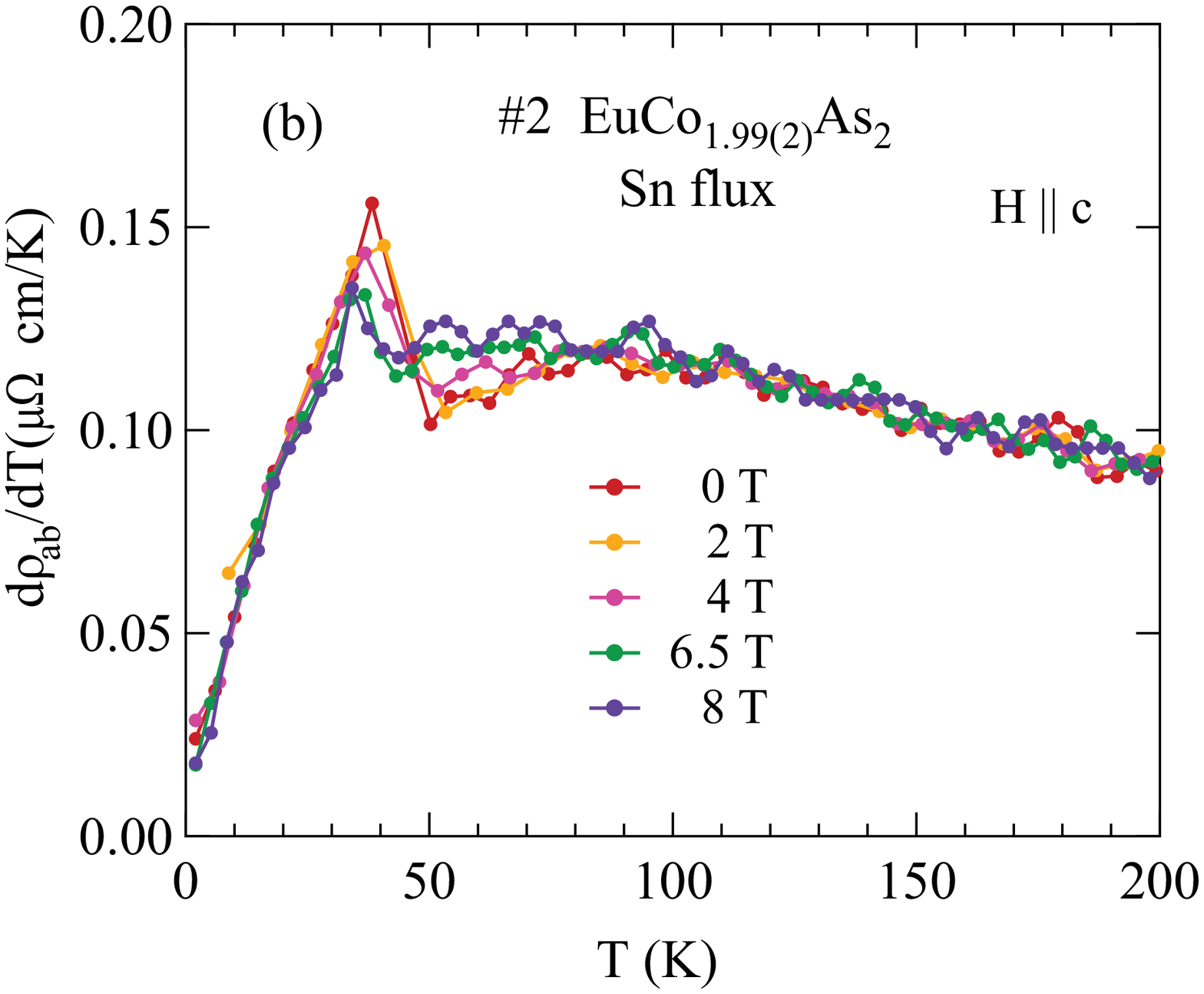}
\includegraphics[width=3.in]{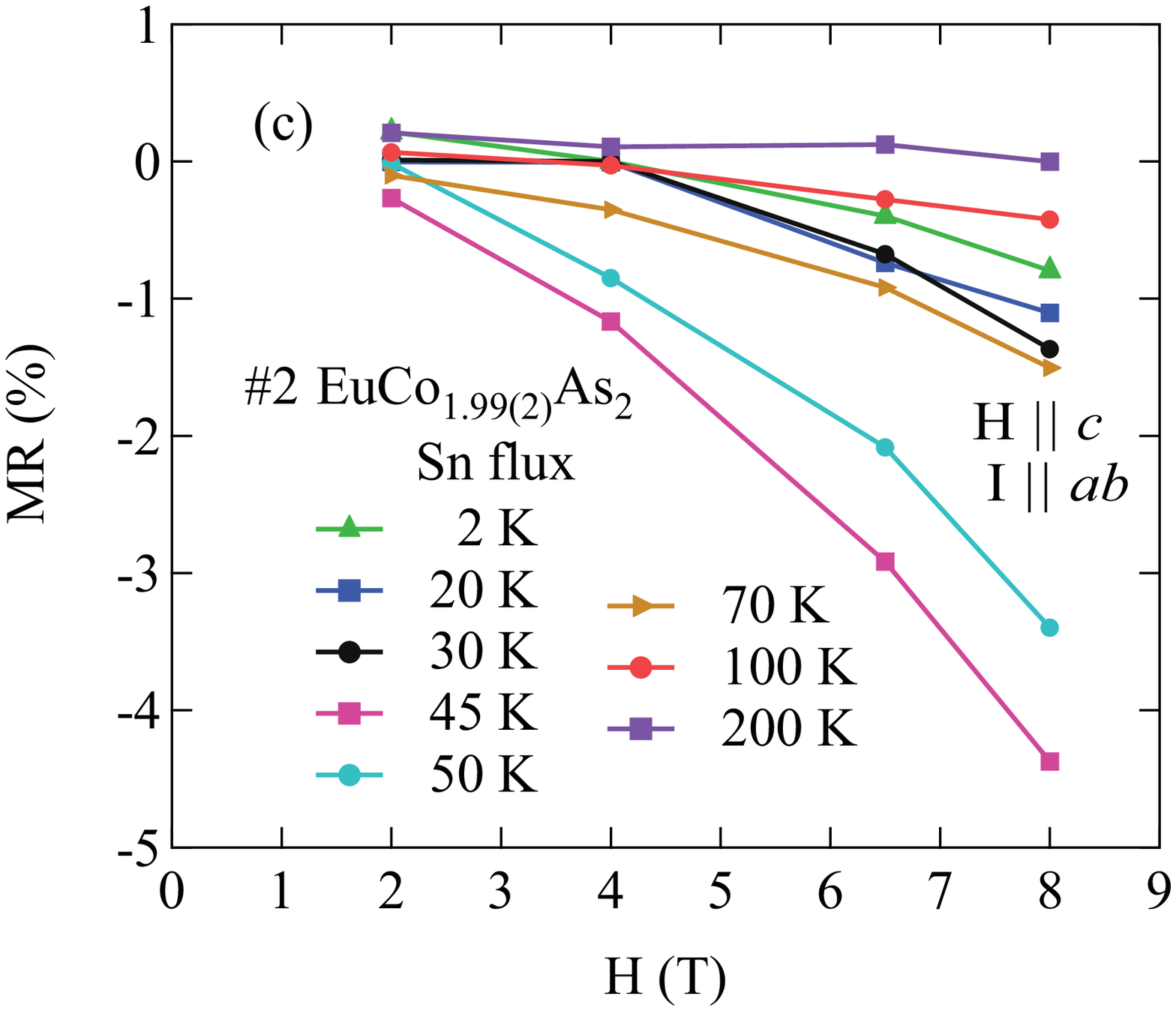}
 \caption{(a) In-plane electrical resistivity $\rho$ of Sn-flux-grown crystal \#2~$\rm EuCo_{1.99(2)}As_2$ as a function of temperature~$T$ measured in the indicated magnetic fields $H\parallel c$. For clarity, the data for successive fields are offset from each other by 2~$\mu\,\Omega$\,cm as indicated.  (b)~Temperature derivative $d\rho/dT$ versus~$T$ obtained from the data in~(a).  (c)~Magnetoresistance MR versus applied field at temperatures ranging from 2 to 20~K for current density $J\parallel ab$ and magnetic fields $H\parallel c$. }
\label{Fig_Rho_H_Snflux}
\end{figure}

\begin{figure}
\includegraphics[width=3.in]{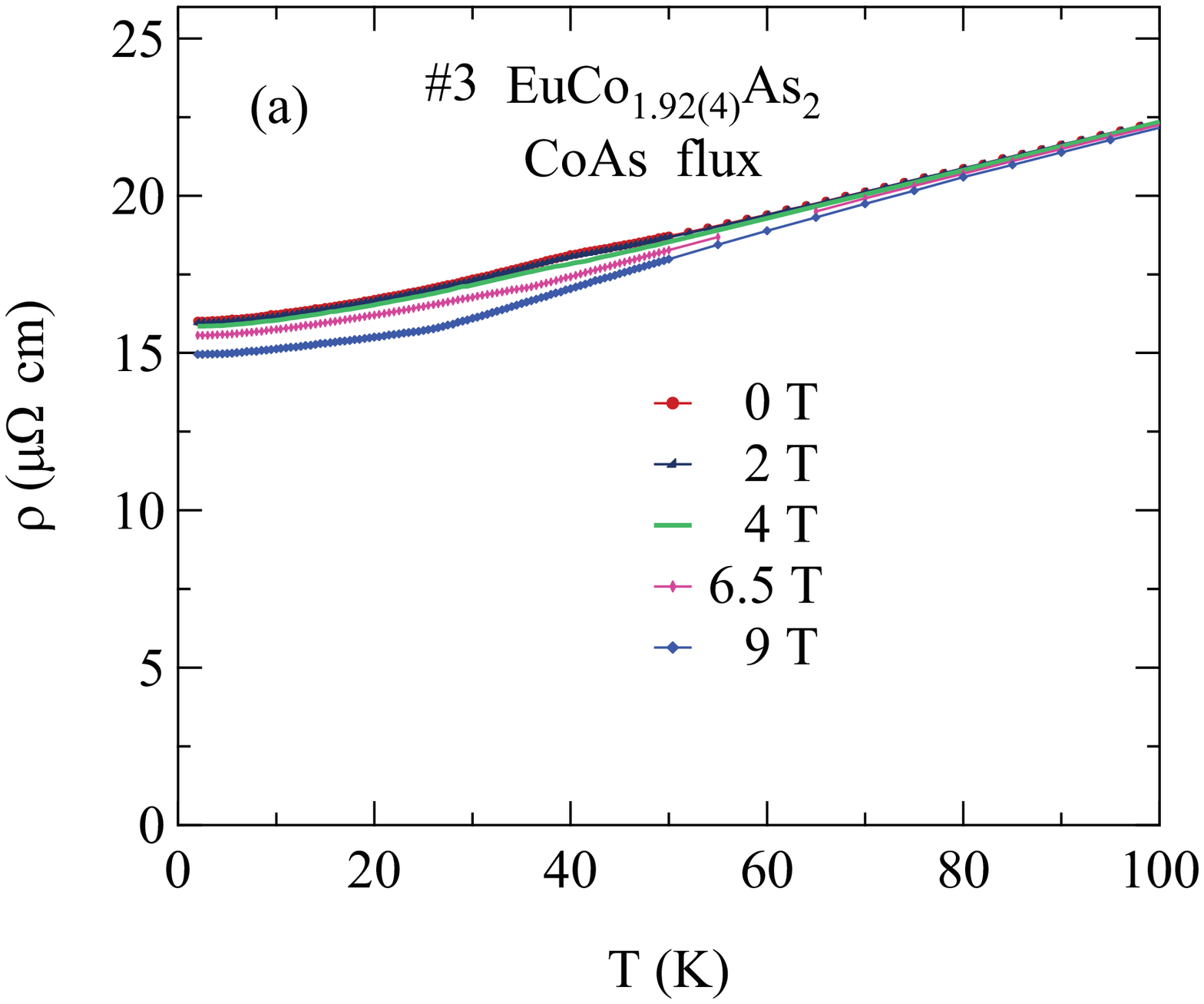}\vspace{0.1in}
\includegraphics[width=3.in]{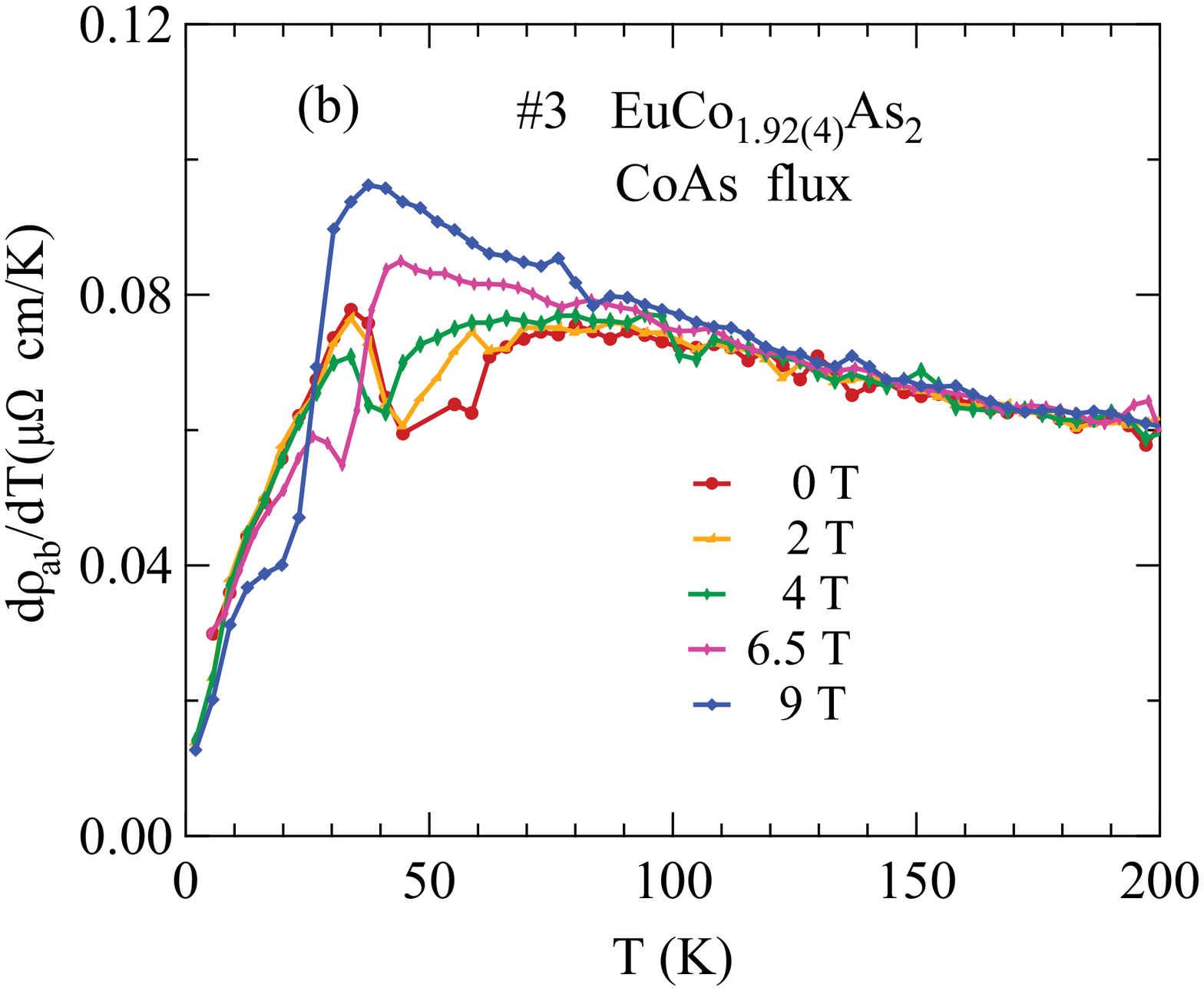}
\includegraphics[width=3.in]{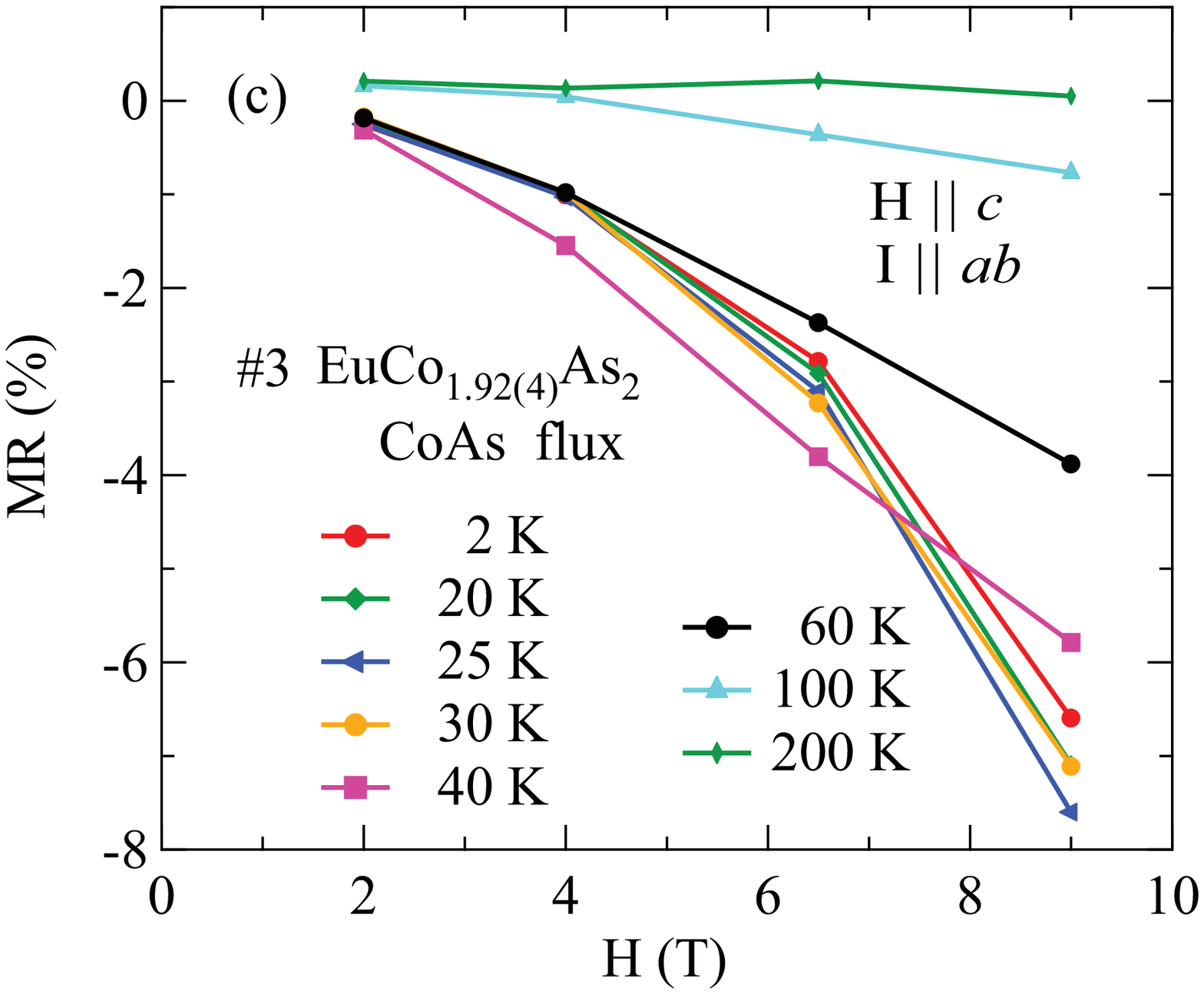}
 \caption{Same as Fig.~\ref{Fig_Rho_H_Snflux} but with CoAs-flux-grown single crystal \#3~$\rm EuCo_{1.92(4)}As_2$ instead.}
\label{Fig_Rho_H_CoAsflux}
\end{figure}

The $\rho(T)$ data at selected magnetic fields applied along the $c$-axis for the \eca\ crystals grown from Sn flux (\#2) and CoAs flux (\#3) are shown in Figs.~\ref{Fig_Rho_H_Snflux}(a) and~\ref{Fig_Rho_H_CoAsflux}(a), respectively. For the Sn-flux-grown crystal, the $d\rho(T)/dT$ data in Fig.~\ref{Fig_Rho_H_Snflux}(a) show that the peak position at $T_{\rm N}$ shifts from 45.0(4)~K at $H=0$ to 35.2(5)~K at $H=8$~T and the transition broadens and smears out progressively with increasing field up to 8~T\@. For the CoAs-grown crystal, $T_{\rm N}$ shifts from 40.0(9)~K to 25.5(3)~K with increasing field up to 9~T\@. The field-dependent $\rho$($H, T$) data for CoAs-flux-grown and Sn-flux-grown crystals show different shapes below $T_{\rm N}$. 

The magnetoresistance (MR), defined as MR($H, T$)$\equiv$ 100\%[$\rho$($H, T$)$-\rho$(0,$T$)]/$\rho$(0,$T$),  calculated from the $\rho$($H$) data are shown in Fig.~\ref{Fig_Rho_H_Snflux}(c) and Fig.~\ref{Fig_Rho_H_CoAsflux}(c). At 2~K, the MR of Sn-flux-grown crystal is negative and attains a maximum negative value of $-0.79\% $ at 8~T whereas for the CoAs-flux-grown crystal, the MR at 9~T is $-6.6\%$ at 2~K\@. The negative curvature in MR versus~$H$ is enhanced as $T_{\rm N}$ is approached, leading to a MR of a $-4.4\%$ for the Sn-flux-grown crystal at $H=$~8~T and $T_{\rm N}=$~45~K, and a MR of $-7.6\%$ for the CoAs-flux-grown crystal at $H=$~9~T and $T_{\rm N}=$~40~K\@. At higher temperatures $T>T_{\rm N}$, the MR shows positive curvature at low fields, and becomes positive at 200~K\@.

In the AFM-ordered state, the exchange interactions tend to align the spins in a different way than an external magnetic field does. As $T_{\rm N}$ is approached from below, the average coupling of the exchange interactions with the conduction electrons is reduced.  Hence the Eu spins become better aligned with the applied field.  This results in a reduction of the spin-disorder scattering, leading to an enhanced negative MR as $T_{\rm N}$ is approached. However, increasing $T$ also results in an increase in spin-disorder scattering due to spin randomization by the thermal energy. Eventually spin-disorder by thermal energy dominates spin alignment by the applied magnetic field, resulting in a positive MR as seen at 200~K \cite{Pippard1989}.

\section{\label{Sec:ElecStruct} Electronic Structure Calculations}

In order to gain further insight on the enhanced Eu moments we performed electronic structure calculations. Our goals were (i) to check  whether there is an enhanced polarization that could justify the observed enhanced effective moment, (ii) if so, to find where it resides, and (iii) how the density of states relates to the measured specific heat.

We performed total energy, and band structure calculations employing the implementation of density functional theory  in the code Dmol$^3$ \cite{Dmol} within Materials Studio. This was done for the stoichiometric 122 system. Since we have permanent magnetic moments due to the $^{8}\text{S}_{7/2}$ configuration of the  Eu $4f$~electrons, we must do spin polarized calculations; otherwise DFT would wrongly split the $4f$~electrons equally over spin-up and spin-down states. We performed a calculation with all Eu spins pointing in the same direction, and another with alternating orientation in consecutive $ab$ plains (from here on referred as configurations F and A, respectively). Although these are only two amongst the infinitely many configurations visited by the system in a paramagnetic state,  such a comparison can give us information on how the relative orientation of the local spins can affect the polarization of the conduction band. This is motivated by the fact that EuCo$_{2-y}$As$_2$ is metallic and it is very likely that exchange interactions between the local moments and the conduction band play a role in the magnetic properties. In addition, the antiferromagnetic ground state should result in zero net polarization of the conduction electrons, while this does not have to be the case for other configurations.

Our calculations included all electrons (i.e., no pseudo-potential was used) in the scalar relativistic approximation. We employed the Perdew-Burke-Ernzerhof exchange correlation functional~\cite{PBE} in the generalized gradient approximation. The Kohn-Sham quasiparticle states were sampled over a {\bf k}-space grid with $7\times 7\times 9$ points and the {\bf k}-space integration for the total energy was done with the tetrahedron method~\cite{Tetra}.  Self-consistency tolerance was set to $2\times 10^{-6}\,\text{Rydberg}$ for the total energy per cell.

\begin{figure}
\includegraphics[width=2.9in]{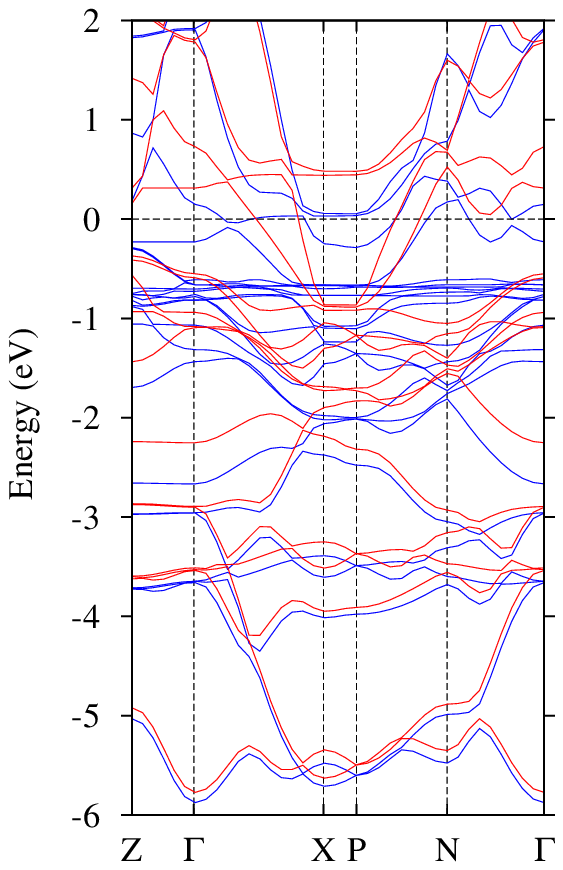}
\includegraphics[width=2.9in]{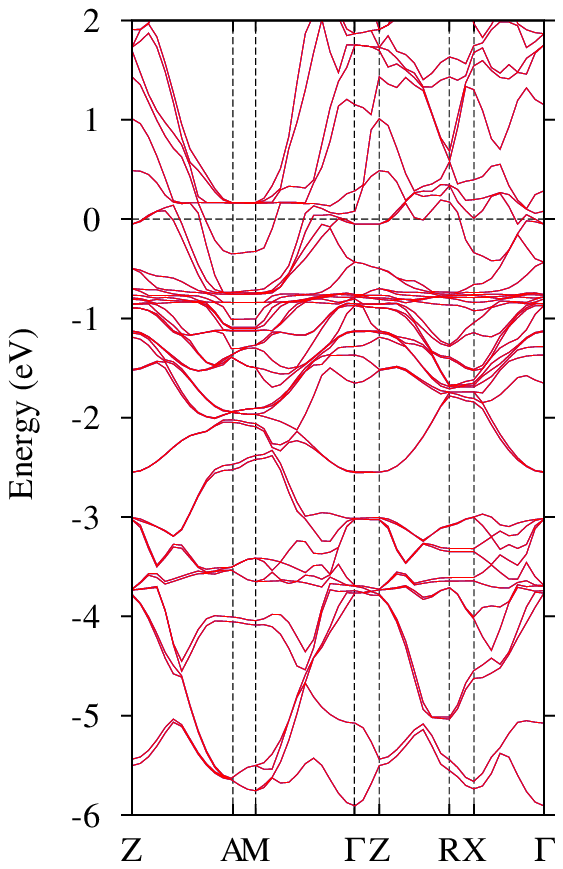}
\caption{Electronic band structure from DFT calculations for $\text{Eu}\text{Co}_2\text{As}_2$ with Eu moments in configuration F (top) and A (bottom). Only states at energies above $-6\, e\text{V}$ (with respect to the Fermi energy $E_{\rm F}$) are shown. These bands are mainly formed by As~$4p$, Co~$3d$, and the localized Eu~$4f$ states which appear around $-0.8$~eV\@.}
\label{bnds}
\end{figure}

\begin{figure}
\includegraphics[width=3.4in]{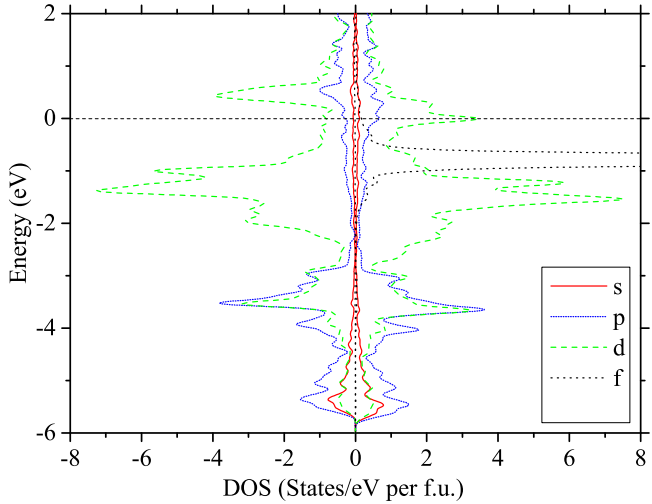}
\caption{Electronic density of states (DOS) from DFT calculations for $\text{Eu}\text{Co}_2\text{As}_2$ with the Eu moments in configuration~F\@. The projection of the $s$ states is shown as the solid red curve, $p$ as the short-dotted blue curve, $d$ as the dashed green curve, and $f$ as dotted-black curve. }
\label{DOSF}
\end{figure}

\begin{figure}
\includegraphics[width=3.4in]{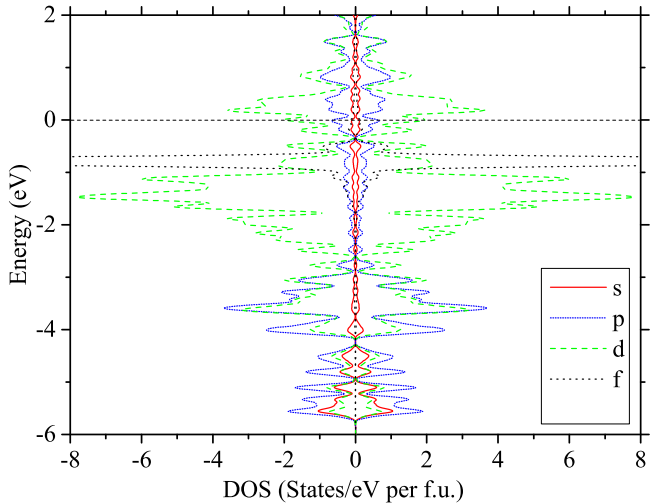}
\caption{Electronic density of states (DOS) from DFT calculations for $\text{Eu}\text{Co}_2\text{As}_2$ with the Eu moments in configuration~A\@. The projection of the $s$ states is shown as the solid red curve, $p$ as the short-dotted blue curve, $d$ as the dashed green curve, and $f$ as dotted-black curve. }
\label{DOSA}
\end{figure}

The band structures in both configurations are shown in Fig.~\ref{bnds}. Projected density of states on atomic orbital type for configurations F and A are shown in Figs.~\ref{DOSF} and~\ref{DOSA}, respectively. One can notice in Fig.~\ref{DOSF} that the polarization induced  by the local Eu moments resides in the $d$~states, which are mainly coming from cobalt atoms. Following the tetrahedral coordination of Co by As; one can roughly divide the $d$ orbitals into two sets, the $e_g$ doublet and the $t_{2g}$ triplet. The former is less affected by the As $4p$ states and appear less hybridized  between $-2.5$~eV and $-1$~eV\@. The $t_{2g}$ states mix more strongly with the As $p$~states resulting in a bonding fraction between $-4$~eV and $-3$~eV (with dominant contribution from As~$p$ orbitals), and an antibonding component at and above the Fermi energy $E_{\rm F}$ (with dominant cobalt~$d$ contribution).  While $d$~states with different spin orientations are shifted with respect to each other at all energies in the F configuration, the $e_g$ states have no net polarization as they appear fully occupied below $E_{\rm F}$\@. The net polarization originates from the $t_{2g}$ states around $E_{\rm F}$\@. States with the same spin orientation as the Eu moments are stabilized (shifted down in energy) and those with the opposite orientation are shifted up (destabilized), resulting in a net enhanced moment per Eu atom.

In configuration F, the projection of the electronic states onto atomic centers gives $7/2$ spin for europium ions and 0.26 for the states belonging to cobalt. In configuration A, the total projected moment on the Eu sites remains as $7/2$ while the cobalt states display a negligible polarization of $\pm 0.01$. This is in agreement with the conclusion from neutron diffraction experiments that Co makes no contribution to the moments in the low-temperature ordered AFM phase.  It is also consistent with the observation that in the paramagnetic state, the fluctuating moments have an enhanced value.  As a very rough estimate, we can consider that having two Co per Eu, which are only polarized half of the time and fully correlated with the orientation of the Eu spins, the effective moment  per Eu turns out to be $\mu_{\text{eff}}\sim 2\sqrt{(7/2+0.26)\times(7/2+0.26+1)}\approx 8.5$. This estimate is suggestively similar to the values obtained from the susceptibility fits.

The total electronic density of states at the Fermi level is predicted to have a very similar value of $D(E_{\rm F})\approx 5\, \text{states}/ e\text{V}$ per f.u.\ for both F and A configurations. This value is comparable to the value of $\approx 6$~states/eV~f.u.\ obtained in Table~\ref{Tab:HC} from the high-temperature fit of Eqs.~(\ref{Eq:Debye_Fit}) to $C_{\rm p}(T)$.   The experimentally-derived value of $D(E_{\rm F})$ is indeed expected to be larger than the band-structure value due to  enhancement of the experimental value by the electron-phonon interaction.

\section{\label{Sec:HeisExchInts} Heisenberg Exchange Interactions}

\begin{figure}[t]
\includegraphics [width=2.in]{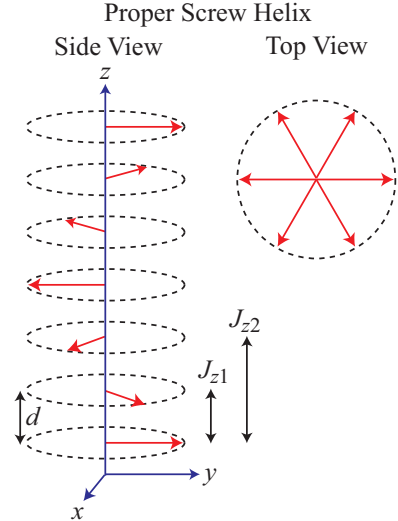}
\caption {Generic helix AFM structure \cite{Johnston2012}.  Each arrow represents a layer of moments perpendicular to the $z$~axis that are ferromagnetically aligned within the $xy$ plane and with interlayer separation $d$.  The wave vector {\bf k} of the helix is directed along the $z$~axis.  The magnetic moment turn angle between adjacent magnetic layers is $kd$.  The exchange interactions $J_{z1}$ and $J_{z2}$ within the $J_0$-$J_{z1}$-$J_{z2}$ Heisenberg MFT model are indicated.}
\label{Fig:J0_Jz1_Jz2_model_helix}
\end{figure}

We now estimate the intralayer and interlayer Heisenberg exchange interactions within the minimal $J_0$-$J_{z1}$-$J_{z2}$ MFT model for a helix in Fig.~\ref{Fig:J0_Jz1_Jz2_model_helix} \cite{Nagamiya1967}, where $J_0$ is the sum of all Heisenberg exchange interactions of a representative spin to all other spins in the same spin layer perpendicular to the helix ($c$) axis, $J_{1z}$ is the sum of all interactions of the spin with spins in an adjacent layer along the helix axis, and $J_{2z}$ is the sum of all interactions of the spin with spins in a second-nearest layer.  Within this model $kd$, $T_{\rm N}$ and $\theta_{{\rm p}}$ are related to these exchange interactions by \cite{Johnston2012, Johnston2015}
\bse
\label{Eqs:J0J1zJ2z}
\bea
&&\cos(kd) = -\frac{J_{z1}}{4J_{z2}},\\*
T_{\rm N} &=& -\frac{S(S+1)}{3k_{\rm B}} \big[J_0 + 2J_{z1}\cos(kd)\nonumber\\*
&& \hspace{0.9in} +\ 2J_{z2}\cos(2kd)\big], \label{eq:TN}\\*
\theta_{\rm p} &=& -\frac{S(S+1)}{3k_{\rm B}} \left(J_0+2J_{z1}+2J_{z2}\right),
\label{eq:thetap}
\eea
\ese
where a positive (negative) $J$ corresponds to an AFM (FM) interaction. The three exchange constants $J_0,\ J_{z1}$ and~$J_{z2}$ are obtained by solving Eqs.~(\ref{Eqs:J0J1zJ2z}) using $S = 7/2$, $kd=0.79\pi$, and the  $T_{\rm N}$ and $\theta_{\rm p}=\theta_{\rm p\,ave}$ values in Table~\ref{Tab:CuriFit},  and the results are listed in Table~\ref{Tab:HEI}.

\begin{figure}
\includegraphics[width=1in]{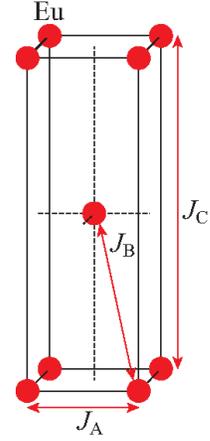}
\caption {Body-centered tetragonal Eu sublattice, where $c/a = 2.93$.  The Heisenberg exchange interactions $J_{\rm A},\ J_{\rm B}$ and $J_{\rm C}$ are defined in the figure.}
\label{Fig:bct_Eu_lattice}
\end{figure}

The classical energy per spin in an ordered spin system in $H=0$ with no anisotropy and containing identical crystallographically-equivalent spins is
\be
E_i = \frac{1}{2} \sum_j J_{ij}{\bf S}({\bf R}_i)\cdot{\bf S}({\bf R}_j),
\label{Eq:Hi}
\ee
where the factor of 1/2 arises because the energy of an interacting spin pair is equally shared between the two spins in the pair, the sum is over the neighboring ordered spins ${\bf S}({\bf R}_j)$ of the given central spin ${\bf S}({\bf R}_i)$ and the $J_{ij}$ are the Heisenberg exchange interactions between each respective spin pair.  
Here we only consider Bravais spin lattices where the position of each spin is a position of inversion symmetry of the spin lattice such as the body-centered-tetragonal (bct) spin lattice in Fig.~\ref{Fig:bct_Eu_lattice}.  We further restrict our attention to coplanar AFMs in which the ordered moments in the ordered AFM state are aligned in the $xy$ plane such as for the coplanar helix.

The expression for the classical ground-state energy per spin obtained from Eq.~(\ref{Eq:Hi}) is
\be
E_i = \frac{S^2}{2} \sum_j J_{ij}\cos\phi_{ji},
\label{Eq:Ei}
\ee
where $\cos\phi_{ji} = \hat{{\bf S}}({\bf R}_i)\cdot\hat{{\bf S}}({\bf R}_j)$ and $\phi_{ji}$ is the azimuthal angle within the $xy$~plane between the ordered spins ${\bf S}({\bf R}_j)$ and ${\bf S}({\bf R}_i)$.  Within the $J_0$-$J_{z1}$-$J_{z2}$ model one obtains
\be
E_i = \frac{S^2}{2} \Big[J_0 + 2J_{z1}\cos(kd) + 2 J_{z2}\cos(2kd)\Big],
\label{Eq:EiHelix}
\ee
where we take the ground-state turn angle to be $kd=0.79\pi$ for all \eca\ samples.  Using $S=7/2$ and the values of $J_0,\ J_{z1}$ and~$J_{z2}$ in Table~\ref{Tab:HEI}, one obtains the classical ground-state energies per spin $E_i$ listed in Table~\ref{Tab:HEI}.  The values are in the range $-46$~K to $-52$~K, with magnitudes that are similar to the N\'eel temperatures themselves as might have been expected.

\begin{table*}
\caption{\label{Tab:HEI} Exchange constants in the $J_0$-$J_{z1}$-$J_{z2}$ model obtained from Eqs.~(\ref{Eqs:J0J1zJ2z}) and the corresponding classical ground-state energies per spin~$E_i$ calculated from Eq.~(\ref{Eq:EiHelix}).  The exchange interactions between Eu spins $J{\rm_A}$, $J{\rm_B}$ and $J{\rm_C}$ obtained using Eq.~(\ref{Eq:JABC}) are also listed. Negative $J$ values are FM and positive values are AFM.}

\begin{ruledtabular}
\begin{tabular}{lccccccc}
Compound& $J_0/k_{\rm B}$ & $J_{z1}/k_{\rm B}$ & $J_{z2}/k_{\rm B}$ & $E_i/k_{\rm B}$ & $J{\rm_A}/k{\rm_B}$ & $J{\rm_B}/k{\rm_B}$ & $J{\rm_C}/k{\rm_B}$  \\
           &(K)     &(K)         &(K)               &(K)        &(K)         &(K)               \\
\hline
\#1 EuCo$_{1.90(1)}$As$_2$\footnotemark[1]	& $-6.85$	& 1.222	&0.387	& $-50.1$	& $-1.712$	& 0.306	& 0.387	 \\
\#2 EuCo$_{1.99(2)}$As$_2$\footnotemark[2]	& $-6.84$	& 1.200	&0.380	& $-49.9$	& $-1.711$ 	& 0.300 	& 0.380 	 \\
\#3 EuCo$_{1.92(4)}$As$_2$\footnotemark[3]	& $-6.58$	& 0.836	&0.265	& $-45.9$	& $-1.645$ 	& 0.209 	& 0.265 	\\
\#4 EuCo$_{1.90(2)}$As$_2$\footnotemark[4]	& $-6.54$	& 0.853	&0.270	& $-45.7$	& $-1.635$ 	& 0.213 	& 0.270 	\\
\#5 EuCo$_{1.92(1)}$As$_2$\footnotemark[4]	& $-6.60$	& 0.755	&0.239	& $-47.0$	& $-1.651$ 	& 0.189 	& 0.239 	\\
EuCo$_2$As$_2$\footnotemark[4] \cite{Ballinger2012} 	& $-6.87$	& 0.606	&0.192	& $-46.1$	& $-1.718$ 	& 0.151 	& 0.192 	 \\
EuCo$_2$As$_2$\footnotemark[5] \cite{Tan2016} 	& $-6.77$	& 1.533	& 0.485	& $-51.7$	&  $-1.693$	& 0.383	& 0.485	\\
\end{tabular}
\end{ruledtabular}

\footnotetext[1]{Grown in Sn flux}
\footnotetext[2]{Grown in Sn flux with H$_2$-treated Co powder}
\footnotetext[3]{Grown in CoAs flux with H$_2$-treated Co powder}
\footnotetext[4]{Grown in CoAs flux}
\footnotetext[5]{Grown in Bi flux}

\end{table*}

The bct Eu sublattice of \ecaa\ is shown in Fig.~\ref{Fig:bct_Eu_lattice}, where the measured ratio $c/a = 2.93$ is to scale.  Assuming that the exchange interactions $J_{\rm A}$, $J_{\rm B}$ and $J_{\rm C}$ in the figure are the only ones present, in terms of the interactions in the $J_0$-$J_{z1}$-$J_{z2}$ model one has
\be
J_0 = 4J_{\rm A}, \quad J_{z1} = 4J_{\rm B}, \quad J_{z2} = J_{\rm C}.
\label{Eq:JABC}
\ee
Then using the values of $J_0$, $J_{z1}$ and $J_{z2}$ in Table~\ref{Tab:HEI} one obtains the $J_{\rm A}$, $J_{\rm B}$, and $J_{\rm C}$ values which are listed in Table~\ref{Tab:HEI}.

\section{\label{Summary} Summary}

Investigations of the physical properties of \eca\ crystals with the \tcs\ structure that were grown in Sn and CoAs fluxes are reported.  For most of our crystals, we find $\approx5$\% vacancies on the Co sites, similar to the value of 7\% vacancies on the Co sites in \cca\ \cite{Anand2014, Quirinale2013}.  

In-plane electrical resistivity $\rho(T)$ measurements indicate metallic behavior of the two crystals studied, with a kink in $\rho(T)$ at the respective $T_{\rm N}$.  High-field $\rho(T)$ data with $H\parallel c$ reveal negative magnetoresistance, reaching $\approx -5$\% at $T=2$~K and $H = 9$~T\@.

\eca\ contains Eu$^{+2}$ ions with expected spin~$S=7/2$ and $g=2$, which exhibit AFM ordering at $\approx 45$~K for the Sn-flux-grown crystals and $\approx 41$~K for the CoAs-flux-grown crystals.  We obtained good fits using molecular-field theory (MFT) to the low-field $ab$-plane magnetic susceptibility of the helical AFM structure below~$T_{\rm N}$ with the Eu moments aligned in the $ab$~plane.  Zero-field heat capacity $C_{\rm p}$ measurements were carried out and the magnetic contribution $C_{\rm mag}(T)$ was extracted.  The $C_{\rm mag}(T)$ data below $T_{\rm N}$ were fitted reasonably well by MFT\@.  The $C_{\rm mag}(T)$ above $T_{\rm N}$ is nonzero, indicating the presence of dynamic short-range AFM ordering above~$T_{\rm N}$.  Thus the molar magnetic entropy $S_{\rm mag}$ at $T_{\rm N}$ is only about 90\% of the completely disordered value $R\ln8$, the remainder being recovered by about 70~K\@.

The high-field magnetization in the $ab$~plane below $T_{\rm N}$ exhibits a spin-flop-like transition followed by a second-order metamagnetic transition to an unknown AFM structure and then a second-order AFM to paramagnetic (PM)  transition, whereas high-field $c$-axis measurements reveal a second-order transition of unknown nature in addition to the expected second-order canted-AFM to PM transition.  High-field $C_{\rm p}(T)$ measurents with $H\parallel c$ only reveal the AFM to PM transition, where the $T_{\rm N}$ and the heat capacity jump at $T_{\rm N}$ both decrease with increasing $H$\@.  Phase diagrams in the $H\parallel ab$ and $H\parallel c$ versus~$T$ planes were constructed from the high-field magnetization and heat capacity results.  

\begin{table*}
\caption{\label{Tab:MusatMueff} Effective moment $\mu_{\rm eff}$ and saturation moment $\mu_{\rm sat}$ at $T=2$~K of \eca\ obtained from Tables~\ref{Tab:CuriFit} and~\ref{Tab:MH}.  The fourth and sixth columns show the deviations of these quantities from the theoretical values in Eqs.~(\ref{Eqs:mueff0musat0}).  Literature data for other compounds are also shown. }

\begin{ruledtabular}
\begin{tabular}{lccccr}
Crystal								& Field			&  $\mu_{\rm eff}$ 	& $\frac{\Delta\mu_{\rm eff}}{\mu_{\rm eff0}}$	& $\mu_{\rm_{sat}}$	& $\frac{\Delta\mu_{\rm sat}}{\mu_{\rm sat0}}$	\\
Designation							& Direction		&  ($\mu_{\rm B}$/Eu)  & (\%)		& ($\mu_{\rm B}$/Eu)&(\%)	\\
\hline
\#1 EuCo$_{1.90(1)}$As$_2$\footnotemark[1]	& $H\parallel ab$ 	& 8.48			& 6.8	& 7.15	& 2.1			\\
									& $H\parallel c$ 	& 8.47			& 6.7	& 7.05	& 0.7			\\
\#2 EuCo$_{1.99(2)}$As$_2$\footnotemark[2]	& $H\parallel ab$ 	& 8.59			& 8.2	& 7.03	& 0.4			\\
									& $H\parallel c$ 	& 8.66			& 9.1	& 7.05	& 0.7		 	\\
\#3 EuCo$_{1.92(4)}$As$_2$\footnotemark[3]	& $H\parallel ab$ 	& 8.59			& 8.1	& 7.59	& 8.4			\\
									& $H\parallel c$ 	& 8.49			& 6.9	& 7.57	& 8.1			\\
\#4 EuCo$_{1.90(2)}$As$_2$\footnotemark[4]	& $H\parallel ab$ 	& 8.51			& 7.2	& 7.34	& 4.9			\\
									& $H\parallel c$ 	& 8.50			& 7.1	& 7.19	& 2.7			\\
\#5 EuCo$_{1.90(2)}$As$_2$\footnotemark[4]	& $H\parallel ab$ 	& 8.56			& 7.8	& 7.50	& 7.1			\\
									& $H\parallel c$ 	& 8.71			& 9.7	& 7.58	& 8.3			\\
${\rm EuCo_2As_2}$ \cite{Tan2016}			&				&				&		& 7.26(8)\footnotemark[6]		&	3.7		\\
${\rm EuCo_2P_2}$ \cite{Sangeetha2016,Reehuis1992}& $H\parallel ab$& 7.83(1)		& $-1.4$	& 6.9(1)\footnotemark[7]	&	$-1.4$			\\
									& $H\parallel c$	& 7.84(1)			& $-1.3$	& 	&				\\
${\rm EuFe_2As_2}$ \cite{Xiao2009}			&				&				& 		& 6.8(3)\footnotemark[7]		&	$-2.9$		\\
${\rm EuPd_2Sb_2}$\footnotemark[5] \cite{Das2010}	&			& 7.61(2)			& $-4.2$		& 		&				\\
${\rm EuCu_2As_2}$ \cite{Anand2015}			& $H\parallel ab$	& 7.72(1)			& $-2.8$		& 6.66	& $-4.9$		\\
									& $H\parallel c$	& 7.82(1)			& $-1.5$		& 6.77	& $-3.3$		\\
${\rm EuCu_{1.82}Sb_2}$\footnotemark[5] \cite{Anand2015,Ryan2015}& $H\parallel ab$	& 7.70(1)		& $-3.0$		& 6.76\footnotemark[8]	& $-3.4$		\\
													& $H\parallel c$	& 7.77(1)		& $-2.1$		& 6.95	& $-0.7$		\\

\end{tabular}
\end{ruledtabular}
\footnotetext[1]{Grown in Sn flux}
\footnotetext[2]{Grown in Sn flux with H$_2$-treated Co powder}
\footnotetext[3]{Grown in CoAs flux with H$_2$-treated Co powder}
\footnotetext[4]{Grown in CoAs flux}
\footnotetext[5]{Primitive-tetragonal ${\rm CaBe_2Ge_2}$ structure with space group $P4/nmm$}
\footnotetext[6]{Crystal grown in Bi flux; no Co vacancies detected; neutron diffraction measurement}
\footnotetext[7]{From neutron-diffraction measurements \cite{Reehuis1992}}
\footnotetext[8]{Neutron-diffraction measurements \cite{Ryan2015} give an ordered moment of 7.08(15)~$\mu_{\rm B}$/Eu}

\end{table*}

A primary goal of the present work was to investigate a possible enhancement of the Eu magnetic moment for crystals of \eca\ prepared under different conditions.  Shown in Table~\ref{Tab:MusatMueff} is a summary of the effective moments $\mu_{\rm eff}$ obtained from modified Curie-Weiss law fits in the paramagnetic state at $T>T_{\rm N}$ for five of the crystals studied here and the corresponding saturation moments $\mu\rm_{sat}$ obtained from high-field $M(H)$ isotherms at $T=2$~K of \eca\ from Tables~\ref{Tab:CuriFit} and~\ref{Tab:MH}, respectively.  These two moments are given in general for a spin with no contribution of orbital moments by
\bse
\label{Eqs:mueffmusat}
\bea
\mu_{\rm eff} &=& g\sqrt{S(S+1)}\,\mu_{\rm B},\\*
\mu_{\rm sat} &=& gS\,\mu_{\rm B}.
\eea
\ese
For spin-only Eu$^{+2}$, one expects $S=7/2$ and~$g\approx 2$, yielding
\bse
\label{Eqs:mueff0musat0}
\bea
\mu_{\rm eff0} &=& 7.94\,\mu_{\rm B}/{\rm Eu},\\*
\mu_{\rm sat0} &=& 7.00\,\mu_{\rm B}/{\rm Eu}
\eea
\ese
Comparing these values with those in Table~\ref{Tab:MusatMueff} shows that both Sn-flux-grown and CoAs-flux-grown crystals show significant  enhancements of $\mu_{\rm eff}$ and/or $\mu_{\rm sat}$.  Also shown in the table are the relative enhancements of the observed moments with respect to the expected moments as expressed by $\Delta\mu/\mu_0\equiv (\mu_{\rm obs}-\mu_0)/\mu_0$.  One sees that the effective moment $\mu_{\rm eff}$ values are all enhanced by 6.7\% to 9.1\% with respect to the unenhanced value.  The saturation moments $\mu_{\rm sat}$ also exhibit enhancements, but the enhancement is more variable, from 0.4\% to 8.4\%.

\begin{figure}
\includegraphics[width=3.3in]{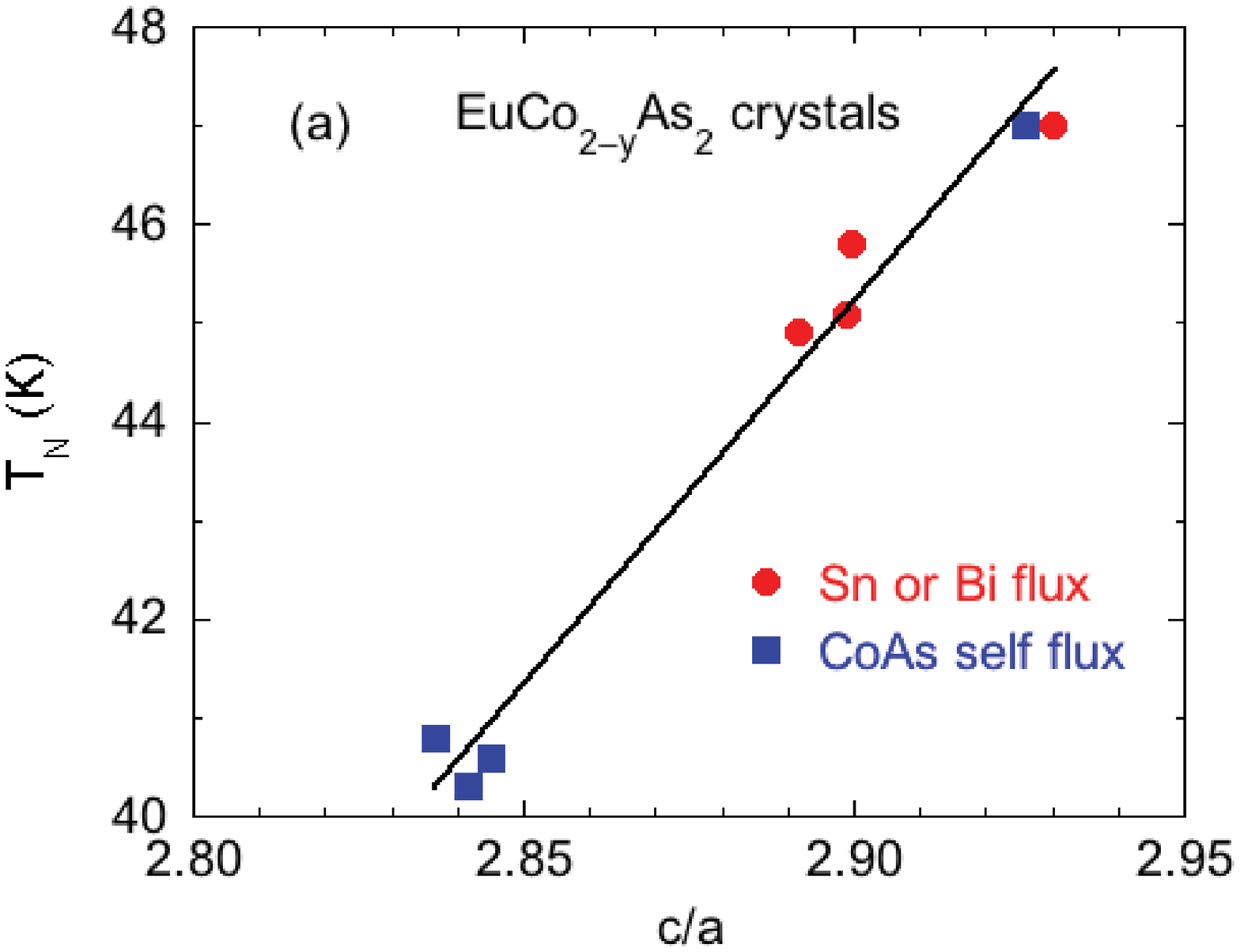}
\includegraphics[width=3.3in]{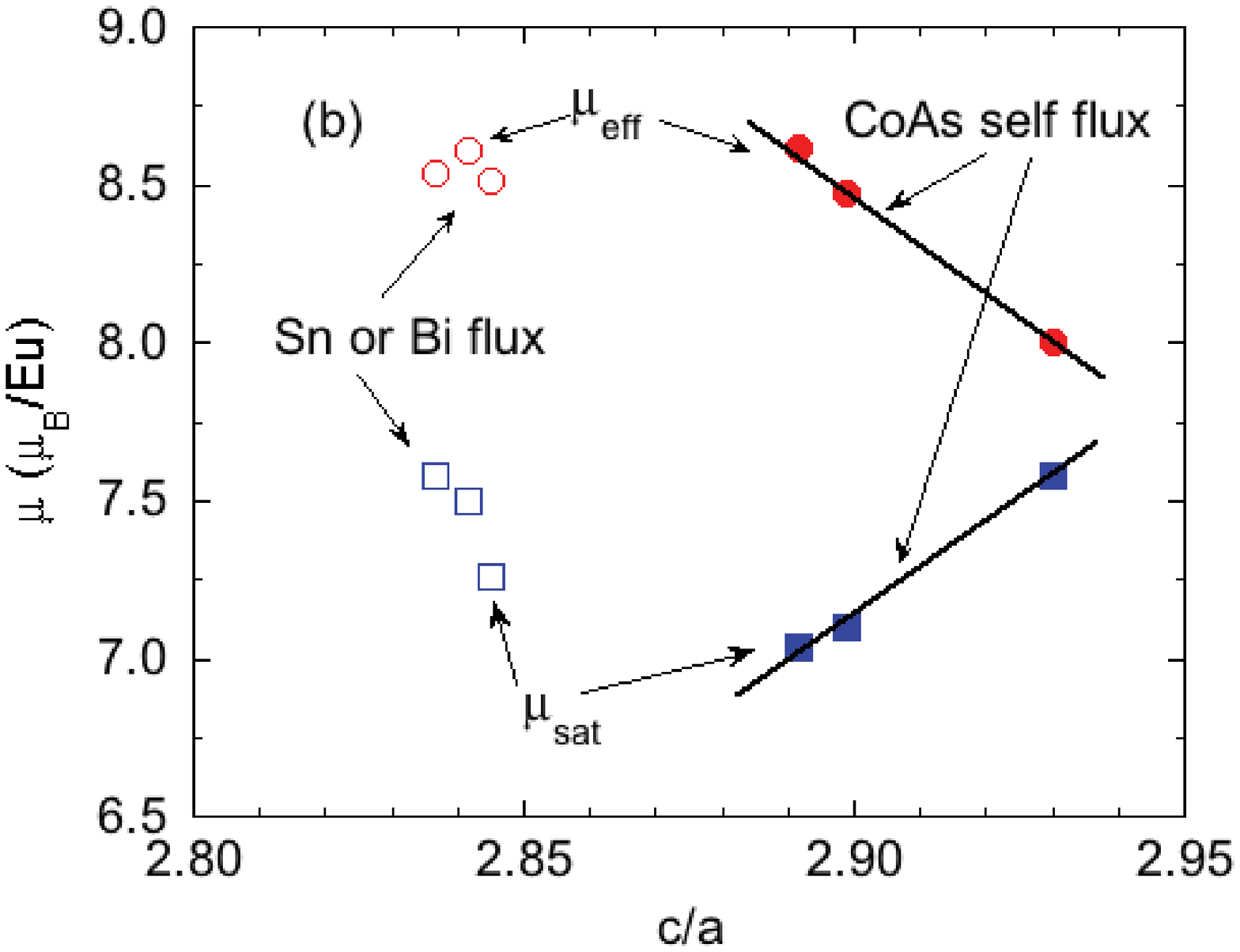}
\caption{(a) N\'eel temperature $T_{\rm N}$ versus crystallographic $c/a$ ratio for \eca\ crystals grown with Sn or Bi flux or with CoAs self-flux.   (b)~Effective moment $\mu_{\rm eff}$ and saturation moment $\mu_{\rm sat}$ versus $c/a$.  Data from Refs.~\cite{Tan2016} and~\cite{Raffius1993} are included.  The lines in (a) and~(b) are guides to the eye.}
\label{Fig:EuCo2As2_Data_Summary}
\end{figure}

Shown in Fig.~\ref{Fig:EuCo2As2_Data_Summary}(a) is a plot of $T_{\rm N}$ versus the tetragonal $c/a$ ratio obtained using the data in Tables~\ref{Tab:ChemAnal} and~\ref{Tab:MusatMueff}.  One sees an approximately linear positive correlation between $T_{\rm N}$ and~$c/a$.  On the other hand, the correlations between $\mu_{\rm eff}$ and $\mu_{\rm sat}$ versus $c/a$ show no clear correlation.

If one does not include a $T$-independent term $\chi_0$ when fitting the paramagnetic-state data by the Curie-Weiss law, negative curvature is usually observed in the $\chi^{-1}(T)$ plots which according to Fig.~\ref{Fig:CC_theta_Snflux} would then be attributed to an effective moment that increases with decreasing temperature.  We calculated an approximate value of $\chi_0$ which is negative but with a magnitude far smaller than the diamagnetic fitted values for our crystals.  This suggests that indeed the Curie constant and hence effective moment may be temperature-dependent, increasing with decreasing temperature.

Table~\ref{Tab:MusatMueff} also contains literature data for $\mu_{\rm eff}$ and $\mu_{\rm sat}$ for several other 122-type compounds containing Eu$^{+2}$ spins.  One sees that the respective values for all these compounds are less than the expected value.  This divergence between the values of the Eu moments in \eca\ and those of the other compounds starkly illustrates the anomalous enhancement of the Eu moments in \eca.

From Eqs.~(\ref{Eqs:mueffmusat}), enhancement of the Eu moment could arise from enhancement of $g$, of~$S$, or both.   Such an enhancement occurs in ferromagnetic Gd metal containing Gd$^{+3}$ ions with $S=7/2$, where the saturation moment at 4.2~K is $7.55(2)~\mu_{\rm B}$/Gd \cite{Nigh1963}.  This enhancement above the expected value $7~\mu_{\rm B}$/Gd was found from electronic structure calculations to arise from polarization of the conduction $d$-band electrons by the Gd spins \cite{Harmon1974}.  The enhancement is  similar to the maximum enhancements of the moment of isoelectronic Eu$^{+2}$ with $S=7/2$ in Table~\ref{Tab:MusatMueff}.  It has been inferred from neutron diffraction studies \cite{Raffius1993} that the Co atoms do not contribute to the ordered moment of \eca\ below $T_{\rm N}$.  It therefore seems likely that the effective spin value is increased by polarization  of the conduction carrier spins by the ordered Eu spins.  This expectation is indeed confirmed by our electronic structure calculations.

\acknowledgments

We thank Shalabh Gupta for the H$_2$ treatment of our Co powder starting material, George Lindemann for his help during the early stages of this work, and Michael Shatruk for sending us his published $\chi(T)$ data for an \ecaa\ crystal \cite{Tan2016}.  The research at Ames was supported by the U.S. Department of Energy, Office of Basic Energy Sciences, Division of Materials Sciences and Engineering.  Ames Laboratory is operated for the U.S. Department of Energy by Iowa State University under Contract No.~DE-AC02-07CH11358.


\end{document}